\documentclass[a4paper,11pt]{article}
\pdfoutput=1 

\usepackage{jheppub} 
\usepackage[T1]{fontenc} 
\usepackage{slashed}
\usepackage{subcaption}
\usepackage{ucs}
\usepackage{hyperref}
\usepackage[utf8x]{inputenc}
\usepackage{amsmath}
\usepackage{amsfonts}
\usepackage{comment}
\usepackage{amssymb}
\usepackage{array}
\usepackage{epsfig}
\usepackage{slashed}
\usepackage{lipsum}
\usepackage{hyperref}
\hypersetup{
	colorlinks=true,
	linkcolor=blue,
	filecolor=red,
	urlcolor=blue,
	citecolor=red
} 

\title{\boldmath Constraining  momentum space correlators  using slightly broken higher spin symmetry  }

\author[a]{Sachin Jain,}
\author[a,b,c]{ Renjan Rajan John,}
\author[a]{Vinay Malvimat}

\affiliation[a]{Indian Institute of Science Education and Research, Homi Bhabha Road, Pashan, Pune 411 008, India}
\affiliation[b]{Universit\`a del Piemonte Orientale, Dipartimento di Scienze e Innovazione Tecnologica, Viale T. Michel 11, I-15121 Alessandria, Italy}
\affiliation[c]{I.\,N.\,F.\,N. - sezione di Torino, Via P. Giuria 1, I-10125 Torino, Italy}

\emailAdd{\{sachin.jain,vinaymm\}@iiserpune.ac.in}
\emailAdd{renjan.rajan@to.infn.it}
\abstract{In this work, building up on \cite{Jain:2020rmw} we present momentum space Ward identities related to broken higher spin symmetry as an alternate approach to computing correlators of spinning operators in interacting theories such as the quasi-fermionic and quasi-bosonic theories. The direct Feynman diagram approach to computing correlation functions is intricate and in general has been performed only in specific kinematic regimes.  
We use higher spin equations to obtain the parity even and parity odd contributions to two-, three- and four-point correlators involving spinning and scalar operators in a general kinematic regime, and match our results with existing results in the literature for cases where they are available. 

One of the interesting facts about higher spin equations is that one can use them away from the conformal fixed point.
We illustrate this by considering mass deformed free boson theory and solving for two-point functions of spinning operators using higher spin equations.

}
\begin{document}
	\maketitle
\raggedbottom
\flushbottom
\section{Introduction}

Conformal Field Theory (CFT) is one of the cornerstones of theoretical physics. Among other things, it plays an important role in the well celebrated gauge-gravity correspondence, in explaining critical phenomena at phase transitions, and in defining  quantum field theories as deformations of theories with conformal symmetry. 

One of the most successful approaches to studying CFT has been the conformal bootstrap program in which one makes use of symmetries and imposes consistency conditions to constrain the space of CFTs. The program was first successfully applied in two dimensions in \cite{BPZ}. After several years of dormancy, the program was revived by the seminal work \cite{Rattazzi:2008pe} in which bootstrap techniques were successfully implemented to constrain CFTs in higher dimensions. Since then a plethora of work has appeared (see \cite{Rychkov_2017,Poland_2019,simmonsduffin2016tasi} and references therein), most of which has focussed on a position space analysis. 

While an understanding of CFTs in momentum space is desirable especially because of their applicability in the context of cosmology \cite{Mata_2013,Ghosh_2014,Kundu_2015,arkanihamed2015cosmological,arkanihamed2017cosmological,arkanihamed2018cosmological,Sleight:2019mgd,Sleight:2019hfp,baumann2019cosmological,baumann2020cosmological}, it is far less developed compared to its position space counterpart. One of the major obstacles is the lack of momentum space analogue of position space cross-ratios \cite{arkanihamed2018cosmological,Bzowski_2020,gillioz2020scattering}. 

Major steps in the study of momentum space correlators were initiated in \cite{Corian__2013,Bzowski_2014}. In these papers and in works that followed, specific spinning correlators were computed by solving momentum space conformal Ward identities. For example, three-point spinning correlators have been studied in \cite{Corian__2013,Bzowski_2014,Bzowski:2015yxv,Bzowski_2016,Bzowski_2018,Corian__2018,Corian__2019,Bautista_2020}. An alternate approach is to use the technique of weight-shifting and spin-raising operators of \cite{Costa:2011dw,Karateev:2017jgd,Costa:2018mcg} adapted to the momentum space context \cite{arkanihamed2017cosmological,baumann2019cosmological,baumann2020cosmological}. In the context of the AdS-CFT correspondence momentum space correlators have been computed in  \cite{Giombi:2009wh,Albayrak:2018tam,Albayrak:2019asr,Albayrak:2020isk,Albayrak:2020bso}. 

Parity odd correlators in momentum space have not received much attention \cite{Bonora:2015nqa,Bonora:2015odi,Bonora:2016ida,sissathesis,Jain:2020rmw}. See also the very recent work \cite{Jain:2021wyn} in which we computed parity odd correlators by solving momentum space conformal Ward identities and using weight-shifting and spin-raising operators. In three dimensions, parity odd correlators are important as they arise in the study of the free fermion theory and theories with slightly broken higher spin symmetry.  A class of theories with slightly broken higher spin symmetry is the Chern-Simons coupled to matter theory \cite{Aharony:2018npf,Maldacena:2011jn,Maldacena:2012sf,Aharony:2012nh,GurAri:2012is,Aharony:2011jz,Giombi:2011kc,Bedhotiya:2015uga,Giombi:2016zwa,Giombi:2017rhm,Yacoby:2018yvy,Skvortsov:2018uru,Aharony:2019mbc,Jain:2019fja,Jain:2012qi,Yokoyama:2012fa,Aharony:2012ns,Jain:2013py,Takimi:2013zca,Yokoyama:2013pxa,Gur-Ari:2015pca,Minwalla:2015sca,Geracie:2015drf,Gur-Ari:2016xff,Choudhury:2018iwf,Dey:2018ykx,Dey:2019ihe,Dandekar:2014era,Jain:2014nza,Inbasekar:2015tsa,Yokoyama:2016sbx,Inbasekar:2017ieo,Inbasekar:2017sqp,Halder:2019foo,Jain:2013gza,Aharony:2018pjn,Radicevic:2015yla,Aharony:2015mjs,Hsin:2016blu,Cordova:2017vab,Cordova:2018qvg,Turiaci:2018nua,Li:2019twz,Giveon:2008zn,Benini:2011mf,Park:2013wta,Aharony:2013dha,Kalloor:2019xjb}. Using momentum space conformal Ward identities to solve for correlators gets inefficient when one has to deal with complicated parity odd three-point correlators, and four or higher point correlators \cite{Isono_2018,Isono_2019,Gillioz:2018mto,Maglio:2019grh,corian2019fourpoint,Albayrak_2019,Bzowski_2020,Coriano:2020ccb,Serino:2020pyu,Coriano:2020ees,Bzowski:2020kfw}. We show that for theories with a (broken) higher spin symmetry, one can use higher spin equations to obtain a large class of spinning correlators including parity odd ones and higher point correlators. 

In \cite{Jain:2020rmw} we computed  momentum space spinning correlators in free boson and free fermion theory in three dimensions using  Ward identities associated with exact higher spin symmetry \cite{Maldacena:2011jn}. In the current work we extend our analysis to the case of interacting theories with slightly broken higher spin symmetry \cite{Maldacena:2012sf,Aharony:2012nh,GurAri:2012is}. Spinning correlators in these theories have both a parity even part and a parity odd part  \cite{Maldacena:2012sf}, and the higher spin currents are not exactly conserved but obey a non-conservation equation. We use the slightly broken Ward identities associated with these currents \cite{Maldacena:2012sf} along with the modified higher spin algebra to solve for the parity even and parity odd parts of two-, three- and four-point spinning correlators.
In our analysis we see that it is crucial to consider contact terms in various correlators to solve the higher spin equations in momentum space. This is in contrast to the case of higher spin equations in position space where one can ignore contact terms by working at non-coincident points. Higher spin equations in position space were used to determine spinning three-point correlators in \cite{Maldacena:2012sf} and for a specific four-point spinning correlator in \cite{Li:2019twz}.
Our results for the spinning correlators match the results obtained by explicit computation in special kinematic regimes using Feynman diagram approach and other CFT techniques in \cite{Aharony:2012nh,GurAri:2012is,Caron-Huot:2017vep,Turiaci:2018nua}. We also present arguments based on conformal bootstrap to determine the five-point function of scalar operators in an interacting theory with slightly broken higher spin symmetry \footnote{Although not fully rigorous we believe the arguments we present shed some light on the structure of the five-point correlator.}.

We then consider theories away from conformality through relevant deformations of the conformal theory. In particular we consider mass deformed free boson theory and compute the two-point function of spinning operators using higher spin equations. We also mention an algorithm to compute arbitrary two-point functions in such theories using higher spin equations. 

The rest of the paper is arranged as follows. In Section \ref{HigherSpinEquations} we review some details of position space higher spin equations in free and interacting theories. We also discuss the higher spin algebra and an example of higher spin equation in momentum space. In Section \ref{freetheories} we consider momentum space higher spin equations for two-, three-, and four-point correlators in the free boson and fermion theories. In Section \ref{InteractingTheories} we consider interacting theories with slightly broken higher spin symmetry such as the critical bosonic, quasi-fermionic and quasi-bosonic theories and use broken higher spin equations to compute spinning correlators in momentum space.  In Section \ref{Massive} we move away from conformality and use higher spin equations to determine two-point functions in the massive free boson theory. In Section \ref{Summary1} 
we give a summary our main results.  We end with a conclusion and discussion in Section \ref{Summary}. 

We collect a lot of technical details in six appendices. In Appendix \ref{notation} we introduce the notations and normalizations that we follow in this paper.  In Appendix \ref{SchoutenIdentity} we give a Schouten identity that was extremely useful in our computations.  In Appendix \ref{5ptscalar} we derive an expression for the momentum space five-point function of the scalar operator in the quasi-fermionic theory. In Appendix \ref{CPWS} we give the derivation of a conformal partial wave in momentum space. In Appendix \ref{parityodd} we derive constraints for specific three- and four-point parity odd spinning correlators in both quasi-bosonic and quasi-fermionic theories. In Appendix \ref{massive-app} we give some details involved in the computation of correlators in the free massive bosonic theory.

\section{Higher Spin Equations}
\label{HigherSpinEquations}
In this section we begin with a brief review of  higher spin equations in  theories with slightly broken higher spin symmetry. 

\subsection{Free Theories}
To begin with let us consider theories with exact higher spin symmetry. The operator spectrum of these theories contain a tower of conserved currents $J_s$ for every integer spin $s>0$ \cite{Maldacena:2011jn}.  Corresponding to each one of these currents there exists a conserved charge $Q_s$ defined as follows :
\begin{align}
Q_s=\int_{x^{+}=\text{const}}\,dx^{-}\,dy\,J_{\underbrace{--\ldots -}_{s\,\text{times}}}
\end{align}
We work in light-cone coordinates in which the metric takes the form: $ds^2=dx^+dx^-+dy^2$. Let us now consider the action of such a charge  $Q_s$ on the correlator $\langle J_{s_1}(x_1)\ldots J_{s_i}(x_i)  \ldots J_{s_n}(x_n)\rangle$
\begin{align}
\sum_{i=1}^{n}\langle J_{s_1}(x_1) \ldots[Q_{s},J_{s_i}(x_i) ] \ldots J_{s_n}(x_n)\rangle=0\label{HSEFreeG}
\end{align}
The R.H.S of the above equation vanishes only in theories with exact higher spin symmetry. The algebra $[Q_{s},J_{s_i}]$  is fixed by twist conservation as well as parity and charge conjugation symmetry. The equation obtained by utilizing the algebra of $Q_s$ with various operators $J_{s_i}$ in the above equation is known as  the higher spin equation. In this article, the theories with exact higher symmetry we will consider are the free bosonic and free fermionic theories. The momentum space higher spin equations in these theories is obtained by performing a Fourier transform of \eqref{HSEFreeG}.

We will use the action of two such conserved charges $Q_3$ and $Q_4$ on various correlators in free theories to derive higher spin equations. While $Q_3$ corresponds to the odd current $J_3$ which changes sign under charge conjugation, $Q_4$ corresponds to the even current $J_4$ which does not\footnote{In free fermion theory for example, charge conjugation is defined by $\psi \rightarrow i \sigma^{2} \psi^{*},\, \psi^{\dagger} \rightarrow i \psi^{T} \sigma^{2}$ (see Appendix of \cite{Gur-Ari:2016xff} for a detailed discussion  on odd spins and charge conjugation).}. Hence the algebra  $[Q_3,J_{s}^{\text{odd}}]$ must contain only even spins where as  $[Q_3,J_{s}^\text{{even}}]$ must contain only odd spins to preserve charge conjugation symmetry. As the notation has it clear, $J_{s}^\text{{odd}}$ and $J_{s}^\text{{even}}$ correspond to spin $s$ operators with odd and even parity respectively. In \cite{Maldacena:2011jn} and in most of the studies that followed, the charge $Q_4$ was considered. In our work, we consider both $Q_3$ and $Q_4$.  

\subsubsection{Free boson theory}
\label{FBTsection}
Free complex bosonic theory in three dimensions has a scalar operator $J_0$ with dimension $\Delta=1$ and an infinite tower of exactly conserved currents $J_s$ with integer spin $s$ and dimension $\Delta=s+1$.
The algebra of charges $Q_3$ and $Q_4$ with  $J_0$ and $J_1$ is given by :
\begin{align}
[Q_3,J_0]&=\,\partial_{-}J_{-}\nonumber\\[5pt]
[Q_3,J_-]&=4\,\partial_{-}T_{--}-\frac{1}{2}\,\partial_{-}^3J_0\label{FBQ3algebra}\\[5pt]
[Q_4,J_0]&=\,\partial_{-}^3J_0-\frac{24}{5}\,\partial_-T_{--}\label{FBQ4algebra}
\end{align}
Let us now come to the free fermion theory in three dimensions.
\subsubsection{Free fermion theory}
Free fermion theory in three dimensions has a pseudo-scalar operator $J_0$ with dimension $\Delta=2$ and an infinite tower of exactly conserved currents $J_s$ with integer spin $s$ and dimension $\Delta=s+1$.
The algebra of charges $Q_3$ and $Q_4$ with  $J_0$ and $J_1$ is as follows :
\begin{align}
\label{FFQ3algebra}
[Q_3,J_0]&=\,\epsilon_{-\mu\nu}\,\partial^{\mu}\,\partial_{-}J^\nu\cr
[Q_3,J_-]&=2\,i\,\partial_{-}T_{--}\notag\\
[Q_4,J_0]&=\,\partial_{-}^3J_0+\,\epsilon_{-\mu\nu}\,\partial_{-}\,\partial^{\mu}\,T^\nu_{-}
\end{align}
Using a dimension and spin analysis, keeping in mind parity and charge conjugation symmetry one can write down the algebra for higher spin charges $Q_s$, for $s>4$. However, we will not use them in our work. Note that in free theories the algebra of the generators with the operators can also be obtained by Wick contraction.

\subsection{Interacting Theories}
In contrast to free theories, for theories with slightly broken higher spin symmetry the currents $J_s$, $s>3$ obey a non-conservation equation and the corresponding charges are not exactly conserved \cite{Maldacena:2012sf,Giombi:2016zwa}. The action of such a charge on  $\langle J_{s_1}(x_1)\ldots J_{s_i}(x_i)  \ldots J_{s_n}(x_n)\rangle$ results in the following equation :
\begin{align}\label{HSEIT}
\sum_{i=1}^{n}\langle J_{s_1}(x_1) \ldots[Q_{s},J_{s_i}(x_i) ] \ldots J_{s_n}(x_n)\rangle=\int_x\langle \partial_{\mu}J^{\mu}_{\underbrace{-\ldots -}_{(s-1)\,\text{times}}}(x)J_{s_1}(x_1)\ldots J_{s_i}(x_i)  \ldots J_{s_n}(x_n)\rangle
\end{align}
In this article we mainly focus on specific interacting conformal field theories with slightly broken higher spin symmetry as discussed in \cite{Maldacena:2012sf}.  As in the case of theories with exact higher spin symmetry, in most of the studies of theories with broken higher spin symmetry \cite{Maldacena:2012sf,Li:2019twz}, the pseudo-charge $Q_4$ was considered.  In this article, we solve for spinning correlators in momentum space using higher spin equations resulting from the action of $Q_3$ and $Q_4$ on specific correlators.

Below we consider interacting theories such as the critical bosonic theory, the quasi-fermionic theory and the quasi-bosonic theory. Quasi-fermionic theory refers to two theories, the regular fermion coupled to Chern-Simons theory and the critical boson coupled to Chern-Simons theory, and the two theories are dual to each other. The strong coupling limit of the quasi-fermionic theory is the critical bosonic theory. Quasi-bosonic theory refers to two dual theories, the regular boson coupled to Chern-Simons theory as well as the critical fermion coupled to Chern-Simons theory. In addition to the dualities, one has in the large $N$ limit that the regular boson coupled to Chern-Simons theory and the critical boson coupled to Chern-Simons theory are related by a Legendre transformation. Similarly, there is a Legendre transformation that relates the regular fermion coupled to Chern-Simons theory and the critical boson coupled to Chern-Simons theory in the large-$N$ limit \cite{Giombi:2016zwa,Aharony:2012nh,GurAri:2012is}.

%
\subsubsection{Critical bosonic theory}
The spectrum of single trace primary operators in the critical bosonic theory contains a spin-0 operator $\widetilde{J}_{0}$ with conformal dimension $\Delta=2+\mathcal O(\frac 1N)$, spin-one and spin-two currents with dimensions 2 and 3 respectively  and an infinite tower of spinning operators $J_s$ with integer spin $s>3$ and dimension $\Delta=s+1+\mathcal O(\frac 1N)$. Here $N$ is the number of scalars in the free theory which we deform with a $\phi^4$ interaction to get the critical theory.

The action of $Q_4$  on the scalar operator $\widetilde{J}_{0}$ is as follows \cite{Giombi:2016zwa} : 
\begin{align}\label{Q4algebraCB}
\left[Q_{4}, \widetilde{J}_{0}\right]&= \partial_{-}^{3} \widetilde{J}_{0}
\end{align}
The non-conservation equation for the current $J_4$ is :
\begin{align}
\partial_{\mu} J_{---}^{\mu}&=\left(s_0\partial_{-} \widetilde{J}_{0} T_{--}+s_1 \widetilde{J}_{0} \partial_{-} T_{--}\right) \label{DivJ4CB}
\end{align}
where $s_0$ and $s_1$ are given by
\begin{align}
s_0=\frac{30}{7} ,~~
s_1=-\frac{12}{7}\label{c0c1CB}
\end{align}
\subsubsection{Quasi-fermionic theory}
The spectrum of single trace primary operators in this theory contains a spin-0 operator $\widetilde{J}_{0}$ with conformal dimension $\Delta=2+\mathcal O(\frac 1N)$, spin-one and spin-two currents with dimensions 2 and 3 respectively  and an infinite tower of spinning operators $J_s$ with integer spin $s>3$ and dimension $\Delta=s+1+\mathcal O(\frac 1N)$. The algebra of $Q_3$ and  $Q_4$ with  $J_0$ and $J_1$ is given by
 \begin{align}\label{QFQ3algebra}
 \left[Q_{3}, J_{0}\right]&=\epsilon_{-\mu \nu}\,\partial^{\mu} \partial_{-} J^{\nu}\notag \\
 \left[Q_{3}, J_{-}\right]&=2\,i\,\partial_{-} T_{--}\notag\\ 
 [Q_4,J_0]&=\,\partial_{-}^3J_0+\,\,\epsilon_{-\mu\nu}\,\partial_{-}\,\partial^{\mu}\,T^\nu_{-}
 \end{align}
 %
 For $Q_4$ we utilize the algebra in \cite{Maldacena:2012sf} except for the normalisation of $J_0$. Our normalisation is such that  $J_0=J_0^{MZ}(1+ \widetilde{\lambda}_{\text{qf}}^2)$ where $J_0^{MZ}$ is the scalar operator in 
 \cite{Maldacena:2012sf}. 
In the limit $\widetilde\lambda_{\text{qf}}\rightarrow \infty$ we get the critical bosonic algebra \eqref{Q4algebraCB}. For this we require $\widetilde J_0=4\pi\lambda_f J_0$.
 
 The current non-conservation equations for  $J_3$ and $J_4$ at the leading order in large $N$ are respectively given by \cite{Giombi:2016zwa}
 \begin{align}
 \partial_{\mu} J^{\mu}_{--}&=-16\,i\,\left(r_0\partial_{-}J_0 J_{-}+r_1\partial_{-}J_{-} J_{0}\right)\label{DivJ3AF}\\
 \partial_{\mu} J^{\mu}_{---}&=\frac{480}{7}\left(c_0\partial_{-}J_0 T_{--}+c_1\partial_{-}T_{--} J_{0}\right)\label{DivJ4AF}
 \end{align}
 where 
 \begin{align}
 r_0&=\frac{3\widetilde\lambda_{\text{qf}}}{5\widetilde N(1+\widetilde\lambda_{\text{qf}}^2)},\quad   r_1=-\frac{2\widetilde\lambda_{\text{qf}}}{5\widetilde N(1+\widetilde\lambda_{\text{qf}}^2)}\\
 c_0&=\frac{\widetilde\lambda_{\text{qf}}}{\widetilde N(1+\widetilde\lambda_{\text{qf}}^2)},\quad c_1=-\frac{2\widetilde\lambda_{\text{qf}}}{5\widetilde N(1+\widetilde\lambda_{\text{qf}}^2)}\label{c0c1}
 \end{align}
 \subsubsection{Quasi-bosonic theory}
 The spectrum of single trace primary operators in this theory contains a spin-0 operator ${J}_{0}$ with conformal dimension $\Delta=1+\mathcal O(\frac 1N)$, spin-one and spin-two currents with dimensions 2 and 3 respectively  and an infinite tower of spinning operators $J_s$ with integer spin $s>3$ and dimension $\Delta=s+1+\mathcal O(\frac 1N)$. 
The algebra of $Q_3$ and  $Q_4$ with  $J_0$ and $J_1$ is given by :
\begin{align}
[Q_3,J_0]&=\partial_{-}J_{-}\nonumber\\[5pt]
[Q_3,J_-]&=4\,\widetilde d_{12}\,\partial_{-}T_{--}-\frac{1}{2}\,\partial_{-}^3J_0\label{QBQ3algebra},\quad\text{where}\quad\widetilde d_{12}=1+\frac{\widetilde\lambda_{\text{qb}}^2}{1+\widetilde\lambda_{\text{qb}}^2}
\\[5pt]
[Q_4,J_0]&=\partial_{-}^3J_0-\frac{24}{5}\,\partial_-T_{--}\label{QBQ4algebra}
\end{align}
The free theory limit ($\widetilde\lambda\rightarrow 0$) of the above algebra gives the algebra in the free boson theory given in Section \ref{FBTsection}. The current non-conservation equation for $J_3$ at the leading order in large $N$ is given by \cite{Giombi:2016zwa}:
\begin{align}
\partial_{\mu} J^{\mu}_{--}=-\frac{\epsilon_{\mu\nu-}}{10}\,\left[f_0\,J^\mu\,\partial^\nu\,\partial_{-}J_0+b_1\,\partial^\nu\,J^\mu\,\partial_{-}J_0+f_1\,\partial_{-}J^\mu\,\partial^\nu J_0+b_2\,J_0\partial_{-}\partial^\nu J^\mu\right]\label{DivJ3QB}
\end{align}
where $b_i$ and $f_i$ are given by :
\begin{align}
b_1=-\frac{512i \widetilde{\lambda}_{\text{qb}}}{3\widetilde N\,(1+ \widetilde{\lambda}_{\text{qb}}^2)},\quad f_1=-\frac{128i \widetilde{\lambda}_{\text{qb}}}{3\widetilde N\,(1+ \widetilde{\lambda}_{\text{qb}}^2)},\quad f_0=\frac{64i \widetilde{\lambda}_{\text{qb}}}{\widetilde N\,(1+ \widetilde{\lambda}_{\text{qb}}^2)},\quad b_2=\frac{128i \widetilde{\lambda}_{\text{qb}}}{3\widetilde N\,(1+ \widetilde{\lambda}_{\text{qb}}^2)}
\end{align}
The current equation for $J_4$ at the leading order in large $N$ is given by \cite{Giombi:2016zwa}:
\begin{align}
\partial_{\mu} J^{\mu}_{---}=\epsilon_{\mu\nu-}\,\left[f_0\,T^\mu_-\,\partial^\nu\,\partial_{-}J_0+b_1\,\partial^\nu\,T^\mu_-\,\partial_{-}J_0+f_1\,\partial_{-}T^\mu_-\,\partial^\nu J_0+b_2\,J_0\partial_{-}\partial^\nu T^\mu_-\right]\label{DivJ4QB}
\end{align}
where $b_i$ and $f_i$ are given by :
\begin{align}
b_1=-\frac{192\,i\,\widetilde{\lambda}_{\text{qb}}}{\widetilde N\,(1+ \widetilde{\lambda}_{\text{qb}}^2)},\quad f_1=-\frac{64\,i\, \widetilde{\lambda}_{\text{qb}}}{\widetilde N\,(1+ \widetilde{\lambda}_{\text{qb}}^2)},\quad f_0=\frac{160\,i\, \widetilde{\lambda}_{\text{qb}}}{\widetilde N\,(1+ \widetilde{\lambda}_{\text{qb}}^2)},\quad b_2=\frac{32\,i\, \widetilde{\lambda}_{\text{qb}}}{\widetilde N\,(1+ \widetilde{\lambda}_{\text{qb}}^2)}
\end{align}
Let us now turn our attention to higher spin equations in momentum space which is the main focus of our work. 
\subsection{Higher spin equations in momentum space}
Momentum space higher spin equations are determined by performing a Fourier transform of the corresponding position space equations \eqref{HSEFreeG} and \eqref{HSEIT}.
As a first simple example we briefly describe one such momentum space higher spin equation  constructed by the action of $Q_4$ on $\langle J_0J_0J_0\rangle$ in the quasi-fermionic theory :
\begin{align}
\langle [Q_4, J_0(x_1)]J_0(x_2)J_0(x_3)\rangle_\text{QF}+1\leftrightarrow 2+1\leftrightarrow 3=\int_x \langle\partial_{\mu} J^{\mu}_{---}(x)J_0(x_1)\,J_0(x_2)\,J_0(x_3)\rangle_\text{QF}
\end{align}
where $1\leftrightarrow 2$ and $1\leftrightarrow 3$ represent exchanges $x_1\leftrightarrow x_2$ and $x_1\leftrightarrow x_3$ respectively.
Upon utilizing the algebra in \eqref{QFQ3algebra} and the current equation \eqref{DivJ3AF} we get :
\begin{align}
\label{Q4onJ000QF}
\bigg[&\langle \partial_{-}^3 J_0(x_1)\,J_0(x_2)\,J_0(x_3)\rangle_\text{QF}
+\epsilon_{-\mu\nu}\,\langle\partial_{-}\partial^\mu T^\nu_{-}(x_1)J_0(x_2) J_0(x_3)\rangle_\text{QF}\bigg]+1\leftrightarrow 2+1\leftrightarrow 3\notag\\&\hspace{3.8cm}=\frac{96\,\widetilde\lambda_{\text{qf}}}{1+\widetilde\lambda_{\text{qf}}^2}\int_x \langle \partial_{-}J_0\,T_{--}(x)J_0(x_1)\,J_0(x_2)\,J_0(x_3)\rangle_\text{QF}
\end{align}
where we performed integration by parts on the R.H.S.
Upon performing the Fourier transform we get the following equation in momentum space :
\begin{align}
&-i\,(k_{1-}^3+k_{2-}^3+k_{3-}^3)\langle \langle J_0(k_1) J_0(k_2) J_0(k_3)\rangle \rangle_{\text{QF}}\notag\\&-\bigg[k_{1-}\epsilon_{-k_1\nu}\langle\langle T^\nu_{-}(k_1)J_0(k_2)J_0(k_3)\rangle \rangle_{\text{QF}}+1\leftrightarrow 2+1\leftrightarrow 3\bigg]\notag\\=
&\frac{96\,\widetilde\lambda_{\text{qf}}}{1+\widetilde\lambda_{\text{qf}}^2}\Bigg[k_{1-}\,\langle J_0(k_1) J_0(-k_1)\rangle_{\text{QF}}\langle \langle  T_{--}(k_1)J_0(k_2)J_0(k_3)\rangle \rangle_{\text{QF}}+1\leftrightarrow 2+1\leftrightarrow 3\Bigg]
\end{align}
where  $1\leftrightarrow 2$ and $1\leftrightarrow 3$ represent exchanges $k_1\leftrightarrow k_2$ and $k_1\leftrightarrow k_3$ respectively. 
We also used the large $N$ factorisation of the five-point correlator that appears on the RHS of \eqref{Q4onJ000QF}. Throughout this paper we will work in the large $N$ limit and at the leading order in $N$. This will help us use such factorisations of higher point correlation functions.
The double bracket in the above equation represents the momentum space correlator with the momentum conserving delta function stripped off. 
\section{Free Theories}
\label{freetheories}
In this section we consider free theories with exact higher spin symmetry. In particular we will consider the action of the conserved charges $Q_3$ and $Q_4$ on various momentum space two-, three-, and four-point correlators in the free bosonic and the free fermionic theories. Using the resulting higher spin equations, and some of the known results from \cite{Jain:2020rmw}  we determine or constrain spinning  correlators in these theories.

\subsection{Free Boson Theory}

The currents $J_s$, in particular $J_3$ and $J_4$ are conserved in the free theory and hence the right hand side of the higher spin equation is always zero.
\subsubsection{Two-point functions}
\label{FB2ptJJeqn}
We begin by considering the simple example of a higher spin equation involving  two-point functions resulting from the action of $Q_3$ on the correlator $\langle J_\mu J_0\rangle$ in the free boson theory. 

\subsubsection*{$\boldsymbol{Q_3}$ on  $\boldsymbol{\langle J_1 J_0 \rangle_\text{FB}}$ : Fixing $\boldsymbol{\langle J_\mu J_\nu \rangle_\text{FB}}$}
Consider the action of $Q_3$ on the two-point function $\langle J_- J_0\rangle$ :
\begin{align}
\langle [Q_3, J_-(x_1)]J_0(x_2)\rangle_\text{FB}+\langle J_-(x_1) [Q_3, J_0(x_2)]\rangle_\text{FB}=0
\end{align}
From the algebra in \eqref{FBQ3algebra}, we get the following higher spin equation :
\begin{align}
-\frac{1}{2}\,\langle \partial_{-}^3J_0(x_1)J_0(x_2)\rangle_\text{FB}+\,\langle J_{-}(x_1)\partial_{-}J_{-}(x_2)\rangle_\text{FB}=0
\end{align}
where we have used the fact that the two-point function $\langle T J_0\rangle=0$. The higher spin equation in momentum space is obtained through a Fourier transform of the above  : 
\begin{align}
\frac{1}{2}k_{1-}^3\langle J_0(k_1) J_0(-k_1)\rangle_{\text{FB}}- k_{1-}\langle J_-(k_1) J_-(-k_1)\rangle_{\text{FB}}=0
\end{align}
Substituting for the scalar correlator $\langle J_0 J_0\rangle_{\text{FB}}$, we  solve this to determine $\langle J_- J_-\rangle_{\text{FB}}$  :
\begin{align}
\langle J_-(k_1) J_-(-k_1)\rangle_{\text{FB}}=\frac{k_{1-}^2}{16 k_1} \label{JmJmFBH}
\end{align}
The transverse Ward identity fixes the form of the correlator to be :
\begin{align}
\langle J_{\mu}(k_1) J_\nu(-k_1)\rangle_{\text{FB}}&=A_1\left(g_{\mu\nu}-\frac{k_{1\mu} k_{1\nu}}{k_1^2}\right)\
\end{align}
We now compare the above equation with \eqref{JmJmFBH} and get the form factor $A_1$ to be :
\begin{align}
A_1=-\frac{1}{16}k_1
\end{align}
This completely fixes the two-point spinning correlator $\langle J_\mu J_\nu\rangle_\text{FB}$ in momentum space.

\subsubsection{Three-point functions}
Recently in \cite{Jain:2020rmw}, we obtained a few momentum space spinning correlators in the free boson theory by solving the higher spin equations resulting from the action of  ${Q_3}$ on  ${\langle J_- J_0 J_0\rangle}$  and ${Q_4}$ on  ${\langle J_0 J_0 J_0\rangle}$.  

\subsubsection*{$\boldsymbol{Q_3}$ on  $\boldsymbol{\langle J_1 J_1 J_1\rangle_\text{FB}}$: Constraining  $\boldsymbol{\langle TJ_1J_1\rangle_\text{FB}}$}

Consider the action of the charge $Q_3$ on $\langle J_- J_- J_-\rangle$ :
\begin{align}
\langle [Q_3, J_-(x_1)]J_-(x_2)J_-(x_3)\rangle_\text{FB}+1\leftrightarrow 2+1\leftrightarrow 3=0
\end{align}
Upon utilizing the higher spin algebra \eqref{FBQ3algebra} we get :
\begin{align}
\bigg[\langle \partial_{-} T_{--}(x_1)J_-(x_2) J_-(x_3)\rangle_\text{FB}-\frac{1}{8 } \langle \partial_{-}^{3} J_{0}(x_1) J_-(x_2)J_-(x_3)\rangle_\text{FB}+1\leftrightarrow 2+1\leftrightarrow 3\bigg]=0
\end{align}
 Upon performing the Fourier transform of the above equation, we get
\begin{align}\label{TJJFB}
\bigg[k_{1-}\langle\langle  T_{--}(k_1)J_-(k_2) J_-(k_3)\rangle\rangle_\text{FB}&+\frac{1}{8 } (k_{1-}^{3}+k_{2-}^{3}+k_{2-}^{3})  \langle\langle J_{0}(k_1) J_-(k_2)J_-(k_3)\rangle\rangle_\text{FB}\notag\\[5pt]
&+1\leftrightarrow 2+1\leftrightarrow 3\bigg]=0
\end{align}
One could check that the above equation is satisfied by the $\langle TJ_1J_1\rangle$ correlator  obtained by explicit computation. However, we cannot solve the above equation for $\langle TJ_1J_1\rangle$ as the unknowns (including those coming from the exchanges) exceed the equations that constrain them. However, we will use the above equation  to simplify the higher spin equation in the interacting theory. In the interacting theory we will once again see that there are not enough number of equations to solve for $\langle TJ_1J_1\rangle $.
\subsubsection{Four-point functions}
In this subsection we derive constraints imposed by higher spin equations on four-point correlators in the free boson theory. Note that the higher spin equation resulting from  action of ${Q_4}$ on  ${\langle J_0 J_0 J_0 J_0\rangle_\text{FB}}$ was demonstrated to be satisfied in \cite{Jain:2020rmw}.  Here we consider a different higher spin equation resulting from action of ${Q_3}$ on  ${\langle J_1 J_0 J_0 J_0\rangle_\text{FB}}$.

\subsubsection*{$\boldsymbol{Q_3}$ on  $\boldsymbol{\langle J_1 J_0 J_0 J_0\rangle_\text{FB}}$: Constraining $\boldsymbol{\langle J_{1}J_1 J_0J_0\rangle_\text{FB}}$}

Consider the action of $Q_3$ on the correlator  $\langle J_-J_0J_0J_0\rangle_\text{FB}$ :
\begin{align}
&\langle [Q_3, J_-(x_1)]J_0(x_2)J_0(x_3)J_0(x_4)\rangle_\text{FB}\notag\\[5pt]
&\hspace{1cm}+\langle J_-(x_1) [Q_3, J_0(x_2)]J_0(x_3)J_0(x_4)\rangle_\text{FB}+2\leftrightarrow 3+2\leftrightarrow 4=0
\end{align}
Upon utilizing the algebra in \eqref{FBQ3algebra} this leads us to
\begin{align}
4 &\langle \partial_{-} T_{--}(x_1)J_0(x_2)J_0(x_3)J_0(x_4)\rangle_\text{FB}-\frac{1}{2}\langle \partial_{-}^{3} J_{0}(x_1)J_0(x_2)J_0(x_3)J_0(x_4)\rangle_\text{FB}\notag\\[5pt]&+ \langle   J_{-}(x_1) \partial_{-} J_{-}(x_2)J_0(x_3)J_0(x_4)\rangle_\text{FB} +2\leftrightarrow 3+2\leftrightarrow 4
=0
\end{align}
Upon performing the Fourier transform we get
\begin{align}
4\,k_{1-}&\langle \langle T_{--}(k_1)J_0(k_2)J_0(k_3)J_0(k_4)\rangle\rangle_\text{FB}+\frac{1}{2}k_{1-}^{3}\langle \langle J_{0}(k_1)J_0(k_2)J_0(k_3)J_0(k_4)\rangle\rangle_\text{FB}\notag\\[5pt]&+ k_{2-}\langle\langle   J_{-}(k_1) J_{-}(k_2)J_0(k_3)J_0(k_4)\rangle\rangle_\text{FB} +2\leftrightarrow 3+2\leftrightarrow 4
=0
\end{align}
The above equation could not be solved but verified to be satisfied in \cite{Jain:2020rmw}. This is because  the number of unknowns were much more than the number of equations. We will see in the next section that this will continue to be true in the interacting theory as well .
\subsection{Free Fermion Theory}
We will now proceed to derive some of the higher spin equations in the  free fermion theory.

\subsubsection{Two-point functions}
\label{2ptfnFF}
\subsubsection*{$\boldsymbol{Q_3}$ on  $\boldsymbol{\langle J_1 J_0 \rangle_\text{FF}}$ : A consistency check}
Consider the action of $Q_3$ on  $\langle J_- J_0 \rangle_\text{FF}$ :
\begin{align}
\langle [Q_3, J_-(x_1)]J_0(x_2)\rangle_\text{FF}+\langle J_-(x_1) [Q_3, J_0(x_2)]\rangle_\text{FF}=0
\end{align}
From the algebra in \eqref{FFQ3algebra}, we have the following higher spin equation
\begin{align}
2\,i\,\langle \partial_{-}T_{--}(x_1)J_0(x_2)\rangle_\text{FF}+\,\epsilon_{-\mu\nu}\,\langle J_-(x_1)\partial^{\mu}\,\partial_{-}J^\nu(x_2)\rangle_\text{FF}=0
\end{align}
The first term is zero as the two-point function $\langle TJ_0\rangle=0$ from conformal invariance. Upon performing the Fourier transform we get :
\begin{align}
\,\epsilon_{-k_1\nu}\,k_{1-}\,\langle J_-(k_1)J^\nu(-k_1)\rangle_\text{FF}=0
\end{align}
The contraction of the correlator with the Levi-Civita tensors ensures that the above higher spin equation is trivially satisfied, and the analysis leads to no constraints. However, it serves as a consistency check of the higher spin algebra  \eqref{FFQ3algebra}.

\subsubsection{Three-point functions}
\label{3ptFFHSE}
A few three-point spinning correlators in the free fermion theory were obtained by explicit computation in \cite{Jain:2020rmw}. 

\subsection*{$\boldsymbol{Q_3}$ on $\boldsymbol{\langle J_1J_0J_0\rangle_\text{FF}}$ : Fixing the contact term in $\boldsymbol{\langle TJ_0J_0\rangle_\text{FF}}$}

The action of $Q_3$ on $\langle J_-J_0J_0\rangle$ gives :
\begin{align}
\langle [Q_3, J_-(x_1)]J_0(x_2)J_0(x_3)\rangle_\text{FF}+\left(\langle J_-(x_1)[Q_3, J_0(x_2)]J_0(x_3)\rangle_\text{FF}+2\leftrightarrow 3\right)=0
\end{align}
Using the algebra in \eqref{FFQ3algebra} and performing a Fourier transform of the resulting equation :
\begin{align}
\label{HSEFFQ3-3pt}
2\,k_{1-}\langle\langle T_{--}(k_1)J_0(k_2)J_0(k_3)\rangle\rangle_{\text{FF}}+\left(\epsilon_{-k_2\nu}\,k_{2-}\,\langle \langle J_-(k_1)J_\nu(k_2) J_0(k_3\rangle\rangle_{\text{FF}}+2\leftrightarrow 3\right)=0
\end{align}
The correlator $\langle TJ_0J_0\rangle_\text{FF}$ has a transverse part, a local part and we allow for a contact term :
\begin{align}
\langle TJ_0J_0\rangle_\text{FF}=\langle TJ_0J_0\rangle_\text{FF}^{\text{transverse}}+ \langle TJ_0J_0\rangle_\text{FF}^{\text{local}}+\langle TJ_0J_0\rangle_\text{FF}^{\text{contact}}
\end{align}
The explicit expressions for the transverse and local parts of the correlator were obtained in \cite{Jain:2020rmw}. Solving the higher spin equation \eqref{HSEFFQ3-3pt} requires :
\begin{align}\label{TOOcontactF}
\langle T_{\mu\nu}J_0J_0\rangle_{\text{FF}}^{\text{contact}}=\frac{1}{16}\,\frac{k_{1\mu}k_{1\nu}}{k_1}
\end{align}
%
%
Thus we see that the higher spin analysis in this case helps fix the contact term in the $\langle TJ_0J_0\rangle$ correlator in the free fermion theory. We note that this contact term is in addition to the contact terms in the correlator which come from conservation Ward identities in momentum space.

\subsection*{$\boldsymbol{Q_3}$ on $\boldsymbol{\langle J_1J_1J_1\rangle_\text{FF}}$ : Constraining $\langle J_1J_1T\rangle$}
Consider the action of $Q_3$ on $\langle J_- J_- J_-\rangle$ in the free fermion theory. This gives us
\begin{align}
\langle[Q_{3}, J_{-}(x_1)]J_-(x_2)J_-(x_3)]\rangle+1\leftrightarrow 2+1\leftrightarrow 3=0
\end{align}
Utilizing the algebra in \eqref{FFQ3algebra} we get
\begin{align}
\langle  \partial_{-} T_{--}(x_1)J_-(x_2) J_-(x_3)\rangle_\text{FF}+1\leftrightarrow 2+1\leftrightarrow 3=0
\end{align}
Upon Fourier transforming we obtain the following equation in the momentum space :
\begin{align}\label{TJJFF}
k_{1-}\langle \langle  T_{--}(k_1)J_-(k_2) J_-(k_3)\rangle\rangle_\text{FF}+1\leftrightarrow 2+1\leftrightarrow 3=0
\end{align}
As in the bosonic theory we will not solve the above equation. Instead, we will utilize it to simplify the parity even part of the corresponding equation in the interacting theory.
\subsubsection{Four-point functions}
Let us now consider four-point functions in the free fermion theory.
\subsection*{$\boldsymbol{Q_4}$ on $\boldsymbol{\langle J_0J_0J_0J_0\rangle_\text{FF}}$ }
Consider the action of $Q_4$ on $\langle J_0J_0J_0 J_0\rangle_\text{FF}$ in the quasi-fermionic theory :
\begin{align}
\langle& [Q_4, J_0(x_1)]J_0(x_2)J_0(x_3)J_0(x_4)\rangle_\text{FF}+1\leftrightarrow 2+1\leftrightarrow 3+1\leftrightarrow 4 =0
\end{align}
Upon utilizing the algebra in \eqref{FFQ3algebra}, we  obtain 
\begin{align}
\label{Q4onJ0000FF}
&\langle \partial_{-}^3 J_0(x_1)\,J_0(x_2)\,J_0(x_3)J_0(x_4)\rangle_\text{FF}+1\leftrightarrow 2+1\leftrightarrow 3+1\leftrightarrow 4\notag\\[6pt]&
+\epsilon_{-\mu\nu}\,\langle\partial_{-}\partial^\mu T^\nu_{-}(x_1)J_0(x_2) J_0(x_3)J_0(x_4)\rangle_\text{FF}+1\leftrightarrow 2+1\leftrightarrow 3+1\leftrightarrow 4=0
\end{align}
Upon Fourier transforming we get the following equation in momentum space :
\begin{align}
&\,(k_{1-}^3+k_{2-}^3+k_{3-}^3+k_{4-}^3)\langle \langle J_0(k_1) J_0(k_2) J_0(k_3)\rangle \rangle_{\text{FF}}+1\leftrightarrow 2+1\leftrightarrow 3+1\leftrightarrow 4\notag\\[6pt]&-i\,\bigg[k_{1-}\epsilon_{-k_1\nu}\langle\langle T^\nu_{-}(k_1)J_0(k_2)J_0(k_3)J_0(k_4)\rangle \rangle_{\text{FF}}+1\leftrightarrow 2+1\leftrightarrow 3+1\leftrightarrow 4\bigg]=0\label{HSEFF4ptT}
\end{align}
We will not solve the above higher spin equation here. However, this equation could be checked to be satisfied from the free fermionic correlators. We will demonstrate that the above higher spin equation is crucial to solving for the $\langle T J_0 J_0J_0\rangle$ in quasi fermionic theory.
\subsection*{$\boldsymbol{Q_3}$ on $\boldsymbol{\langle J_1J_0J_0J_0\rangle_\text{FF}}$ : Constraining $\boldsymbol{\langle J_1J_1J_0J_0\rangle_\text{FF}}$ }

Consider action of $Q_3$ on the correlator  $\langle J_-J_0J_0J_0\rangle$ :
\begin{align}
&\langle \left[Q_{3}, J_{-}(x_1)\right]J_0(x_2)J_0(x_3)J_0(x_4)\rangle\notag\\[5pt]
&\hspace{2cm}+\langle  J_{-}(x_1)\left[Q_{3}, J_{0}(x_2)\right]J_0(x_3)J_0(x_4)\rangle+2\leftrightarrow 3+2\leftrightarrow 4=0
\end{align}
Upon utilizing the algebra in \eqref{FFQ3algebra} we obtain
\begin{align}
2 &\langle \partial_{-} T_{--}(x_1)J_0(x_2)J_0(x_3)J_0(x_4)\rangle_\text{FF}\notag\\[5pt]&-i\,\epsilon_{-\mu \nu}\langle   J_{-}(x_1) \partial^{\mu} \partial_{-} J^{\nu}(x_2)J_0(x_3)J_0(x_4)\rangle_\text{FF} +2\leftrightarrow 3+2\leftrightarrow 4
=0
\end{align}
Fourier transforming the above equation we get
\begin{align}
\bigg[2\,i&\,k_{1-} \langle\langle T_{--}(k_1)J_0(k_2)J_0(k_3)J_0(k_4)\rangle\rangle_\text{FF}\notag\\[5pt]&+i\, \epsilon_{-k_2 \nu}k_{2-}\langle \langle J_{-}(k_1)  J^{\nu}(k_2)J_0(k_3)J_0(k_4)\rangle\rangle_\text{FF} +2\leftrightarrow 3+2\leftrightarrow 4\bigg]
=0
\end{align}
We will not solve the above equation here. However we will use  it to simplify the parity even part of the  corresponding higher spin equation in the interacting theory.

\section{Interacting Theories}
\label{InteractingTheories}
Having obtained the higher spin equations for various spinning correlators in the free theory we will now proceed to do the  corresponding analysis in  interacting theories with slightly broken higher spin symmetry 
These theories were studied in \cite{Maldacena:2012sf} and approximate higher spin equations were used to determine the structure of spinning correlators in position space. The authors showed that three-point correlators in an interacting theory with broken higher spin symmetry can be fixed in terms of three-point functions of the free fermion theory, free boson theory and a parity odd part and they
fixed the coefficients of these terms. In our analysis we look at correlation functions in the momentum space. Momentum space correlators unlike position space correlators are difficult to solve for in general. We use the higher spin equations to solve for these correlators.  One of the new ingredients in our analysis that was missing in \cite{Maldacena:2012sf}  is the local and contact terms \footnote{In two-point functions contact terms are polynomials in the momentum associated to the correlator. As \cite{sissathesis} clearly notes, the modulus $p$ of the momentum is not a contact term since it comes under a square root, $p=\sqrt{g_{\mu\nu}p^\mu\,p^\nu}$. In three-point functions one could have semi-local or local terms. Semi-local terms are those terms that are polynomial in only one of the two momenta whereas local terms are polynomial in both momenta associated to the correlator \cite{sissathesis}.} in correlators which turn out to be crucial in solving higher spin equations in momentum space. While in our analysis we fix the contact terms for specific spinning correlators we hope that our method can be systematised to obtain contact terms for any spinning correlator. We leave this for future work.

In our discussion on interacting theories we will first consider the critical bosonic theory which is obtained by a deformation of the free boson theory with a $\lambda_4 \phi^4$ interaction term. Following that we will use higher spin equations to determine momentum space correlators in the quasi-fermionic and the quasi-bosonic theories. We first consider two-point spinning correlators and fix the parity odd contribution whose coefficient  turns out to be universal. We then move on to solving three and four-point correlators in these theories. In this process we recover results previously obtained by explicit Feynman diagram computations which could only be done in specific kinematic regimes \cite{Aharony:2012nh,GurAri:2012is} and also obtain new results. 

\subsection{Critical Bosonic Theory}
In this subsection as a first example of the interacting theory with a slightly broken higher spin symmetry, we will consider the critical bosonic theory obtained by  deforming the free boson theory with a $\phi^4$ interaction. \subsubsection{Three-point function }
We begin by deriving a higher spin equation for the three-point correlator in this theory constructed through the action of $Q_4$ on the three-point function $\langle\widetilde{J}_0\widetilde{J}_0\widetilde{J}_0\rangle$.
\subsection*{$\boldsymbol{Q_4}$ on $\boldsymbol{\langle \widetilde{J}_0\widetilde{J}_0\widetilde{J}_0\rangle_\text{CB}}$ : Constraining  $\boldsymbol{\langle T\widetilde{J}_0\widetilde{J}_0\rangle_\text{CB}}$}
Let us begin by considering the action of the  charge $Q_4$ on the three-point function $\widetilde{J}_0\widetilde{J}_0\widetilde{J}_0$.
\begin{align}
\langle[Q_{4}, \widetilde{J}_{0}(x_1)]\widetilde{J}_{0}(x_2)\widetilde{J}_{0}(x_3)\rangle+1\leftrightarrow 2+1\leftrightarrow 3=\int_x \langle \partial_{\mu} J_{---}^{\mu}\widetilde{J}_{0}(x_1)\widetilde{J}_{0}(x_2)\widetilde{J}_{0}(x_3)\rangle
\end{align}
Upon utilizing the algebra in \eqref{Q4algebraCB} and the current equation in \eqref{DivJ4CB} we obtain the high spin equation in position space to be as follows
\begin{align}
\label{CBeqn111}
\langle \partial_{-}^{3}\widetilde{J}_{0}(x_1)\widetilde{J}_{0}(x_2)\widetilde{J}_{0}(x_3)\rangle+(1\leftrightarrow 2)+(1\leftrightarrow 3)&=
6 \int_x \langle\partial_{-}\widetilde{J}_{0} T_{--}(x)\widetilde{J}_{0}(x_1)\widetilde{J}_{0}(x_2)\widetilde{J}_{0}(x_3)\rangle
\end{align}
Performing a Fourier transform of the above equation and using large $N$ factorisation we get the following equation in momentum space :
\begin{align}
(k_{1-}^{3}&+k_{2-}^3+k_{3-}^3)\langle\langle\widetilde{J}_{0}(k_1)\widetilde{J}_{0}(k_2)\widetilde{J}_{0}(k_3)\rangle\rangle_\text{CB}\notag\\[5pt]&-6\,k_{1-}\langle\widetilde{J}_{0}(k_1)\widetilde{J}_{0}(-k_1)\rangle_\text{CB}\langle\langle T_{--}(k_1)\widetilde{J}_{0}(k_2)\widetilde{J}_{0}(k_3)\rangle\rangle_\text{CB}+(1\leftrightarrow 2)+(1\leftrightarrow 3)=0
\end{align}
Although we will not solve this higher spin equation here, one could straightforwardly check the above equation by obtaining the correlators in the critical bosonic theory by a Legendre transformation of the free boson theory as described in \cite{Bedhotiya:2015uga}.
\subsubsection{Four-point functions}
Having obtained a higher spin equation we now proceed to derive the higher spin equation for the four-point function which we will show to be crucial for solving the same in quasi-fermionic theory in a subsequent subsection.
\subsection*{$\boldsymbol{Q_4}$ on $\boldsymbol{\langle \widetilde{J}_0\widetilde{J}_0\widetilde{J}_0\widetilde{J}_0\rangle_\text{CB}}$ : Constraining $\boldsymbol{\langle \widetilde{T}\widetilde{J}_0\widetilde{J}_0\widetilde{J}_0\rangle_\text{CB}}$  }
Let us consider the action of $Q_4$ on the four-point function $\langle\widetilde{J}_0\widetilde{J}_0\widetilde{J}_0\widetilde{J}_0\rangle$ :
\begin{align}
&\langle[Q_{4}, \widetilde{J}_{0}(x_1)]\widetilde{J}_{0}(x_2)\widetilde{J}_{0}(x_3)\widetilde{J}_{0}(x_4)\rangle+(1\leftrightarrow 2)+(1\leftrightarrow 3)+(1\leftrightarrow 4)\notag\\
&\hspace{.5cm}=\int_x \langle \partial_{\mu} J_{---}^{\mu}\widetilde{J}_{0}(x_1)\widetilde{J}_{0}(x_2)\widetilde{J}_{0}(x_3)\widetilde{J}_{0}(x_4)\rangle
\end{align}
Upon utilizing the algebra in \eqref{Q4algebraCB}, the current equation in \eqref{DivJ4CB} and large $N$ factorization, we obtain the high spin equation in position space to be as follows
\begin{align}
\langle &\partial_{-}^{3}\widetilde{J}_{0}(x_1)\widetilde{J}_{0}(x_2)\widetilde{J}_{0}(x_3)\widetilde{J}_{0}(x_4)\rangle+(1\leftrightarrow 2)+(1\leftrightarrow 3)\notag\\[5pt]&
=6\int_x \langle\partial_{-}\widetilde{J}_{0}(x)\widetilde{J}_{0}(x_1)\rangle\langle T_{--}(x)\widetilde{J}_{0}(x_2)\widetilde{J}_{0}(x_3)\widetilde{J}_{0}(x_4)\rangle+(1\leftrightarrow 2)+(1\leftrightarrow 3)+(1\leftrightarrow 4)\notag\\[5pt]&
\hspace{.25cm}+6\int_x\langle\partial_{-} \widetilde{J}_{0}(x) \widetilde{J}_{0}(x_{1}) \widetilde{J}_{0}(x_{2}) \rangle\langle T_{--}(x)\widetilde{J}_{0}(x_{3}) \widetilde{J}_{0}(x_{4})\rangle\notag \\[5pt]&\hspace{1cm}+(1 \leftrightarrow 3)+(1 \leftrightarrow 4)+(2 \leftrightarrow 3)+(2 \leftrightarrow 4)+(1,2\leftrightarrow 3,4)
\end{align}
A Fourier transform of the above equation gives in momentum space the following :
\begin{align}
(k_{1-}^{3}&+k_{2-}^3+k_3^3+k_4^3)\langle\langle  \widetilde{J}_{0}(k_1)  \widetilde{J}_{0}(k_2)  \widetilde{J}_{0}(k_3)  \widetilde{J}_{0}(k_4)\rangle\rangle_\text{CB}\notag\\[7pt]&=6\,\bigg[ k_{1-}\langle  \widetilde{J}_0(k_1) \widetilde{J}_0(-k_1)\rangle_\text{CB}\langle\langle  T_{--}(k_1) \widetilde{J}_0(k_2) \widetilde{J}_{0}(k_3)  \widetilde{J}_{0}(k_4)\rangle\rangle_\text{CB}\notag\\[4pt]&\hspace{1.5cm}+(1 \leftrightarrow 2)+(1 \leftrightarrow 3)+(1 \leftrightarrow 4)\bigg]\notag\\&~~~+6\,\bigg[(k_{1-}+k_{2-}) \langle\widetilde{J}_{0}(-k_1-k_2) \widetilde{J}_{0}(k_1) \widetilde{J}_{0}(k_2) \rangle_\text{CB}\langle T_{--}(-k_3-k_4)\widetilde{J}_{0}(k_3) \widetilde{J}_{0}(k_4)\rangle_\text{CB} \notag\\&\hspace{1.5cm}+(1 \leftrightarrow 3)+(1 \leftrightarrow 4)+(2 \leftrightarrow 3)+(2 \leftrightarrow 4)+(1,2\leftrightarrow 3,4)\bigg]\label{HSECB}
\end{align}
We will not solve this higher-spin equation in the present article but we will demonstrate that the above equation is crucial to the solve the corresponding higher spin equation in the quasi-fermionic theory.
\subsection{Quasi-fermionic Theory}
We now turn our attention to the quasi-fermionic theory.
We first consider the two-point function of spinning operators and show how higher spin equations recover the known parity-odd contributions \cite{Witten:2003ya,Leigh:2003ez}.  

\subsubsection{Two-point functions}
\label{2ptQFsection}

\subsubsection*{$\boldsymbol{\langle J_1 \,J_1 \rangle_\text{QF}}$}
Consider the action of the spin 2 generator $Q_3$ on $\langle J_- J_0\rangle_{\text{QF}}$. From the algebra in \eqref{QFQ3algebra}, we see that this generates $\langle J_-J_-\rangle_{\text{QF}}$ which can then be obtained by solving the corresponding higher spin equation. In position space the higher spin equation takes the following form :
\begin{align}
\langle [Q_3, J_-(x_1)]J_0(x_2)\rangle_\text{QF}+\langle J_-(x_1) [Q_3, J_0(x_2)]\rangle_\text{QF}=\int_x\langle \partial_{\mu} J^{\mu}_{--}(x)J_-(x_1)J_0(x_2)\rangle_\text{QF}
\end{align}
Using the algebra in \eqref{QFQ3algebra} and the current equation \eqref{DivJ3AF} this becomes :
\begin{align}
\epsilon_{-\mu\nu}\,\langle J_-(x_1)\partial^{\mu}\,\partial_{-}J^\nu(x_2)\rangle_\text{QF}=-16\,i\,(r_0-r_1)\int_x\langle \partial_{-}J_0 J_{-}(x)J_-(x_1)J_0(x_2)\rangle_\text{QF}
\end{align}
In momentum space the equation takes the form :
\begin{align}
\label{HSEstep2Q3QF}
\epsilon_{-k_1\nu}\langle J_-(k_1)J^\nu(-k_1)\rangle_\text{QF}=\frac{16\,i\,\widetilde\lambda_{\text{qf}}}{\widetilde N(1+\widetilde\lambda_{\text{qf}}^2)}\langle J_-(k_1)J_{-}(-k_1) \rangle_\text{QF} \langle J_0(k_1)J_{0}(-k_1)\rangle_\text{QF}
\end{align}
where on the RHS we have kept the leading order term from the large $N$ factorization. 

The two-point function of the scalar operator in quasi fermionic theory is \cite{Maldacena:2012sf} :
\begin{align}
\label{J0J0QF2pt}
\langle J_0(k)J_0(-k)\rangle_\text{QF}&=\frac{\widetilde{N}}{2}(1+\widetilde\lambda_{\text{qf}}^2)\langle J_{0}(k)J_{0}(-k)\rangle_\text{FF}\nonumber\\[5pt]
&=-\frac{\widetilde{N}}{2}(1+\widetilde\lambda_{\text{qf}}^2)\times\frac k8
\end{align}
The two-point function of the spin-one current in quasi fermionic theory has the following structure  :
\begin{align}
\label{JmuJnuQF}
\langle J_\mu(k)\,J_\nu(-k)\rangle_\text{QF}&=\frac{\widetilde N}{2}\,\left[\langle J_\mu(k)\,J_\nu(-k)\rangle_{\text{even}}+\langle J_\mu(k)\,J_\nu(-k)\rangle_{\text{odd}}\right]\nonumber\\[5pt]
&=\frac{\widetilde N}{2}\,\left[\frac{k_{\mu}\,k_{\nu}-g_{\mu\nu}\,k^2}{16\,k}+c_{123}\,\epsilon_{\mu\nu k}\right]
\end{align}
where $c_{123}$ is undetermined.
We substitute the above expressions for the two-point functions in the momentum space higher spin equation \eqref{HSEstep2Q3QF} and obtain
%
the following for the coefficient of the parity odd term in \eqref{JmuJnuQF} :
\begin{align}
\label{ct01}
c_{123}=\frac{\widetilde\lambda_{\text{qf}}}{16}
\end{align}
Thus we see that the higher spin equation \eqref{HSEstep2Q3QF} fixes the coefficient of the parity-odd contact term in the two-point function of the spin-one current in the quasi-fermionic theory.  Notice that this term vanishes in the free theory limit, $\widetilde\lambda\rightarrow 0$ limit,  consistent with the free theory analysis in Section \ref{2ptfnFF}.

\subsubsection*{$\boldsymbol{\langle T \,T \rangle_\text{QF}}$}

We will now look at the two-point function of spin-two currents. From the algebra in \eqref{QFQ3algebra}, we see that the action of $Q_4$ on $\langle J_0\,T_{--}\rangle_{\text{QF}}$ generates $\langle T_{--}\,T_{--}\rangle_{\text{QF}}$. The latter can then be obtained by solving the resulting higher spin equation. Using the higher spin algebra in \eqref{QFQ3algebra} and the current equation \eqref{DivJ4AF} we get the following equation \footnote{The algebra of $Q_4$ with the stress-tensor takes the form $[Q_4,T_{--}]\propto\partial_{-}J_{----}+\partial_{-}^3T_{--}$. It is easy to eliminate by a dimension, spin, and parity analysis the possibility of a term proportional to the pseudo-scalar $J_0$. As a result the LHS of \eqref{HSEQ4QF2ptTT} does not get any contribution from $\langle J_0\,[Q_4,T_{--}]\rangle$.} :
\begin{align}
\label{HSEQ4QF2ptTT}
\epsilon_{-\mu\nu}\partial_{-}\,\partial^\mu\,\langle T^\nu_{-}\,T_{--}\rangle_{\text{QF}}=\frac{96\widetilde\lambda_{\text{qf}}}{\widetilde N(1+\widetilde\lambda_{\text{qf}}^2)}\,\langle\partial_{-}J_0\,J_0\rangle_{\text{QF}}\,\langle T_{--}\,T_{--}\rangle_{\text{QF}}
\end{align}
After taking a Fourier transform on both sides of the above equation we get the following equation in momentum space :
\begin{align}
\label{HSEQ4QF2ptTTMS}
\epsilon_{-k_1\nu}\langle T^\nu_{-}\,T_{--}\rangle_{\text{QF}}=-\frac{96\,i\,\widetilde\lambda_{\text{qf}}}{\widetilde N(1+\widetilde\lambda_{\text{qf}}^2)}\,\langle J_0\,J_0\rangle_{\text{QF}}\,\langle T_{--}\,T_{--}\rangle_{\text{QF}}
\end{align}
The two-point function of stress-tensor in quasi-fermionic theory takes the form \cite{Leigh:2003ez}:
\begin{align}
\label{TT2pt}
\langle T_{\mu\nu}({k})\,T_{\rho\sigma}(-{k})\rangle_{\text{QF}}&=\frac{\widetilde N}{2}\Bigg[\Pi_{\mu\nu\rho\sigma}(\boldsymbol{k})\,A(k)+\Pi_{\mu\nu}(\boldsymbol{k})\,\Pi_{\rho\sigma}(\boldsymbol{k})B(k)\cr
&\hspace{-1.5cm}+\frac{c_{TT}}{4}\left(\epsilon_{\mu\rho\lambda}\Pi_{\nu\sigma}(\boldsymbol{k})+\epsilon_{\nu\rho\lambda}\Pi_{\mu\sigma}(\boldsymbol{k})+\epsilon_{\mu\sigma\lambda}\Pi_{\nu\rho}(\boldsymbol{k})+\epsilon_{\nu\sigma\lambda}\Pi_{\mu\rho}(\boldsymbol{k})\right)\,k^2\,k^\lambda\Bigg]
\end{align}
where  $\Pi_{\mu\nu\rho\sigma}(\boldsymbol{k})$ and $\Pi_{\mu\nu}(\boldsymbol{k})$ are the orthogonal projectors and 
\begin{align}
A(k)=k^3,\quad B(k)=0
\end{align}
It can be easily seen that the trace Ward identities satisfied by the even and the odd parts of the correlator \eqref{TT2pt} are trivial.
We substitute the forms of the two-point functions of the stress-tensor \eqref{TT2pt} and the scalar operator \eqref{J0J0QF2pt} in the higher spin equation \eqref{HSEQ4QF2ptTTMS} and then separate the resulting equation into two independent equations : one proportional to the Levi-Civita tensor and one without that. It can be easily checked that the former is trivially satisfied while the latter fixes the coefficient of the parity odd contact term in the $\langle T\,T \rangle$ correlator \eqref{TT2pt} to be :
\begin{align}
\label{cttoddqf}
c_{TT}=-6\,\widetilde\lambda_{\text{qf}}
\end{align}
We will now extend the above analysis to spin-4 currents.
%
%
\subsubsection*{$\boldsymbol{\langle J_4 \,J_4 \rangle_\text{QF}}$}
To get a non-trivial equation we consider the action of $Q_4$ on $\langle J_{-y}\,J_{----}\rangle_{\text{QF}}$ instead of $\langle J_{--}\,J_{----}\rangle_{\text{QF}}$. 

From the higher spin algebra of the charge $Q_4$ with spin $s>2$ operators $J_s$ \cite{Maldacena:2012sf} we obtain the following higher spin equation in momentum space : 
%
%
%
\begin{align}
\label{j4j4hseqf}
k_{-}\langle J_{---y}(k)\,J_{----}(-k)\rangle_{\text{QF}}+k_{-}^5\langle T_{-y}(k)\,T_{--}(-k)\rangle_{\text{QF}}=0
\end{align}
The RHS of the above higher spin equation is zero as the large $N$ factorization leads to two-point functions of operators with different conformal dimensions. 

To obtain the parity odd part of $\langle J_{---y}\,J_{----}\rangle_{\text{QF}}$ we look at the part of the higher spin equation \eqref{j4j4hseqf} which contains the Levi-Civita tensor :
\begin{align}
\label{j4j4hseqfodd}
k_{-}\langle J_{---y}(k)\,J_{----}(-k)\rangle_{\text{QF,odd}}+k_{-}^5\langle T_{-y}(k)\,T_{--}(-k)\rangle_{\text{QF,odd}}=0
\end{align}
To solve this equation, we use the following result from the analysis of the $\langle TT \rangle_{\text{QF}}$ correlator \eqref{TT2pt} and \eqref{cttoddqf} :
\begin{align}
\label{TTparityoddQF}
\langle T_{y-}(k)\,T_{--}(-k)\rangle_{\text{QF,odd}}=3\,\widetilde\lambda_{\text{qf}}\,k_{-}^2\,\epsilon_{y-k}
\end{align}
%
%
%
Substituting the above correlator in \eqref{j4j4hseqfodd} and solving for $\langle J_4J_4\rangle$ we obtain :
%
\begin{align}
\langle J_{---y}(k)\,J_{----}(-k)\rangle_{\text{QF,odd}}=3\,\widetilde\lambda_{\text{qf}}\,k_{-}^6\,\epsilon_{y-k}
\end{align}
Our analysis can be easily extended to obtain the parity odd term in the two-point function of arbitrary higher spin currents. 

We see that the coupling constant dependence of the coefficient of the parity-odd contact term in the two-point function of higher spin currents is universal. We expect such a universality because while the kinematic dependence of correlators is fixed by conformal invariance, their coupling constant dependence is fixed by higher spin equations, which in turn relate the two-point functions of various spinning operators.

This can be seen from the large frequency limit of transport coefficients. For example, we expect a relation between the parity-odd part of shear viscosity i.e. Hall viscosity and the parity-odd part of conductivity i.e. Hall conductivity. In fact this is indeed the case as was shown in \cite{Geracie:2015drf} (see equations 5.8 and 6.10 therein).

\subsubsection{Three-point functions}
\label{3ptQFsection}
Three-point correlators in the quasi-fermionic theory are given by the sum of contributions from the free boson theory (FB), the free fermion theory (FF) and a parity odd piece (odd) \cite{Maldacena:2011jn,Maldacena:2012sf} \footnote{Following the general literature, we label the part of a correlator that does not come from the free theory answer as odd. However this label need not always correspond to expressions that are odd under parity.}  :
\begin{align}
\langle J_{s_{1}} J_{s_{2}} J_{s_{3}}\rangle=& \alpha_{s_{1} s_{2} s_{3}}\langle J_{s_{1}} J_{s_{2}} J_{s_{3}}\rangle_\text{FB}+\beta_{s_{1} s_{2} s_{3}}\langle J_{s_{1}} J_{s_{2}} J_{s_{3}}\rangle_\text{FF}+\gamma_{s_{1} s_{2} s_{3}}\langle J_{s_{1}} J_{s_{2}} J_{s_{3}}\rangle_\text{odd}
\end{align}
where $s_i$ represents the spin of the operators. For the correlators that we encounter in our computations these relations read as : 
\begin{align}
\langle T_{\mu\nu}(k_1)J_0(k_2)J_0(k_3)\rangle_{\text{QF}}&=\beta_{200}\left[\langle T_{\mu\nu}(k_1)J_0(k_2)J_0(k_3)\rangle_{\text{FF}}+\widetilde c_{TOO}\,\frac{k_{1\mu}\,k_{1\nu}}{16k_1}\right] \label{TJ0J0QF}\\[6pt]
\langle J_{\mu}(k_1)J_{\nu}(k_2)J_0(k_3)\rangle_{\text{QF}}&=\beta_{110}\langle J_{\mu}(k_1)J_\nu(k_2)J_0(k_3)\rangle_{\text{FF}}\nonumber\\[6pt]
&\hspace{-1.5cm}+\gamma_{110}\langle J_{\mu}(k_1)J_\nu(k_2)J_0(k_3)\rangle_{\text{odd}}+\gamma_{110}\langle J_{\mu}(k_1)J_\nu(k_2)J_0(k_3)\rangle_{\text{contact}}\label{J1J1J0QF}\\[6pt]
\langle J_0(k_1) J_0(k_2) J_0(k_3)\rangle_{\text{QF}}&= (1+\widetilde\lambda_{\text{qf}}^2)\,\frac{ \widetilde N\widetilde\lambda_{\text{qf}}}{8}\,\label{J0J0J0QF}
\end{align}
Note that unlike in the free fermion theory where the momentum space three-point correlator of the scalar operator vanishes, in the quasi-fermionic theory it is a contact term \eqref{J0J0J0QF} as shown by explicit computations in \cite{GurAri:2012is}. The coefficients $\beta_{200}$, $\beta_{110}$, and $\gamma_{110}$ are given by :
\begin{align}
\beta_{200}=\frac{\widetilde N}{2}(1+ \widetilde{\lambda}_{\text{qf}}^2),\quad
\beta_{110}=\frac{\widetilde N}{2},\quad
\gamma_{110}=\frac{\widetilde N  \widetilde{\lambda}_{\text{qf}}}{2}\label{JJJ0QFcoef}
\end{align}
and the details of the free fermion correlators are given in \eqref{Tj0j0reconstructionF}, \eqref{jmujnu3pful} and \eqref{jmujnu3p}.

Note that in $\langle TJ_0J_0\rangle_{\text{QF}}$ in \eqref{TJ0J0QF} we have included a contact term that corresponds to the contact term in \eqref{TOOcontactF} that was required to solve the higher spin equation in the free fermion theory in Section \ref{3ptFFHSE}. We will fix the coefficient of this term, viz. $\widetilde c_{TOO}$ shortly.
In $\langle J_1J_1J_0\rangle_{\text{QF}}$ in \eqref{J1J1J0QF} we allow for a contact term which takes the following form \footnote{In \eqref{J1J1J0QF} we have stripped off a factor of $\gamma_{110}$ for convenience.} :
\begin{align}
\label{jmujnuj0qfcontact}
\langle\langle J_\mu(k_1)\,J_\nu(k_2)\,J_0(k_3)\rangle\rangle_{\text{QF}}^{\text{contact}}=c_{JJO}\,g_{\mu\nu}
\end{align}
%

\subsubsection*{$\boldsymbol{\langle J_1J_1 J_0\rangle_{\text{QF,odd}}}$ and fixing the contact term in $\boldsymbol{{\langle TJ_0 J_0\rangle_\text{QF}}}$}
We will now obtain an expression for $\langle J_{\mu}J_\nu J_0\rangle_{\text{odd}}$ in the quasi fermionic theory by solving a relevant higher spin equation in momentum space.
Let us consider the Ward identity arising from the action 
of  $Q_3$ on $\langle J_-J_0J_0\rangle_\text{QF}$ :
\begin{align}
\langle& [Q_3, J_-(x_1)]J_0(x_2)J_0(x_3)\rangle_\text{QF}+\langle J_-(x_1) [Q_3, J_0(x_2)]J_0(x_3)\rangle_\text{QF}+\langle J_-(x_1)J_0(x_2) [Q_3, J_0(x_3)]\rangle_\text{QF}\notag\\[5pt]
&\hspace{4cm}=\int_x\langle \partial_{\mu} J^{\mu}_{--}(x)J_-(x_1)J_0(x_2)J_0(x_3)\rangle_\text{QF}
\end{align}
Using the higher spin algebra \eqref{QFQ3algebra}, the current equation for the spin-3 current in \eqref{DivJ3AF}, and the large $N$ factorization of correlators, we get after performing a Fourier transform the following equation in momentum space :
\begin{align}
\label{HSEQFQ3-3pt}
&\left(\epsilon_{-k_2\nu}\,k_{2-}\,\langle J_-(k_1)J^\nu(k_2) J_0(k_3)\rangle_{\text{QF}}+\epsilon_{-k_3\nu}\,k_{3-}\,\langle J_-(k_1) J_0(k_2)J^\nu(k_3)\rangle_{\text{QF}}\right)\notag\\[5pt]
&+2\,i\,k_{1-}\langle \langle T_{--}(k_1)J_0(k_2)J_0(k_3)\rangle \rangle_{\text{QF}}\notag\\[5pt]
&=-\frac{16\,\widetilde\lambda_{\text{qf}}}{\widetilde N(1+\widetilde\lambda_{\text{qf}}^2)}\bigg[k_{1-}\langle J_{-}(k_1)J_{-}(-k_1)\rangle_{\text{QF}}\langle \langle J_0(k_1)J_0(k_2)J_0(k_3)\rangle\rangle_{\text{QF}}\notag\\[5pt]&\hspace{2.5cm}+k_{2-}\langle J_{0}(k_2)J_{0}(-k_2)\rangle_{\text{QF}}\langle\langle J_{-}(k_1)J_{-}(k_2)J_0(k_3)\rangle\rangle_{\text{QF}}\notag\\[5pt]&\hspace{2.5cm}+k_{3-}\langle J_{0}(k_3)J_{0}(-k_3)\rangle_{\text{QF}}\langle\langle J_{-}(k_1)J_0(k_2)J_{-}(k_3)\rangle\rangle_{\text{QF}}\bigg]
\end{align}
We consider the following ansatz for $\langle J_{\mu}J_\nu J_0\rangle_\text{QF}^{\text{odd}}$ :
\begin{align}
\langle\langle J_{\mu}(k_1)J_\nu(k_2)J_0(k_3)\rangle\rangle_\text{QF}^{\text{odd}}&=\langle\langle J_{\mu}(k_1)J_\nu(k_2)\widetilde{J}_0(k_3)\rangle\rangle_{\text{CB}}\notag\\[5pt]
&=-k_3\langle\langle J_{\mu}(k_1)J_\nu(k_2)J_0(k_3)\rangle\rangle_{\text{FB}}\label{JJJ0QFodd}
\end{align}
This form of the odd part of  $\langle J_\mu J_\nu J_0\rangle_\text{QF}$ is consistent with the result obtained in \cite{GurAri:2012is} in the specific kinematic regime considered by the authors. This is also consistent with what one expects from duality.

Upon utilizing the result for three-point correlators given in \eqref{TJ0J0QF}, \eqref{J1J1J0QF} and \eqref{J0J0J0QF} and the result for two-point correlators given in \eqref{J0J0QF2pt} and \eqref{JmuJnuQF}, we observe that the ansatz for the odd part of $\langle J_1J_1J_0\rangle_\text{QF}$ in \eqref{JJJ0QFodd} solves the higher spin equation \eqref{HSEQFQ3-3pt} provided :
\begin{align}
\label{EvendataQF}
\widetilde c_{TOO}=\frac{1}{1+\widetilde\lambda^2}
\end{align}
%
%
%
This coefficient for $\langle TJ_0J_0\rangle$ in \eqref{TJ0J0QF} in the quasi-fermionic theory is in agreement with the free fermion theory result \eqref{TOOcontactF}. 

Thus we see that the analysis via. momentum space higher spin equations helped us verify the ansatz \eqref{JJJ0QFodd} for the odd part of $\langle J_1J_1J_0\rangle_{\text{QF}}$ and fix the coefficient of the contact term in $\langle TJ_0J_0\rangle_{\text{QF}}$ \eqref{TJ0J0QF}. 
\subsubsection*{Further check of the contact term in $\langle TJ_0J_0\rangle$}

Let us consider the action of the charge $Q_4$ on $\langle J_0\,J_0\,J_0\rangle$. 
The higher spin equation is : 
\begin{align}
\langle [Q_4, J_0(x_1)]J_0(x_2)J_0(x_3)\rangle_\text{QF}+1\leftrightarrow 2+1\leftrightarrow 3=\int_x \langle\partial_{\mu} J^{\mu}_{---}(x)J_0(x_1)\,J_0(x_2)\,J_0(x_3)\rangle_\text{QF}
\end{align}
Upon utilizing the algebra in \eqref{QFQ3algebra}, the current equation \eqref{DivJ4AF} and the large $N$ factorisation we get :
\begin{align}
\label{Q4onJ0J0J0QF}
\bigg[&\langle \partial_{-}^3 J_0(x_1)\,J_0(x_2)\,J_0(x_3)\rangle_\text{QF}
+\epsilon_{-\mu\nu}\,\langle\partial_{-}\partial^\mu T^\nu_{-}(x_1)J_0(x_2) J_0(x_3)\rangle_\text{QF}\bigg]+1\leftrightarrow 2+1\leftrightarrow 3\notag\\[5pt]&\hspace{3.5cm}=\frac{96\widetilde\lambda_{\text{qf}}}{\widetilde N(1+\widetilde\lambda_{\text{qf}}^2)}\,\int_x \langle \partial_{-}J_0\,T_{--}(x)J_0(x_1)\,J_0(x_2)\,J_0(x_3)\rangle_\text{QF}
\end{align}
In momentum space the equation takes the following form :
\begin{align}
&\,(k_{1-}^3+k_{2-}^3+k_{3-}^3)\langle \langle J_0(k_1) J_0(k_2) J_0(k_3)\rangle \rangle_{\text{QF}}-\bigg[i\,k_{1-}\epsilon_{-k_1\nu}\langle\langle T^\nu_{-}(k_1)J_0(k_2)J_0(k_3)\rangle \rangle_{\text{QF}}+1\leftrightarrow 2+1\leftrightarrow 3\bigg]\notag\\[5pt]=
&-\frac{96\widetilde\lambda_{\text{qf}}}{\widetilde N(1+\widetilde\lambda_{\text{qf}}^2)}\Bigg[k_{1-}\,\langle J_0(k_1)\,J_0(-k_1)\rangle_{\text{QF}}\langle\langle  T_{--}(k_1)J_0(k_2)J_0(k_3)\rangle\rangle_{\text{QF}}+1\leftrightarrow 2+1\leftrightarrow 3\Bigg]
\end{align}
It can be easily checked using the expressions for the three-point correlators in  \eqref{TJ0J0QF} and \eqref{J0J0J0QF} that the above higher spin equation is satisified.
%
This analysis serves as a consistency check of the contact term in $\langle TJ_0J_0\rangle$ \eqref{TJ0J0QF}.
\subsubsection*{Fixing the contact term in $\boldsymbol{\langle J_1 J_1 J_0\rangle_\text{QF}}$}
Using higher spin equations we verified our ansatz in \eqref{JJJ0QFodd} that the odd part of $\langle J_1J_1J_0\rangle$ in the quasi fermionic theory is given by the correlator in the critical bosonic theory. We will complete our analysis of $\langle J_1J_1J_0\rangle_\text{QF}$ by  computing the possible contact term in \eqref{jmujnuj0qfcontact}.

To compute the contact term we analyse the higher spin equation obtained by the action of the generator $Q_3$ on $\langle J_{\mu}J_0J_0\rangle$. This requires us to generalize the algebra in \eqref{QFQ3algebra} to the case where the generator acts on $J_\mu$ :
\begin{align}
[Q_3,J_\mu]&=\widetilde a_1\,\epsilon_{-\mu\nu}\,\partial_{-}\,\partial^{\nu}\,J_0+\widetilde a_2\,\partial_\mu\,T_{--}+\widetilde a_3\,\partial_{-}\,T_{\mu -}+\widetilde a_4\,\eta_{-\mu}\,\partial_{-}\,T^\rho_\rho+\widetilde a_5\,\eta_{-\mu}\,\partial^\alpha\,T_{\alpha -}
\end{align}
%
In our analysis we will not need to fix the coefficients $\widetilde a_i$ that appear in the above algebra. But it turns out from explicit computation that $\widetilde a_2=\widetilde a_4=0$. 

Upon using the current equation \eqref{DivJ3AF} and the large $N$ factorisation, the part of the higher spin equation in momentum space which contains the Levi-Civita tensor takes the form :
\begin{align}
&\widetilde a_1\,\epsilon_{\mu-k_1}\,k_{1-}\,\langle\langle J_0(k_1)\,J_0(k_2)\,J_0(k_3)\rangle\rangle_{\text{QF}}-k_{2-}\epsilon_{\nu - k_2}\langle\langle J_\mu(k_1) J^\nu(k_2) J_0(k_3)\rangle\rangle_\text{QF}^{\text{odd}}\nonumber\\[8pt]
&\hspace{1cm}-k_{3-}\epsilon_{\nu - k_3}\langle\langle J_\mu(k_1) J_0(k_2) J^\nu(k_3)\rangle \rangle_\text{QF}^{\text{odd}}\nonumber\\[8pt]
&=16(r_0-r_1)\,\left[k_{1-}\,\langle\langle J_0(k_1)\,J_0(k_2)\,J_0(k_3)\rangle\rangle_{\text{QF}}\langle J_\mu(k_1)\,J_-(-k_1)\rangle_{\text{QF}}^{\text{even}}\right.\nonumber\\[8pt]
&\left.\hspace{2.5cm}-k_{3-}\,\langle\langle J_\mu(k_1)\,J_0(k_2)\,J_-(k_3)\rangle\rangle^{\text{even}}_{\text{QF}}\langle J_0(k_3)\,J_0(-k_3)\rangle_{\text{QF}}\right.\nonumber\\[8pt]
&\left.\hspace{2.5cm}-k_{2-}\,\langle\langle J_\mu(k_1)\,J_-(k_3)\,J_0(k_2)\,\rangle\rangle^{\text{even}}_{\text{QF}}\langle J_0(k_2)\,J_0(-k_2)\rangle_{\text{QF}}\right]
\end{align}
To determine the possible contact term \eqref{jmujnuj0qfcontact} we project out the transverse parts of the correlators by dotting the above equation with $k_1^\mu$. This retains only the local part of the correlators, and we see that the equation is satisfied iff $c_{JJO}=-\frac 14$  (\eqref{J1J1J0QF} and \eqref{jmujnuj0qfcontact}):
\begin{align}
\label{ct}
\langle J_\mu(k_1) J_\nu(k_2) J_0(k_3)\rangle_{\text{QF}}^{\text{contact}}=-\frac{g_{\mu\nu}}{4}
\end{align}
%
We note that this contact term is in addition to the usual contact terms that come from conservation Ward identities in momentum space.  If one explicitly computes the three-point function using Feynman diagram techniques, one obtains precisely such an additional contact term \cite{GurAri:2012is}. Such contact terms can removed by adding appropriate terms in the action.
 
The part of the higher spin equation without the Levi-Civita tensor is also solved if $\langle J_1J_1J_0\rangle_\text{QF}$ has the contact term in \eqref{ct}. 
%

%
\subsubsection{Four-point functions}

We will now turn our attention to four-point correlators in the quasi-fermionic theory.
\subsubsection*{$\langle \boldsymbol{J_0J_0J_0J_0\rangle_\text{QF}} $ }
The scalar four-point function in the quasi-fermionic theory in position space was determined using the double discontinuity technique in \cite{Turiaci:2018nua} to be \footnote{Note that in \cite{Turiaci:2018nua} the authors have a factor of $1/\widetilde{N}$ instead of $\frac{\widetilde{N}}{2}(1+\widetilde{\lambda_{\text{qf}}^2})^2$. This is because of our definition of $J_0$ which is related to theirs ($J_0^{TZ}$) through a rescaling $J_0=J_0^{TZ}\frac{\sqrt{\widetilde N(1+\widetilde\lambda_{\text{qf}}^2)}}{2^{1/4}}$.} :
\begin{align}
\langle J_0(x_1) J_0(x_2) J_0(x_3) J_0(x_4)\rangle_\text{QF}=\frac{\widetilde{N}}{2}(1+\widetilde{\lambda}_{\text{qf}}^2)^2\langle &J_0(x_1) J_0(x_2) J_0(x_3) J_0(x_4)\rangle_\text{FF}
\end{align}
The momentum space correlation function is straightforwardly obtained by taking the Fourier transform on both sides of the above equation : 
\begin{align}\label{J04ptqf}
\langle \langle  J_{0}( k_1) J_{0}( k_2) J_0(k_3)J_0(k_4)\rangle\rangle_\text{QF}=\frac{\widetilde{N}}{2}(1+\widetilde{\lambda}_{\text{qf}}^2)^2\langle \langle  J_{0}( k_1) J_{0}( k_2) J_0( k_3)J_0( k_4)\rangle\rangle_\text{FF}
\end{align}
where the free fermion correlator was obtained by explicit computation in \cite{Jain:2020rmw}. This matches the result for the same obtained in a specific kinematic regime in \cite{Bedhotiya:2015uga}. 

\subsubsection*{$\langle \boldsymbol{TJ_0J_0J_0 \rangle_\text{QF} }$ }
\label{Tj0j0j0QF}
In this section we discuss the $\langle T_{\mu\nu}J_0J_0J_0 \rangle$ correlator in the quasi-fermionic theory. A direct computation of this correlator using the Feynman diagram approach can only be done in specific kinematic regimes \cite{Kalloor:2019xjb} and a general analysis is out of reach. In the following we illustrate how higher spin equations in momentum space can be used to obtain the correlator.

Recently in \cite{Li:2019twz} (see also \cite{Turiaci:2018nua} and \cite{Silva:2021ece}) the correlator  $\langle {T_{\mu\nu}J_0J_0J_0} \rangle_\text{QF} $ was studied using higher spin equations in position space and it was shown that :
\begin{align}\label{J24ptqf}
&\langle\langle T_{\mu\nu}(x_1) J_0(x_2) J_0(x_3) J_0(x_4)\rangle\rangle_\text{QF}\nonumber\\[6pt]
&\hspace{-.5cm}=\beta_{2000}\langle \langle T_{\mu\nu}(x_1) J_0(x_2) J_0(x_3) J_0(x_4)\rangle\rangle_\text{even}+\gamma_{2000}\langle \langle  T_{\mu\nu}(x_1) J_0(x_2) J_0(x_3) J_0(x_4)\rangle\rangle_\text{odd}
\end{align}
where 
\begin{align}
\beta_{2000}=\frac{\widetilde{N}}{2}(1+ \widetilde{\lambda}_{\text{qf}}^2),\quad
\gamma_{2000}=\frac{\widetilde{N}}{2} \widetilde{\lambda}_{\text{qf}}(1+ \widetilde{\lambda}_{\text{qf}}^2)\,.
\end{align}
The even and odd parts \footnote{Throughout this paper, to be consistent with the literature we call the part of the correlator multiplied by $\beta_{2000}$ as even and the part multiplied by $\gamma_{2000}$ as odd.} of the correlator are given by \cite{Li:2019twz} :
\begin{align}
\label{ansatztj0j0j0QF}
\langle T_{\mu\nu}(x_1) J_0(x_2) J_0(x_3) J_0(x_4)\rangle_\text{even}&=\langle T_{\mu\nu}(x_1) J_0(x_2) J_0(x_3) J_0(x_4)\rangle_\text{FF}\nonumber\\[6pt]
\langle  T_{\mu\nu}(x_1) J_0(x_2) J_0(x_3) J_0(x_4)\rangle_\text{odd}&= \langle  T_{\mu\nu}(x_1) J_0(x_2) J_0(x_3) J_0(x_4)\rangle_\text{CB}
\end{align} 
The position space analysis of \cite{Li:2019twz} was quite involved and in the following we will verify the results for the correlator using a rather simple analysis via momentum space higher spin equations \footnote{\label{footnote10}
We feel that one can extend the arguments of \cite{Li:2019twz} to any almost conserved  higher spin current with even spin in the quasi-fermionic theory, i.e.
$$\langle\langle J_{2s}(k_1) J_0(k_2) J_0(k_3) J_0(k_4)\rangle\rangle_\text{QF}=\beta_{2000}\langle\langle J_{2s}(k_1) J_0(k_2) J_0(k_3) J_0(k_4)\rangle\rangle_\text{FF}+\gamma_{2000}\langle\langle J_{2s}(k_1) J_0(k_2) J_0(k_3) J_0(k_4)\rangle\rangle_\text{CB}$$ However, we have not verified this for $s>1$.}.

Let us consider the action of $Q_4$ on the four-point correlator $\langle J_0J_0J_0J_0\rangle_\text{QF}$ :
\begin{align}
&\langle [Q_4, J_0(x_1)]J_0(x_2)J_0(x_3)J_0(x_4)\rangle_\text{QF}+1\leftrightarrow 2+1\leftrightarrow 3+1\leftrightarrow 4\nonumber\\[6pt]
&\hspace{1cm}=\int_x \langle\partial_{\mu} J^{\mu}_{---}(x)J_0(x_1)\,J_0(x_2)\,J_0(x_3)J_0(x_4)\rangle_\text{QF}
\end{align}
Upon utilizing the algebra in \eqref{QFQ3algebra}, the current equation for $J_4$ in \eqref{DivJ4AF} and the large $N$ factorisation we obtain :
\begin{align}
\partial_{-}^{3}\langle J_{0}(x_{1}) J_{0}(x_{2})& J_{0}(x_{3}) J_{0}(x_{4})\rangle_\text{QF}+ \epsilon_{-\mu \nu} \partial_{-} \partial^{\mu}\langle T_{-}^{\nu}(x_{1}) J_{0}(x_{2}) J_{0}(x_{3}) J_{0}(x_{4})\rangle_\text{QF} \notag\\[7pt]&\hspace{-1cm}+(1 \leftrightarrow 2)+(1 \leftrightarrow 3)+(1 \leftrightarrow 4)
\notag\\&\hspace{-3cm}=\frac{96 \widetilde{\lambda}_{\text{qf}}}{\widetilde{N}(1+  \widetilde{\lambda}_{\text{qf}}^{2})}\Bigg[\int_x\langle\partial_{-} J_{0}(x) J_{0}(x_{1})\rangle\langle T_{--}(x) J_{0}(x_{2}) J_{0}(x_{3}) J_{0}(x_{4})\rangle_\text{QF}\notag \\[5pt]&\hspace{-.5cm} +(1 \leftrightarrow 2)+(1 \leftrightarrow 3)+(1 \leftrightarrow 4)\notag\\&\hspace{-.5cm} +\int_x\langle\partial_{-} J_{0}(x) J_{0}(x_{1}) J_{0}(x_{2}) \rangle_\text{QF}\langle T_{--}(x)J_{0}(x_{3}) J_{0}(x_{4})\rangle_\text{QF} \notag\\[5pt]&\hspace{-.5cm}+(1 \leftrightarrow 3)+(1 \leftrightarrow 4)+(2 \leftrightarrow 3)+(2 \leftrightarrow 4)+(1,2\leftrightarrow 3,4)\Bigg]
\end{align}
The above equation in momentum space takes the form :
\begin{align}
(&k_{1-}^{3}+k_{2-}^3+k_3^3+k_4^3)\langle\langle J_{0}(k_1) J_{0}(k_2) J_{0}(k_3) J_{0}(k_4)\rangle\rangle_\text{QF}\notag\\[7pt]&-i\,\epsilon_{-\mu \nu}k_{1-}k_{1}^{\mu}\langle\langle T_{-}^{\nu}(k_1) J_{0}(k_2) J_{0}(k_3) J_{0}(k_4)\rangle\rangle_\text{QF} +(1 \leftrightarrow 2)+(1 \leftrightarrow 3)+(1 \leftrightarrow 4)\notag\\[5pt]&=-\frac{96\widetilde{\lambda}_{\text{qf}}}{\widetilde{N}(1+  \widetilde{\lambda}_{\text{qf}}^{2})}\bigg[k_{1-}\langle J_0(k_1)J_0(-k_1)\rangle_\text{QF}\langle\langle T_{--}(k_1)J_0(k_2)J_{0}(k_3) J_{0}(k_4)\rangle\rangle_\text{QF}\notag\\[5pt]&\hspace{3cm}+(1 \leftrightarrow 2)+(1 \leftrightarrow 3)+(1 \leftrightarrow 4)\notag\\[5pt]&+(k_{1-}+k_{2-})\langle J_{0}(-k_1-k_2) J_{0}(k_1) J_{0}(k_2) \rangle_\text{QF}\langle T_{--}(-k_3-k_4)J_{0}(k_3) J_{0}(k_4)\rangle_\text{QF} \notag\\&\hspace{3.5cm}+(1 \leftrightarrow 3)+(1 \leftrightarrow 4)+(2 \leftrightarrow 3)+(2 \leftrightarrow 4)+(1,2\leftrightarrow 3,4)\bigg]\label{HSEQ4QF}
\end{align}
Using the fact that the free fermionic and critical bosonic scalar four-point functions are exactly equal as shown in \cite{Bedhotiya:2015uga} and using \eqref{J04ptqf} we re-express the scalar four-point function in the quasi-fermionic theory in the following form which will  turn out to be useful :
\begin{align}
\langle\langle& J_{0}(k_1) J_{0}(k_2) J_{0}(k_3) J_{0}(k_4)\rangle\rangle_\text{QF}\notag\\[5pt]&
=\frac{\widetilde{N}}{2}\bigg[(1+\widetilde{\lambda}_{\text{qf}}^2)\langle\langle J_{0}(k_1) J_{0}(k_2) J_{0}(k_3) J_{0}(k_4)\rangle\rangle_\text{FF}+\widetilde{\lambda}_{\text{qf}}^2(1+\widetilde{\lambda}_{\text{qf}}^2)\langle\langle \widetilde J_{0}(k_1) \widetilde J_{0}(k_2) \widetilde J_{0}(k_3) \widetilde J_{0}(k_4)\rangle\rangle_\text{CB}\bigg]\label{ScQF2}
\end{align}
%
%
%
We substitute \eqref{ScQF2} and the ansatz  \eqref{J24ptqf} for the correlator $\langle T J_0J_0J_0\rangle$ in \eqref{HSEQ4QF} and after a bit of rearranging the terms we get :
\begin{align*}
\frac{\widetilde{N}}{2}(1+\widetilde{\lambda}_{\text{qf}}^2)\bigg[(k_{1-}^{3}&+k_{2-}^3+k_{3-}^3+k_{4-}^3)\langle\langle J_{0}(k_1) J_{0}(k_2) J_{0}(k_3) J_{0}(k_4)\rangle\rangle_\text{FF}\notag\\&\hspace{-.75cm}-i\epsilon_{-\mu \nu}k_{1-}k_{1}^{\mu}\langle\langle T_{-}^{\nu}(k_1) J_{0}(k_2) J_{0}(k_3) J_{0}(k_4)\rangle\rangle_\text{even}+(1 \leftrightarrow 2)+(1 \leftrightarrow 3)+(1 \leftrightarrow 4)\bigg]\notag\\+\frac{\widetilde{N}}{2}\widetilde{\lambda}_{\text{qf}}(1+\widetilde{\lambda}_{\text{qf}}^2)\bigg[(k_{1-}^{3}&+k_{2-}^3+k_{3-}^3+k_{4-}^3)\langle\langle J_{0}(k_1) J_{0}(k_2) J_{0}(k_3) J_{0}(k_4)\rangle\rangle_\text{CB}\notag\\&\hspace{-.75cm}-i\epsilon_{-\mu \nu}k_{1-}k_{1}^{\mu}\langle\langle T_{-}^{\nu}(k_1) J_{0}(k_2) J_{0}(k_3) J_{0}(k_4)\rangle\rangle_\text{odd}+(1 \leftrightarrow 2)+(1 \leftrightarrow 3)+(1 \leftrightarrow 4)\bigg]\notag\\
=\frac{96\widetilde{\lambda}_{\text{qf}}}{\widetilde{N}(1+  \widetilde{\lambda}_{\text{qf}}^{2})}&\bigg[k_{1-}\beta_{00}\beta_{2000}\langle J_0(k_1) J_0(-k_1)\rangle_\text{FF}\langle\langle T_{--}(k_1)J_0(k_2)J_{0}(k_3) J_{0}(k_4)\rangle\rangle_\text{even}\notag\\&\hspace{.5cm}+(1 \leftrightarrow 2)+(1 \leftrightarrow 3)+(1 \leftrightarrow 4)\bigg]\notag\\
+\frac{96\widetilde{\lambda}_{\text{qf}}}{\widetilde{N}(1+  \widetilde{\lambda}_{\text{qf}}^{2})}&\bigg[ \frac{k_{1-}}{8}\beta_{00}\gamma_{2000}\langle \widetilde J_0(k_1) \widetilde J_0(-k_1)\rangle_\text{CB}\langle\langle T_{--}(k_1)J_0(k_2)J_{0}(k_3) J_{0}(k_4)\rangle\rangle_\text{odd}\notag\\
&\hspace{.5cm}+(1 \leftrightarrow 2)+(1 \leftrightarrow 3)+(1 \leftrightarrow 4)\bigg]
\end{align*}
\begin{align}
\label{HSECom}
+\frac{96 \widetilde{\lambda}_{\text{qf}}}{\widetilde{N}(1+ \widetilde{\lambda}_{\text{qf}}^{2})}&\bigg[\langle J_{0}(-k_1-k_2) J_{0}(k_1) J_{0}(k_2) \rangle_\text{QF}\langle T_{--}(-k_3-k_4)J_{0}(k_3) J_{0}(k_4)\rangle_\text{QF} \notag\\&\hspace{.5cm}+(1 \leftrightarrow 3)+(1 \leftrightarrow 4)+(2 \leftrightarrow 3)+(2 \leftrightarrow 4)+(1,2\leftrightarrow 3,4)\bigg]
\end{align}
%
The  higher spin equation \eqref{HSECom} splits into two equations that correspond to the parity odd and parity even terms. We will also utilize the relation between three-point functions in quasi-fermionic and critical bosonic theories, which are as follows in our normalization\footnote{Note that for the critical bosonic theory we have used the following normalization for the scalar two-point  function  $$\langle \widetilde J_0(k_1) \widetilde J_0(-k_1)\rangle_\text{CB}=-k_1$$ and the three-point function is a constant given as $$\langle \widetilde J_0(k_1) \widetilde J_0(k_2)\widetilde J_0(-k_1-k_2)\rangle_\text{CB}=\frac{1}{2}$$.} :
\begin{align}
\label{QFCBrln}
\langle J_{0}(k_1) J_{0}(k_2) J_{0}(-k_1-k_2) \rangle_\text{QF}
&= (1+\widetilde\lambda_{\text{qf}}^2)\,\frac{ \widetilde N\widetilde\lambda_{\text{qf}}}{8}\langle\widetilde J_{0}(k_1) \widetilde J_{0}(k_2)\widetilde J_{0}(-k_1-k_2) \rangle_\text{CB}\notag\\[5pt]
\langle
T_{--}(-k_3-k_4)J_{0}(k_3) J_{0}(k_4)\rangle_\text{QF}&=\frac{\widetilde N}{2}(1+ \widetilde{\lambda}_{\text{qf}}^2)\langle T_{--}(-k_3-k_4)\widetilde{J}_{0}(k_3) \widetilde{J}_{0}(k_4)\rangle_\text{CB}\notag\\[5pt]
\langle J_0(k_1) J_0(-k_1)\rangle_\text{QF}&=\beta_{00}\langle  J_0(k_1)  J_0(-k_1)\rangle_\text{FF}\notag\\[5pt]
&=\frac{\widetilde{N}(1+\widetilde{\lambda}_{\text{qf}}^2)}{16}\langle \widetilde J_0(k_1) \widetilde J_0(-k_1)\rangle_\text{CB}
\end{align}
The parity even part of the higher spin equation  \eqref{HSECom}  is given by
\begin{align}
\frac{\widetilde{N}}{2}(1+\widetilde{\lambda}_{\text{qf}}^2)\bigg[(k_{1-}^{3}&+k_{2-}^3+k_{3-}^3+k_{4-}^3)\langle\langle J_{0}(k_1) J_{0}(k_2) J_{0}(k_3) J_{0}(k_4)\rangle\rangle_\text{FF}\notag\\[5pt]&\hspace{-.75cm}-i\epsilon_{-\mu \nu}k_{1-}k_{1}^{\mu}\langle\langle T_{-}^{\nu}(k_1) J_{0}(k_2) J_{0}(k_3) J_{0}(k_4)\rangle\rangle_\text{FF}+(1 \leftrightarrow 2)+(1 \leftrightarrow 3)+(1 \leftrightarrow 4)\bigg]\notag\\[5pt]+\frac{\widetilde{N}}{2}\widetilde{\lambda}_{\text{qf}}(1+\widetilde{\lambda}_{\text{qf}}^2)\bigg[(k_{1-}^{3}&+k_{2-}^3+k_{3-}^3+k_{4-}^3)\langle\langle J_{0}(k_1) J_{0}(k_2) J_{0}(k_3) J_{0}(k_4)\rangle\rangle_\text{CB}\notag\\[5pt]&\hspace{-.75cm}-6k_{1-}\langle \widetilde J_0(k_1) \widetilde J_0(-k_1)\rangle_\text{CB}\langle\langle T_{--}(k_1) \widetilde{J}_0(k_2)\widetilde J_{0}(k_3) \widetilde J_{0}(k_4)\rangle\rangle_\text{CB}\notag\\[5pt]&\hspace{3cm}+(1 \leftrightarrow 2)+(1 \leftrightarrow 3)+(1 \leftrightarrow 4)\notag\\[5pt]&\hspace{-.75cm}-6\langle \widetilde{J}_{0}(-k_1-k_2) \widetilde J_{0}(k_1) \widetilde J_{0}(k_2) \rangle_\text{CB}\langle T_{--}(-k_3-k_4)\widetilde J_{0}(k_3) \widetilde{J}_{0}(k_4)\rangle_\text{CB} \notag\\[5pt]&\hspace{1cm}+(1 \leftrightarrow 3)+(1 \leftrightarrow 4)+(2 \leftrightarrow 3)+(2 \leftrightarrow 4)+(1,2\leftrightarrow 3,4)\bigg]=0\label{HSECom1}
\end{align}
The first two lines of the above equation correspond to the higher spin equation in the free fermion theory \eqref{HSEFF4ptT} whereas the last five lines correspond to the higher spin equation of the critical bosonic theory given in \eqref{HSECB}. We thus see that the momentum space ansatz for $\langle TJ_0J_0J_0\rangle_\text{QF}$ motivated by its position space counter-part \cite{Li:2019twz} in \eqref{J24ptqf}  satisfies the higher spin equation corresponding to the even part.


The  parity odd  part of the higher spin equation  \eqref{HSECom}  is given by :
\begin{align}
&\epsilon_{-\mu \nu}k_{1-}k_{1}^{\mu}\langle\langle  T_{-}^{\nu}(k_1) \widetilde{J}_0(k_2)\widetilde{J}_{0}(k_3) \widetilde{J}_{0}(k_4)\rangle\rangle_\text{CB}+(1 \leftrightarrow 2)+(1 \leftrightarrow 3)+(1 \leftrightarrow 4)\notag\\[5pt]&=192\,i\,k_{1-}\langle J_0(k_1)J_0(-k_1)\rangle_\text{FF}\langle\langle T_{--}(k_1)J_0(k_2)J_{0}(k_3) J_{0}(k_4)\rangle\rangle_\text{FF}\notag\\[5pt]
&\hspace{.5cm}+(1 \leftrightarrow 2)+(1 \leftrightarrow 3)+(1 \leftrightarrow 4)
\end{align}
It can be easily checked that the correlators of the free fermionic and the critical bosonic theory satisfy this equation.

Thus through a simple algebraic manipulation of the higher spin equations we have verified the ansatz \eqref{J24ptqf} for $\langle TJ_0J_0J_0\rangle_\text{QF}$ in momentum space.

\subsection{Quasi-bosonic Theory}
We will now consider the quasi-bosonic theory. We will first discuss the two-point function of the spin-one current.

\subsubsection{Two-point functions}

\subsubsection*{ $\boldsymbol{\langle J_\mu J_\nu \rangle_\text{QB}}$  }
In this subsection we utilize the higher spin equations in the quasibosonic theory to determine the correlator $\langle J_{\mu}J_{\nu} \rangle_\text{QB}$.
Let us consider the action of $Q_3$ on $\langle J_- J_0 \rangle$ :
\begin{align}
\langle [Q_3, J_-(x_1)]J_0(x_2)\rangle_\text{QB}+\langle J_-(x_1) [Q_3, J_0(x_2)]\rangle_\text{QB}=\int_x\langle \partial_{\mu} J^{\mu}_{--}(x)J_-(x_1)J_0(x_2)\rangle_\text{QB}
\end{align}
By using the algebra given by \eqref{QBQ3algebra} we get 
\begin{align}
-\frac{1}{2}\,\langle\partial_{-}^3J_0(x_1)J_0(x_2)\rangle_\text{QB}+\langle J_-(x_1)\partial_{-}J_{-}(x_2)\rangle_\text{QB}=\int_x\langle \partial_{\mu} J^{\mu}_{--}(x)J_-(x_1)J_0(x_2)\rangle_\text{QB}
\end{align}
Making use of the current equation \eqref{DivJ3QB} and the large-$N$ factorisation in the above equation, after Fourier transforming we get the following equation in momentum space :
\begin{align}
\frac{1}{2}&\,k_{1-}^3\langle J_0(k_1)J_0(-k_1)\rangle_{\text{QB}}-k_{1-}\,\langle J_-(k_1)J_-(-k_1)\rangle_{\text{QB}}\notag\\[5pt]&=\frac{32\,\widetilde\lambda_{\text{qb}}}{\widetilde{N}(1+\widetilde\lambda_{\text{qb}}^2)}\epsilon_{\mu\nu-} k_{1-}k_{1}^{\nu}\langle J_0(k_1)J_0(-k_1)\rangle_\text{QB}\langle J^\mu(k_1)J_-(-k_1)\rangle_\text{QB}\label{HSE2QBM}
\end{align}
The two-point function of the scalar operator in quasi bosonic theory is \cite{Maldacena:2012sf} :
\begin{align}
\label{J0J0QB2pt}
\langle J_0(k)J_0(-k)\rangle_\text{QB}&=\frac{\widetilde{N}}{2}(1+\widetilde\lambda_{\text{qb}}^2)\langle J_{0}(k)J_{0}(-k)\rangle_\text{FB}\nonumber\\[5pt]
&=\frac{\widetilde{N}}{2}(1+\widetilde\lambda_{\text{qb}}^2)\times\frac{1}{8k}
\end{align}
The two-point function of the spin-one current in quasi bosonic theory has the following structure  :
\begin{align}
\langle J_\mu(k) J_\nu(-k)\rangle_\text{QB}&=\frac{\widetilde{N}}{2}\bigg[ \langle J_\mu(k) J_\nu(-k)\rangle_\text{FB}+d_{123} ~\epsilon_{\mu\nu k}\bigg]\nonumber\\[5pt]
&=\frac{\widetilde N}{2}\,\left[\frac{k_{\mu}\,k_{\nu}-g_{\mu\nu}\,k^2}{16\,k}+d_{123}\,\epsilon_{\mu\nu k}\right]
\end{align}
The first term on the RHS corresponds to the correlator in the free theory which was determined by solving the higher spin equation in the free theory in Section \ref{FB2ptJJeqn}, and the second term is a parity odd contact term whose coefficient $d_{123}$ is undetermined. Solving the higher spin equation  \eqref{HSE2QBM} we get :
\begin{align}
d_{123}=\frac{\widetilde\lambda_{\text{qb}}}{32}
\end{align}
%
In the free theory limit ($\widetilde\lambda_{\text{qb}}\rightarrow 0$), the coefficient of the parity odd term $d_{123}$ vanishes as expected \footnote{Parity odd contribution to the 2-point function of the spin-one current is zero in the free theory.}.
\subsubsection{Three-point functions}
In this subsection we utilize higher spin equations to determine $\langle J_1J_1J_0\rangle_{\text{QB}}$ in the quasi-bosonic theory. In \cite{Aharony:2012nh}, the authors obtained this correlator in a specific kinematic regime through Feynman diagram methods. The result we obtain precisely matches theirs in the specific kinematic regime explored by the authors.

As in quasi fermionic theory, three-point correlators in quasi-bosonic theory are given by the sum of contributions from the free boson theory (FB), the free fermion theory (FF) and a parity odd piece (odd) \cite{Maldacena:2011jn,Maldacena:2012sf} :
\begin{align}
\langle J_{s_{1}} J_{s_{2}} J_{s_{3}}\rangle=& \alpha_{s_{1} s_{2} s_{3}}\langle J_{s_{1}} J_{s_{2}} J_{s_{3}}\rangle_\text{FB}+\beta_{s_{1} s_{2} s_{3}}\langle J_{s_{1}} J_{s_{2}} J_{s_{3}}\rangle_\text{FF}+\gamma_{s_{1} s_{2} s_{3}}\langle J_{s_{1}} J_{s_{2}} J_{s_{3}}\rangle_\text{odd}
\end{align}
where $s_i$ represents the spin of the operator insertions. For the correlators of interest to us we have : 
\begin{align}
\langle\langle T_{\mu\nu}(k_1)J_0(k_2)J_0(k_3)\rangle\rangle_{\text{QB}}&=\alpha_{200}\langle\langle T_{\mu\nu}(k_1)J_0(k_2)J_0(k_3)\rangle \rangle_{\text{FB}} \label{TJ0J0QB}\\[8pt]
\langle\langle J_{\mu}(k_1)J_{\nu}(k_2)J_0(k_3)\rangle\rangle_{\text{QB}}&=\alpha_{110}\langle\langle J_{\mu}(k_1)J_\nu(k_2)J_0(k_3)\rangle\rangle_{\text{FB}}+\gamma_{110}\langle J_{\mu}(k_1)J_\nu(k_2)J_0(k_3)\rangle \rangle_{\text{QB}}^\text{odd}\notag\\[8pt]&\hspace{1cm}+\langle\langle J_{\mu}(k_1)J_{\nu}(k_2)J_0(k_3) \rangle\rangle_\text{QB}^{\text{contact}}\label{J1J1J0QB}\\[8pt]
\langle \langle J_0(k_1) J_0(k_2) J_0(k_3) \rangle\rangle_{\text{QB}}&=\alpha_{000}\langle\langle J_0(k_1) J_0(k_2) J_0(k_3)\rangle\rangle_{\text{FB}}\,\label{J0J0J0QB}
\end{align}
where  the coefficients $\beta_{200}$, $\alpha_{110}$, and $\gamma_{110}$ are given by :
\begin{align}
\alpha_{200}=\frac{\widetilde N}{2}\,\,(1+\widetilde\lambda_{\text{qb}}^2),\quad\alpha_{110}=\frac{\widetilde N}{2},\quad\gamma_{110}=\frac{\widetilde N \widetilde\lambda_{\text{qb}}}{2},\quad\alpha_{000}=\frac{\widetilde N}{2}\,(1+\widetilde\lambda_{\text{qb}}^2)
\end{align}
The explicit form of the free boson theory correlators appearing on the RHS of the equations above can be found in \cite{Jain:2020rmw} (see also \eqref{Tj0j0reconstructionF}, \eqref{jmujnu3pful}  and \eqref{3ptJ0FB} of Appendix \ref{notation}). Note that we allowed for a contact term in $\langle J_1J_1J_0\rangle_{\text{QB}}$. In the following we will demonstrate that the higher spin equation could be utilized to fix the parity odd term as well as the contact term in \eqref{J1J1J0QB}.
\subsubsection*{$\boldsymbol{\langle J_\mu J_\nu J_0\rangle_\text{QB}}$  }
Let us look at the Ward-identity  corresponding to the action of $Q_3$ on $\langle J_{-}J_0J_0\rangle$ :
\begin{align}
\label{eqnstep1}
\langle[Q_3,J_-( x_1)]J_0( x_2)J_0( x_3)\rangle&+\langle J_{-}( x_1)[Q_3,J_0( x_2)]J_0( x_3)\rangle+\langle J_{-} (x_1) J_0( x_2)[Q_3,J_0( x_3)]\rangle\nonumber\\[5pt]
&=\int _x \langle \partial_{\mu} J^{\mu}_{--}(x)J_-( x_1)J_0( x_2)J_0( x_3)\rangle
\end{align}
This leads to the following higher spin equation upon utilizing the algebra in \eqref{QBQ3algebra} 
\begin{align}
4&\widetilde d_{12}\,\langle\partial_{-}T_{--}( x_1)\,J_0( x_2)\,J_0( x_3)\rangle_\text{QB}-\frac{1}{2}\,\langle\partial_{-}^3J_{0}( x_1)\,J_0( x_2)\,J_0( x_3)\rangle_\text{QB}\notag\\[5pt]&+\left(\langle J_{-}( x_1)\,\partial_{-}J_{-}( x_2)\,J_0( x_3)\rangle_\text{QB}+2\leftrightarrow 3\right)=\int _x \langle \partial_{\mu} J^{\mu}_{--}(x)J_-(x_1)J_0(x_2)J_0(x_3)\rangle_\text{QB}
\end{align}
We now use the current equation \eqref{DivJ3QB} and the large $N$ factorisation, and obtain the following equation in momentum space  after performing a Fourier transform of the above equation :
\begin{align}
4\,&\widetilde d_{12}\,k_{1-}\langle\langle T_{--}(k_1)J_{0}(k_2)J_{0}(k_3)\rangle\rangle_{\text{QB}}+\frac{1}{2}k_{1-}^3\langle\langle J_0(k_1)J_0(k_2)J_0(k_3)\rangle\rangle_{\text{QB}}
\notag\\[5pt]&+\left(k_{2-}\langle\langle J_{-}(k_1)J_{-}(k_2)J_{0}(k_3)\rangle\rangle_{\text{QB}}+k_{3-}\langle\langle J_{-}(k_1)J_{0}(k_2)J_{-}(k_3)\rangle\rangle_{\text{QB}}\right)
\notag\\[5pt]&=\frac{32\,\widetilde\lambda_{\text{qb}}}{\widetilde N(1+\widetilde\lambda_{\text{qb}}^2)}\,\bigg[\epsilon_{\mu k_1 -}k_{1-}\langle\langle J_0(k_1)\,J_0(k_2)\,J_0(k_3)\rangle\rangle_{\text{QB}}\langle J_{-}( k_1)J^{\mu}(-k_1)\rangle_{\text{QB}}
\notag\\[5pt]&
\hspace{1cm}+(\epsilon_{\mu k_2 -}k_{2-}\langle J_0(k_2)\,J_0(-k_2)\rangle_{\text{QB}}\langle\langle J_{-}(k_1)J^{\mu}(k_2)J_0(k_3)\rangle\rangle_{\text{QB}}+2\leftrightarrow 3)\bigg]\label{HSSBHSE}
\end{align}
In this equation we substitute the form of the correlators $\langle TJ_0J_0\rangle_\text{QB}$, $\langle J_1J_1J_0\rangle_\text{QB}$ and $\langle J_0J_0J_0\rangle_\text{QB}$ given in \eqref{TJ0J0QB}, \eqref{J1J1J0QB} and \eqref{J0J0J0QB} respectively. We consider the following ansatz for $\langle J_1 J_1J_0\rangle_\text{QB}^{\text{odd}}$ which could be justified by a Legendre transformation :
\begin{align}
\langle\langle J_{\mu}(k_1)J_\nu(k_2)J_0(k_3)\rangle\rangle_{\text{QB}}^\text{odd}&=\frac{1}{k_3}\langle\langle J_{\mu}(k_1)J_\nu(k_2)J_0(k_3)\rangle\rangle_{\text{FF}}\label{JJJ0QBodd}
\end{align}
The higher spin equation in \eqref{HSSBHSE} splits into two separate equations, one proportional to the Levi-Civita tensor and the other without the Levi-Civita tensor.  We need to satisfy both the equations so that the higher spin equation \eqref{HSSBHSE} is solved consistently.
Note that the odd part of the correlator $\langle J_1J_1J_0\rangle_\text{{QB}}$ in \eqref{J1J1J0QB} contains  a single Levi Cevita tensor as it is determined in terms of the free fermionic correlator $\langle J_1J_1J_0\rangle_\text{FF}$ \eqref{JJJ0QBodd}. Hence, the part of  \eqref{HSSBHSE}  with a Levi-Civita tensor gives the following equation :
\begin{align}
&\gamma_{110}\,k_{2-}\langle\langle J_{-}(k_1)J_{-}(k_2)J_{0}(k_3)\rangle\rangle_\text{{QB}}^{\text{odd}}+2\leftrightarrow 3\notag\\[5pt]
&\hspace{-.5cm}=\frac{32\,\widetilde\lambda_{\text{qb}}}{\widetilde{N}(1+\widetilde\lambda_{\text{qb}}^2)}\,\bigg[\epsilon_{\mu k_2-}k_{2-}\alpha_{110}\langle\langle  J_{-}(k_1)J_\mu(k_2) J_{0}(k_3)\rangle\rangle_{\text{FB}}\langle J_0(k_2) J_0(-k_2)\rangle_{\text{QB}}+2\leftrightarrow 3\bigg]\label{HSSBHSEodd}
\end{align}
In this equation, we use the known expressions for $\langle  J_1J_1J_{0}\rangle_{\text{FB}}$  from \cite{Jain:2020rmw} and for $\langle  J_{0}J_{0}\rangle_{\text{QB}}$ from \eqref{J0J0QB2pt}. For $\langle J_1J_1J_0\rangle_\text{{QB}}^{\text{odd}}$ we substitute the ansatz from \eqref{JJJ0QBodd}. 
Upon making these substitutions we see that the higher spin equation \eqref{HSSBHSEodd} is exactly satisfied. 

Let us now consider the parity even part of the higher spin equation \eqref{HSSBHSE} which contains terms without the Levi-Civita tensor :
\begin{align}
&4\,\widetilde d_{12}\,k_{1-}\alpha_{200}\langle\langle T_{--}(k_1)J_{0}(k_2)J_{0}(k_3)\rangle\rangle_{\text{FB}}+\frac{1}{2}\,k_{1-}^3\alpha_{000} \langle\langle J_{0}(k_1)J_0(k_2)J_{0}(k_3)\rangle\rangle_{\text{FB}}\notag\\[5pt]
&+\left[k_{2-}\alpha_{110}\langle\langle J_{-}(k_1)J_{-}(k_2)J_{0}(k_3)\rangle\rangle_{\text{FB}}+2\leftrightarrow3\right]\notag\\[5pt]
&=\frac{32\,\widetilde\lambda_{\text{qb}}}{\widetilde{N}(1+\widetilde\lambda_{\text{qb}}^2)}\,\bigg[\epsilon_{\mu k_2-}k_{2-}\gamma_{110}\langle\langle J_{-}(k_1)J_\mu(k_2) J_{0}(k_3)\rangle\rangle_{\text{QB}}^\text{odd}\alpha_{00}\langle J_{0}(k_2)J_0(-k_2)\rangle_{\text{FB}}+2\leftrightarrow 3\notag\\[5pt]
&\hspace{2.7cm}-\alpha_{11}k_{1-}^3\,d_{123}\,\langle\langle J_0(k_1)J_0(k_2)J_0(k_3)\rangle\rangle\bigg]\label{HSSBHSEeven}
\end{align}
We substitute the form of the correlators $\langle TJ_0J_0\rangle_\text{QB}$ and $\langle J_1J_1J_0\rangle_\text{QB}$ given in \eqref{TJ0J0QB}  and \eqref{J1J1J0QB} respectively.
For correlators in the free boson theory we utilize the results  from \cite{Jain:2020rmw}. With these substitutions we see that the higher spin equation \eqref{HSSBHSEeven} is satisfied.

The above analysis for  $\langle J_{\mu}J_{\nu}J_0\rangle_{\text{QB}}$ is blind to any contact term in the correlator  that is proportional to $g_{\mu\nu}$  as we worked with  $\langle J_{-}J_{-}J_0\rangle$. In the following we compute this contact term in $\langle J_{\mu}J_{\nu}J_0\rangle_{\text{QB}}$.
%
\subsubsection*{Fixing contact term in $\boldsymbol{\langle J_\mu J_\nu J_0\rangle_\text{QB}}$  }
In this section we will demonstrate that higher spin equation is sensitive to the contact term in $\langle J_\mu J_\nu J_0\rangle_\text{QB}$ and it could be used as a tool to fix such contact terms. To do so let us first consider the algebra  in quasi-bosonic theory
\begin{align}
[Q_3,J_0]&=\,   \partial_{-}J_{-}\label{QBQ31algebra}\nonumber\\[5pt]
[Q_3,J_\mu]&= -\frac{1}{2}  \partial_{\mu}\partial_{-}^2J_0+ 4\,\widetilde d_{12}\,\partial_{-} T_{\mu-}-4\,\widetilde d_{12}\,g_{\mu-} \partial^\alpha T_{\alpha -}+\frac{\widetilde d_{10}}{2}  g_{\mu-} \Box \partial_{-}J_0
\end{align}
Let us consider  the action of $Q_3$ on $\langle J_\mu J_0 J_0\rangle_\text{QB}$
\begin{align}
\langle& [Q_3, J_\mu(x_1)]J_0(x_2)J_0(x_3)\rangle_\text{QB}+\langle J_\mu(x_1) [Q_3, J_0(x_2)]J_0(x_3)\rangle_\text{QB}+\langle J_\mu(x_1)J_0(x_2) [Q_3, J_0(x_3)]\rangle_\text{QB}\notag\\[5pt]&\hspace{4cm}=\int_x \langle\partial_{\mu} J^{\mu}_{--}(x)J_\mu(x_1)J_0(x_2)J_0(x_3)\rangle_\text{QB}
\end{align}
Utilizing the algebra in \eqref{QBQ31algebra}, the current equation \eqref{DivJ3QB} and the large $N$ factorisation, the above equation gives the following in momentum space after Fourier transforming :
\begin{align}
&\frac{1}{2}   k_{1\mu}k_{1-}^2\langle\langle J_0(k_1)J_0(k_2)J_0(k_3)\rangle\rangle_\text{QB} +4\,k_{1-}\langle\langle T_{\mu-}(k_1)J_0(k_2)J_0(k_3)\rangle\rangle_\text{QB}-4 g_{\mu-}k_{1}^\alpha\langle\langle T_{\alpha -}(k_1)J_0(k_2)J_0(k_3)\rangle\rangle_\text{QB}\notag\\[5pt]&-\frac{1}{2} \,g_{\mu-}k_{1}^2k_{1-}\langle\langle J_0(k_1)J_0(k_2)J_0(k_3)\rangle\rangle_\text{QB}+(k_{2-}\langle\langle J_{\mu}(k_1)J_-(k_2)J_0(k_3)\rangle\rangle_\text{QB}+2\leftrightarrow 3)\notag\\[5pt]&=\frac{32\,\widetilde\lambda_{\text{qb}}}{\widetilde N(1+\widetilde\lambda_{\text{qb}}^2)}\bigg[\epsilon_{\alpha k_1-}k_{1}^{\nu}k_{1-}\langle J_{\mu}(k_1)J^{\alpha}(-k_1) \rangle_\text{QB}\langle\langle J_0(k_1)J_0(k_2)J_0(k_3)\rangle\rangle_\text{QB}\notag\\[1pt]&\hspace{2.5cm}+\epsilon_{\alpha k_2-}k_{2-}\langle\langle J_{\mu}(k_1)J^{\alpha}(k_2)J_0(k_3) \rangle\rangle_\text{QB}\langle J_0(k_2)J_0(-k_2)\rangle_\text{QB}+2\leftrightarrow 3\bigg]\label{HSEQ3JmuJ0J0}
\end{align}
The equation can be split into even and odd parts depending on the presence or absence of the Levi-Civita tensor. 
The parity odd part of the  above higher spin equation is given by :
\begin{align}
\label{generalmuHSE}
&(k_{2-}\langle\langle J_{\mu}(k_1)J_-(k_2)J_0(k_3)\rangle\rangle_\text{QB}^{\text{odd}}+2\leftrightarrow 3)\notag\\[5pt]&=\frac{32\,i\,\widetilde{\lambda}_{\text{qb}}}{\widetilde N\,(1+ \widetilde{\lambda}_{\text{qb}}^2)}\,\epsilon_{\alpha\nu-}\bigg[k_{1}^{\nu}k_{1-}\langle J_{\mu}(k_1)J^{\alpha}(-k_1)\rangle_\text{QB}^{\text{even}}\langle\langle J_0(k_1)J_0(k_2)J_0(k_3)\rangle\rangle_\text{QB}\notag\\&\hspace{2.5cm}+k_{2}^{\nu}k_{2-}\langle\langle J_{\mu}(k_1)J^{\alpha}(k_2)J_0(k_3) \rangle\rangle_\text{QB}^{\text{even}}\langle J_0(k_2)J_0(-k_2)\rangle_\text{QB}+2\leftrightarrow 3\bigg]
\end{align}
Since we  have already determined the form of $\langle J_1J_1J_0\rangle_{\text{FB}}$ and $\langle J_1J_1J_0\rangle_{\text{QB}}^\text{odd}$ and are now interested in determining $\langle J_1J_1J_0\rangle_\text{QB}^{\text{contact}}$ we project out the transverse parts of the correlators and keep only the local terms by contracting \eqref{generalmuHSE} with $k_{1}^\mu$. This leads to :
\begin{align}
&( k_{2-}k_{1}^{\mu}\langle\langle J_{\mu}(k_1)J_-(k_2)J_0(k_3)\rangle\rangle_\text{QB}^{\text{odd}}+2\leftrightarrow 3)\notag\\[5pt]&=\frac{32\,i\,\widetilde{\lambda}_{\text{qb}}}{\widetilde N\,(1+ \widetilde{\lambda}_{\text{qb}}^2)}\,\epsilon_{\alpha\nu-}\bigg[k_{1}^{\nu}k_{1-}k_{1}^{\mu}\langle J_{\mu}(k_1)J^{\alpha}(-k_1)\rangle_\text{QB}\langle\langle J_0(k_1)J_0(k_2)J_0(k_3)\rangle\rangle_\text{QB}\notag\\[5pt]&\hspace{2cm}+k_{2}^{\nu}\,k_{2-}\,k_{1}^{\mu}\,\langle\langle J_{\mu}(k_1)J^{\alpha}(k_2)J_0(k_3) \rangle\rangle_\text{QB}^{\text{even}}\langle J_0(k_2)J_0(-k_2)\rangle_\text{QB}+2\leftrightarrow 3\bigg]
\end{align}
We substituted the form of various correlators that appear in the above equation and found that the ansatz in \eqref{JJJ0QBodd} for $\langle J_{\mu}J_{\nu}J_0 \rangle_\text{QB}^{\text{odd}}$ solves the above equation provided we allow for a contact term of the following form :
\begin{align}
\langle\langle J_{\mu}(k_1)J_{\nu}(k_2)J_0(k_3) \rangle\rangle_\text{QB}^{\text{contact}}&=-2\,\gamma_{110}\frac{g_{\mu\nu}}{k_3}\label{JmuJ0J0QBoddct}
\end{align}

The even part of the higher spin equation  \eqref{HSEQ3JmuJ0J0} is given by :
\begin{align}
&\frac{i}{2}\,k_{1\mu}k_{1-}^2\langle\langle J_0(k_1)J_0(k_2)J_0(k_3)\rangle\rangle_\text{QB} +4\,i\,k_{1-}\langle\langle T_{\mu-}(k_1)J_0(k_2)J_0(k_3)\rangle\rangle_\text{QB}\notag\\[5pt]&-4\,g_{\mu-}k_{1}^\alpha\langle\langle T_{\alpha -}(k_1)J_0(k_2)J_0(k_3)\rangle\rangle_\text{QB}-\frac{i}{2}\,g_{\mu-}k_{1}^2k_{1-}\langle\langle J_0(k_1)J_0(k_2)J_0(k_3)\rangle\rangle_\text{QB}\notag\\[5pt]&+  k_{2-}\langle\langle J_{\mu}(k_1)J_-(k_2)J_0(k_3)\rangle\rangle^{\text{even}}_\text{QB}+k_{3-}\langle\langle J_{\mu}(k_1)J_0(k_2)J_-(k_3)\rangle\rangle^{\text{even}}_\text{QB}\notag\\[5pt]&
=\frac{32\,i\,\widetilde{\lambda}_{\text{qb}}}{\widetilde N\,(1+ \widetilde{\lambda}_{\text{qb}}^2)}\epsilon_{\alpha\nu-}\bigg[\epsilon_{\alpha k_1-}k_{1-}\langle J_{\mu}(k_1)J^{\alpha}(-k_1) \rangle_\text{QB}^{\text{contact}}\langle\langle J_0(k_1)J_0(k_2)J_0(k_3)\rangle\rangle_\text{QB}\notag\\[5pt]&\hspace{2cm}+\epsilon_{\alpha k_2-}k_{2-}\langle J_0(k_2)J_0(-k_2)\rangle_\text{QB}\langle\langle J_{\mu}(k_1)J^{\alpha}(k_2)J_0(k_3) \rangle\rangle_\text{QB}^{\text{odd}}+2\leftrightarrow 3\bigg]
\end{align}
Once again since we are only interested in determining the contact terms, we project out the transverse parts of the correlators and keep only the local terms by contracting the above equation with $k_{1}^\mu$. This gives :
\begin{align}
& \frac{i}{2}   k_{1}^2k_{1-}^2\langle\langle J_0(k_1)J_0(k_2)J_0(k_3)\rangle\rangle_\text{QB} +4\,i\,k_{1-}k_{1}^{\mu}\langle\langle T_{\mu-}(k_1)J_0(k_2)J_0(k_3)\rangle\rangle_\text{QB}\notag\\&-4 k_{1-}k_{1}^\alpha\langle\langle T_{\alpha -}(k_1)J_0(k_2)J_0(k_3)\rangle\rangle_\text{QB}-\frac{i}{2}  k_{1}^2k_{1-}^2\langle\langle J_0(k_1)J_0(k_2)J_0(k_3)\rangle\rangle_\text{QB}\notag\\[5pt]&+  k_{2-}k_{1}^{\mu}\langle\langle J_{\mu}(k_1)J_-(k_2)J_0(k_3)\rangle\rangle_\text{QB}+k_{3-}k_{1}^{\mu}\langle\langle J_{\mu}(k_1)J_0(k_2)J_-(k_3)\rangle\rangle_\text{QB}\notag\\&
=\frac{32\,i\,\widetilde{\lambda}_{\text{qb}}}{\widetilde N\,(1+ \widetilde{\lambda}_{\text{qb}}^2)}\bigg[\epsilon_{\alpha k_1-}k_{1-}k_{1}^\mu\langle J_{\mu}(k_1)J^{\alpha}(-k_1)\rangle_\text{QB}^{\text{contact}}\langle\langle J_0(k_1)J_0(k_2)J_0(k_3)\rangle\rangle_\text{QB}\notag\\&\hspace{2cm}+\epsilon_{\alpha k_2-}k_{2-}k_{1}^\mu\langle\langle J_{\mu}(k_1)J^{\alpha}(k_2)J_0(k_3) \rangle\rangle_\text{QB}^{\text{odd}}\langle J_0(k_2)J_0(-k_2)\rangle_\text{QB}+2\leftrightarrow 3\bigg]
\end{align}
We checked that the above equation is satisfied by the ansatz  in \eqref{JJJ0QBodd} for $\langle J_{\mu}J_{\nu}J_0 \rangle_\text{QB}^{\text{odd}}$ provided we have also considered the contact term in \eqref{JmuJ0J0QBoddct}.

%

To summarize, we verified the ansatz for $\langle J_1J_1J_0\rangle_{\text{QB}}^\text{odd}$ in \eqref{JJJ0QBodd} by showing that it solves the higher spin equation constructed from the action of $Q_3$ on $\langle J_{-}J_0J_0\rangle$. We also fixed the contact term in $\langle J_{\mu}J_{\nu}J_0\rangle_{\text{QB}}$  \eqref{JmuJ0J0QBoddct} utilizing the action of $Q_3$ on $\langle J_{\mu}J_0J_0\rangle_\text{QB}$. Our result matches the results in  \cite{Aharony:2012nh} obtained by explicit Feynman diagram techniques in a specific kinematic regime.

\subsubsection*{$\boldsymbol{Q_4\,on\,\langle J_0J_0J_0\rangle_\text{QB}}$}
Let us now consider the action of $Q_4$ on $\langle J_0J_0J_0\rangle$. This gives :
\begin{align}
\label{eqnstep11}
\langle[Q_4,J_0( x_1)]J_0( x_2)J_0( x_3)\rangle&+\langle J_{0}( x_1)[Q_4,J_0( x_2)]J_0( x_3)\rangle+\langle J_{0} (x_1) J_0( x_2)[Q_4,J_0( x_3)]\rangle\nonumber\\[5pt]
&=\int _x \langle \partial_{\mu} J^{\mu}_{---}(x)J_-( x_1)J_0( x_2)J_0( x_3)\rangle
\end{align}
This leads to the following higher spin equation upon utilizing the algebra in \eqref{QBQ4algebra} 
\begin{align}
&\langle\partial_{-}^3J_0( x_1)\,J_0( x_2)\,J_0( x_3)\rangle_\text{QB}-\frac{24}{5}\langle\partial_-T_{--}(x_1)\,J_0( x_2)\,J_0( x_3)\rangle_\text{QB}+1\leftrightarrow 2+1\leftrightarrow 3\notag\\[5pt]
&\hspace{2cm}=\int _x \langle \partial_{\mu} J^{\mu}_{---}(x)J_-(x_1)J_0(x_2)J_0(x_3)\rangle_\text{QB}
\end{align}
We now use the current equation \eqref{DivJ4QB} and large $N$ factorisation, and obtain the following equation in momentum space  after performing a Fourier transform of the above equation :
\begin{align}
&\left(k_{1-}^3+k_{2-}^3+k_{3-}^3\right)\langle\langle J_0(k_1)\,J_0(k_2)\,J_0(k_3)\rangle\rangle_{\text{QB}}\notag\\[5pt]
&\hspace{2cm}+\left[\frac{24}{5}\langle\langle T_{--}(k_1)\,J_0(k_2)\,J_0(k_3)\rangle\rangle_{\text{QB}}+1\leftrightarrow 2+1\leftrightarrow 3\right]=0
\end{align}
In this equation we substitute the form of the correlators $\langle TJ_0J_0\rangle_\text{QB}$ and $\langle J_0J_0J_0\rangle_\text{QB}$ given in \eqref{TJ0J0QB} and \eqref{J0J0J0QB} respectively and see that the equation is satisfied.
\subsubsection{Four-point function}

In this subsection we compute in momentum space the four-point correlation function of the scalar primary operator $J_0=\bar{\phi}\phi$ in the quasi-bosonic theory utilizing the corresponding results in position space \cite{Turiaci:2018nua}.
\subsubsection*{$\boldsymbol{\langle J_0J_0J_0J_0\rangle_\text{QB}}$ }
In \cite{Jain:2020rmw} we obtained the momentum space four-point function of the scalar operator $J_0$ in the free boson theory. Here we will determine the same for the quasi-bosonic theory. The correlator has been studied in a specific kinematic and parameter regime in \cite{Bedhotiya:2015uga,Yacoby:2018yvy}.
Recently, it was demonstrated that in the large-$N$ limit, this correlation function in position space is  determined by the free theory answer upto a conformal partial wave which is given by a $D$ function as follows \cite{Turiaci:2018nua} :
\begin{align}\label{RBP4}
\langle &J_0(x_1) J_0(x_2) J_0(x_3) J_0(x_4)\rangle_\text{QB}=\widetilde{N}(1+\widetilde{\lambda}_{\text{qb}}^2)^2\frac{f_{q b}(u, v)}{x_{13}^{2} x_{24}^{2}}+disc
\end{align}
where
\begin{align}
f_{q b}(u, v)&=f_{f b}(u, v)-\frac{\widetilde{\lambda}_{q b}^{2}}{1+\widetilde{\lambda}_{q b}^{2}} \frac{8}{\pi^{5 / 2}}\left(\bar{D}_{1,1, \frac{1}{2}, \frac{1}{2}}(u, v)+\bar{D}_{1,1, \frac{1}{2}, \frac{1}{2}}(v, u)+\frac{1}{u} \bar{D}_{1,1, \frac{1}{2}, \frac{1}{2}}\left(\frac{1}{u}, \frac{v}{u}\right)\right)\label{f43qb}
\end{align}
and $disc$ represents the disconnected part. The difference in the prefactor with \cite{Turiaci:2018nua} is due to the normalization that we have chosen for the scalar operator in \eqref{2ptJ0QB}. The function $f_{f b}(u, v)$  in the above equation is the corresponding function in the free boson theory defined through the corresponding four-point correlator as follows
\begin{align}
\langle &J_0(x_1) J_0(x_2) J_0(x_3) J_0(x_4)\rangle_\text{FB}= \frac{f_{fb}(u, v)}{x_{13}^{2} x_{24}^{2}}+disc 
\end{align}
The  $ \bar{D}$ function in \eqref{f43qb} is related to the  conformal partial wave $\Psi_{\Delta, 0}^{\Delta^{t}}$ as follows
\begin{align}
\bar{D}_{\frac{d-\Delta}{2} \frac{d-\Delta}{2} \frac{\Delta}{2} \frac{\Delta}{2}}(u, v)= x_{12}^{2 \Delta_{i}} x_{34}^{2 \Delta_{i}}\frac{\Gamma\left(\frac{\Delta}{2}\right)^{2} \Gamma\left(\frac{d-\Delta}{2}\right)^{2}}{\pi^{d / 2}}u^{-\frac{d-\Delta}{2}}\Psi_{\Delta, 0}^{\Delta^{t}}
\end{align}
where $\Delta_{i}$ corresponds to the scaling dimension of the external operator whereas $\Delta$ corresponds to the scaling dimension of the exchange operator. The conformal partial wave $ \Psi_{\Delta, 0}^{\Delta^{t}}$ is defined as :
\begin{equation}\label{CPW2}
\Psi^{\Delta^i}_{\Delta,0}=\int d^dx\langle\mathcal{O}_1(x_1)\mathcal{O}_2(x_2)  \mathcal{O}(x)\rangle\langle \widetilde{\mathcal{O}}(x)\mathcal{O}_3(x_3)\mathcal{O}_4(x_4)\rangle
\end{equation}
where $\mathcal{O}$ corresponds to the exchange/intermediate operator with dimension $\Delta$ whereas $\widetilde{\mathcal{O}}$ corresponds to the corresponding shadow operator of dimension $\tilde{\Delta}=d-\Delta$ defined in Appendix \ref{CPWS}.
Since we are interested in the scalar operator of the bosonic theory, $\Delta_{i}=1$ and it was shown in \cite{Turiaci:2018nua} that the exchange that has to be added corresponds to $\Delta=1$ and $d=3$, and therefore we get the relation between the required partial wave and the $\bar{D}_{11\frac{1}{2}\frac{1}{2}}$ to be as follows :
\begin{eqnarray}\label{CPWF2}
\frac{1}{x_{13}^{2}x_{24}^{2}}\bar{D}_{11\frac{1}{2}\frac{1}{2}}(u,v)&=\frac{1}{\pi^{\frac{1}{2}}}\Psi^{1}_{1,0} 
\end{eqnarray}
Note that the first term in \eqref{f43qb} corresponds to that of the free boson theory and hence upon Fourier transform it just turns into the momentum space correlation function of the free boson theory. On the other hand we see from the above equation  that  the Fourier transform of the  $\bar{D}$ function is simply related to the Fourier transform of the conformal partial wave $\Psi^{1}_{1,0} $ which we denote as $\widetilde{\Psi}^{1}_{1,0}$. Note that  $\bar{D}$ function does not have a simple closed form expression. However as we show in the Appendix \ref{CPWS} the conformal partial wave in momentum space is very simple  and is equal to the product of two three-point functions multiplied by a two-point function. 

Substituting \eqref{CPWF2} in \eqref{RBP4} and performing a Fourier transform of the   $\bar{D}$ functions  as shown in the Appendix \ref{CPWS}, we get the final result for the four-point scalar correlator of the quasi-bosonic theory to be as follows :
\begin{align}
&\langle\langle J_{0}( k_1) J_{0}( k_2) J_0(k_3)J_0(k_4)\rangle\rangle_\text{QB}\notag\\&~~=\widetilde{N}^2(1+\widetilde{\lambda}_{\text{qb}}^2)^2\bigg[\frac{1}{\widetilde{N}}\langle\langle J_{0}( k_1) J_{0}( k_2) J_0( k_3)J_0( k_4)\rangle\rangle_\text{FB}\notag\\&\hspace{3cm}-\frac{8}{\pi^{3}}\frac{\widetilde{\lambda}_{q b}^{2}}{\widetilde{N}(1+\widetilde{\lambda}_{q b}^{2})} 
\bigg(\widetilde{\Psi}^{1}_{1,0}(k_1,k_2,k_3,k_4)+(2\leftrightarrow3)+(2\leftrightarrow4)\bigg)\bigg]
\end{align}
where
\begin{align}
\widetilde{\Psi}^{1}_{1,0}(p_i)
= \frac{\pi^2}{|p_3+p_4||p_1-p_2| |p_3-p_4| |2p_1+p_2||2p_2+p_1||2p_3+p_4||2p_4+p_3|}. \label{CPWFin}
\end{align}
has been derived in  \eqref{CPWFin}, and for $\langle J_{0}(k_1) J_{0}( k_2) J_0( k_3)J_0( k_4)\rangle$ see \cite{Jain:2020rmw}.  In Appendix \ref{5ptJ0} we attempt a derivation of the five-point function of the scalar operator in the quasi-fermionic theory based on conformal block expansion.

\section{Higher spin equations away from conformal fixed point}
\label{Massive}
In  previous sections of this article we focussed on solving higher spin equation explicitly at the  conformal fixed point. We have seen that, at the conformal fixed point  higher spin symmetry is highly constraining. However one can ask how effective is higher spin equation when we deform CFT with relevant or marginal deformations away from fixed point. In this section we examine this interesting question by considering a prototypical example of the massive free boson theory.  We will demonstrate that one may use the higher spin equation to determine some of  the two-point functions of conserved higher spin operators  in this theory. We demonstrate this through an example where we determine $\langle J_{\mu} J_\nu\rangle $ by using   $\langle T_{--} J_0\rangle $  and $\langle J_{0} J_0\rangle $ in the free massive bosonic theory. We also comment on how one could determine several such two-point correlators. This leads us to propose that the corresponding higher spin equations could be used to constrain  spinning correlators even in theories away from conformal fixed point.

\subsection{Massive Free Boson}
Let us begin by considering the action  of $Q_4$ on $\langle J_0 J_0\rangle $ in the free massive bosonic theory
\begin{align}
\langle [Q_4, J_0(x_1)]J_0(x_2)\rangle_{\text{FMB}}+\langle J_0(x_1) [Q_4, J_0(x_2)]\rangle_{\text{FMB}}=0\label{HSEMBQ3J0J0}
\end{align}
The subscript $FMB$ in the above equation represent that the correlator is computed in the free massive bosonic theory. It may be observed that algebra for the massive case remains the same as that of the free theory. Utilizing the algebra in \eqref{FBQ3algebra} the higher spin equation takes the form :
\begin{align}
\langle \partial_{-}^3J_0(x_1)J_0(x_2)\rangle_{\text{FMB}}-\frac{24}{5}\,\langle \partial_-T_{--}(x_1)J_0(x_2)\rangle_{\text{FMB}}+1\leftrightarrow 2=0
\end{align}
Note that the due to the absence of conformal symmetry the two-point correlators of difference spins need not be zero in general. In momentum space this leads to
\begin{align}
(k_{1-}^3+k_{2-}^3)\langle J_0(k_1)J_0(k_2)\rangle_{\text{FMB}}+\frac{24}{5}(k_{1-}\langle T_{--}(k_1)J_0(k_2)\rangle_{\text{FMB}}+k_{2-}\langle J_0(k_1) T_{--}(k_2)\rangle_{\text{FMB}})=0
\end{align}
Utilizing momentum conservation we see that the above equation
is trivially satisfied.  

As the above higher spin equation did not constrain the correlators, let us consider  a different higher spin equation obtained by the action of $Q_3$ on $\langle J_{-}J_0 \rangle$ :
\begin{align}
\langle [Q_3, J_-(x_1)]J_0(x_2)\rangle_{\text{FMB}}+\langle J_-(x_1) [Q_3, J_0(x_2)]\rangle_{\text{FMB}}=0
\end{align}
Utilizing the algebra in \eqref{FBQ3algebra} the momentum space higher spin equation takes the form :
\begin{align}
4 k_{1-} \langle T_{--}(k_1) J_0(-k_1)\rangle_{\text{FMB}}+\frac{1}{2}k_{1-}^3\langle J_0(k_1) J_0(-k_1)\rangle_{\text{FMB}}- k_{1-}\langle J_-(k_1) J_-(-k_1)\rangle_{\text{FMB}}=0\label{HSEMBQ3JmJ0}
\end{align}
We can use the above equation to determine the $\langle J_{\mu} J_{\nu}\rangle_{\text{FMB}}$ correlator by using the expressions for $\langle J_{0} J_0\rangle_{\text{FMB}}$ and $\langle T_{--} J_0\rangle_{\text{FMB}}$ obtained by direct computation given as follows (see Appendix \ref{massive-app}) :
\begin{align}\label{expltrst}
\langle J_{0}(k) J_0(-k)\rangle_{\text{FMB}}&=\frac{1}{4\pi k}\arctan\left(\frac{k}{2m}\right)\notag\\
\langle T_{--}(k) J_0(-k) \rangle_{\text{FMB}}&=\frac{k_-^2}{16\pi k^3}\bigg[2m^2\arctan\left(\frac{k}{2m}\right)-k\,m\bigg]\notag\\
\langle J_{-}(k) J_0(-k)\rangle_{\text{FMB}}&=0
\end{align}
Upon utilizing \eqref{HSEMBQ3JmJ0} and \eqref{expltrst} the above correlators we obtain 
\begin{align}
\langle J_{-}(k) J_-(-k)\rangle_{\text{FMB}}&=\frac{k_-^2}{8\pi k^3}\bigg[(k^2+4m^2)\arctan\left(\frac{k}{2m}\right)-2\,k\,m\bigg]\label{JmJmFMB}
\end{align}
We could use this to determine $\langle J_{\mu} J_\nu\rangle_{\text{FMB}}$. This because $J_\mu J_\nu$ is a conserved current and hence the reconstruction formula fixes the ansatz for this to be
\begin{align}
\langle J_{\mu}(k) J_\nu(-k)\rangle_{\text{FMB}}&=\langle J_{\mu}(k) J_\nu(-k)\rangle^\text{transverse}_{\text{FMB}}+\langle J_{\mu}(k) J_\nu(-k)\rangle^\text{local}_{\text{FMB}}\label{JmJmFMB2}
\end{align}
\begin{align}
\langle J_{\mu}(k) J_\nu(-k)\rangle^\text{transverse}_{\text{FMB}}&=A\left[g_{\mu\nu}-\frac{k_\mu k_\nu}{k^2}\right]\\[5pt]
\langle J_{\mu}(k) J_\nu(-k)\rangle^\text{local}_{\text{FMB}}&=2 \frac{k_{\mu}k_{\nu}}{k^2}\langle J_0(k)\rangle_{FMB}\notag\\[5pt]
&= -\frac{m}{2k^2\pi}k_{\mu}k_{\nu}
\end{align}
Comparing \eqref{JmJmFMB} and \eqref{JmJmFMB2}, the form factor $A$ is  fixed from higher spin equation to be
\begin{align}
A=-\frac{1}{8\pi k}\bigg[(k^2+4m^2)~\arctan\left(\frac{k}{2m}\right)+2\,k\,m\bigg]
\end{align}
The form of the correlator thus fixed by  the higher spin equation exactly matches with that obtained by explicit computation details of which are given in the appendix.

This precisely matches the result obtained from direct computation. We emphasize here one could do a similar analysis to obtain some of the higher  spin correlators through the correlators of the lower spin to build up a lot of data. For example one could determine  the correlator $\langle J_3 J_1\rangle $ through a similar analysis as above  by  considering the action of $Q_4$ on $\langle J_\mu J_\nu\rangle$ which results in a higher spin equation involving   $\langle J_3 J_1\rangle $ and  $\langle J_1 J_1\rangle $.

\section{Summary}
In this section we summarise the results we obtained for various correlators and comment on duality.
\label{Summary1}
\subsection{$\langle J_1J_1J_0\rangle$ in Quasi-fermionic theory}
In the quasi-fermionic theory we have 
\begin{align}
\langle J_{\mu}(k_1)J_{\nu}(k_2)J_0(k_3)\rangle_{\text{QF}}&=\frac{\widetilde N}{2}\langle J_{\mu}(k_1)J_\nu(k_2)J_0(k_3)\rangle_{\text{FF}}+\frac{\widetilde N  \widetilde{\lambda}_{\text{qf}}}{2}\left(\langle J_{\mu}(k_1)J_\nu(k_2)J_0(k_3)\rangle_{\text{odd}}-\frac{g_{\mu\nu}}{4}\right)
\end{align}
where \footnote{As noted in Section \ref{3ptQFsection}, following the general literature, we label the part of a correlator that does not come from the free theory answer as odd. However this label need not always correspond to expressions that are odd under parity. This terms is multiplied by $\widetilde\lambda_{qf}$ which is odd under parity, and hence the labelling odd for this term in the correlator.}
\begin{align}
\langle J_{\mu}(k_1)J_\nu(k_2)J_0(k_3)\rangle_\text{odd}&=\langle J_{\mu}(k_1)J_\nu(k_2)\widetilde{J}_0(k_3)\rangle_{\text{CB}}
\end{align}
We will now express the above in terms of $N_f$ and $\lambda_f$ which are given by the following relations : 
\begin{align}
\widetilde N=2N_f\frac{\sin(\pi\lambda_f)}{\pi\lambda_f}\,\quad \widetilde\lambda_{\text{qf}}=\tan\left(\pi\lambda_f/2\right)
\end{align}
We then have :
\begin{align}
\label{JJOlambdaNQF}
\langle J_{\mu}(k_1)J_{\nu}(k_2)J_0(k_3)\rangle_{\text{QF}}&=N_f\frac{\sin(\pi\lambda_f)}{\pi\lambda_f}\langle J_{\mu}(k_1)J_\nu(k_2)J_0(k_3)\rangle_{\text{FF}}\nonumber\\[5pt]
&\hspace{.5cm}+\frac{2N_f\sin^2\left(\frac{\pi\lambda_f}{2}\right)}{\pi\lambda_f}\left(\langle J_{\mu}(k_1)J_\nu(k_2)J_0(k_3)\rangle_{\text{odd}}-\frac{g_{\mu\nu}}{4}\right)
\end{align}
It can be checked that this result (including the contact term) matches the result for the correlator available in the kinematic regime where the $+$ and $-$ components of momenta are zero (see equations 40 and 41 of \cite{GurAri:2012is}). The contact term can however be removed by adding suitable counter-terms to the action.
\subsubsection*{Duality with critical bosonic theory}
The parameters of the quasi-fermionic theory are related to those of the critical bosonic theory as follows :
\begin{align}
\widetilde{N}=2 N_{f} \frac{\sin \left(\pi \lambda_{f}\right)}{\pi \lambda_{f}}=2 N_{b} \frac{\sin \left(\pi \lambda_{b}\right)}{\pi \lambda_{b}}\\[5pt]
\widetilde{\lambda}_{\text{qf}}=\tan \left(\frac{\pi \lambda_{f}}{2}\right)=-\cot \left(\frac{\pi \lambda_{b}}{2}\right)
\end{align}
The duality transformation on the parameters is given by
\begin{align}
\lambda_f=\lambda_b-1,\quad \kappa_f=-\kappa_b
\end{align}
where $\kappa_{f/b}=\frac{N_{f/b}}{\lambda_{f/b}}$.
Implementing these transformations on the correlator \eqref{JJOlambdaNQF} one can check that it gives the critical bosonic theory answer  \cite{Aharony:2012nh} for the special kinematic regime considered therein.
\subsection{$\langle J_1J_1J_0\rangle$ in Quasi-bosonic theory}
In the quasi-bosonic theory we have
\begin{align}
&\langle J_{\mu}(k_1)J_\nu(k_2)J_0(k_3)\rangle_{\text{QB}}=\frac{\widetilde N}{2}\langle J_{\mu}(k_1)J_\nu(k_2)J_0(k_3)\rangle_{\text{FB}}\nonumber\\[5pt]
&\hspace{2.0cm}+\frac{\widetilde N\widetilde\lambda_{\text{qb}}}{2}\left[\frac{1}{k_3}\langle J_{\mu}(k_1)J_\nu(k_2)J_0(k_3)\rangle_{\text{FF}}-2\,\frac{g_{\mu\nu}}{k_3}\right]
\end{align}
We will now express the above in terms of $N_b$ and $\lambda_b$ which are given by the following relations : 
\begin{align}
\widetilde N=2N_b\frac{\sin(\pi\lambda_b)}{\pi\lambda_b}\,\quad \widetilde\lambda_{\text{qb}}=\tan\left(\pi\lambda_b/2\right)
\end{align}
We then obtain :
\begin{align}
&\langle J_{\mu}(k_1)J_\nu(k_2)J_0(k_3)\rangle_{\text{QB}}=N_b\frac{\sin(\pi\lambda_b)}{\pi\lambda_b}\langle J_{\mu}(k_1)J_\nu(k_2)J_0(k_3)\rangle_{\text{FB}}\nonumber\\[5pt]
&\hspace{2.0cm}+\frac{2N_b\sin^2\left(\frac{\pi\lambda_b}{2}\right)}{\pi\lambda_b}\left[\frac{1}{k_3}\langle J_{\mu}(k_1)J_\nu(k_2)J_0(k_3)\rangle_{\text{FF}}-2\,\frac{g_{\mu\nu}}{k_3}\right]
\end{align}
It can be checked that this matches the results for the correlator available in the kinematic regime where the $+$ and $-$ components of momenta are zero (see equations 54 and 56 of \cite{Aharony:2012nh}). An exactly similar analysis as in the quasi-fermionic theory can be performed here to obtain the critical fermionic theory results.

\subsection{Summary of remaining correlators}
In this section we summarise the results for the rest of the correlators we study in this paper.
\subsection*{Quasi-fermionic theory}
The two-point function of spin-one currents in the quasi-fermionic theory is given by :
\begin{align}
\langle J_\mu(k)\,J_\nu(-k)\rangle_\text{QF}=\frac{\widetilde N}{2}\,\left[\frac{k_{\mu}\,k_{\nu}-g_{\mu\nu}\,k^2}{16\,k}+\frac{\widetilde\lambda_{\text{qf}}}{16}\,\epsilon_{\mu\nu\rho}\,k^\rho\right]
\end{align}
For two-point function of even-spin currents there are both parity even and parity odd contributions :
\begin{align}
\langle J_{2s}(k)\,J_{2s}(-k)\rangle=c_1\,\langle J_{2s}(k)\,J_{2s}(-k)\rangle_\text{even}+c_2\,\langle J_{2s}(k)\,J_{2s}(-k)\rangle_\text{odd}
\end{align}
where $c_1$ and $c_2$ are given by :
\begin{align}
c_1=\frac{\widetilde N}{32},\quad c_2=-\frac{3}{2}\widetilde N\,\widetilde\lambda_{\text{qf}}\,.
\end{align}
The three-point spinning correlators $\langle TJ_0J_0\rangle$ is given by : 
\begin{align}
\langle T_{\mu\nu}(k_1)J_0(k_2)J_0(k_3)\rangle_\text{QF}&=\frac{\widetilde N}{2}\left[(1+ \widetilde{\lambda}_{\text{qf}}^2)\,\langle T_{\mu\nu}(k_1)J_0(k_2)J_0(k_3)\rangle_{\text{FF}}+\frac{k_{1\mu}\,k_{1\nu}}{{16k_1}}\right]
\end{align}
%
%
The scalar four-point function $\langle J_0J_0 J_0 J_0\rangle$ is given by :
\begin{align}
 \langle  J_{0}( k_1) J_{0}( k_2) J_0(k_3)J_0(k_4)\rangle_\text{QF}=\frac{\widetilde{N}}{2}\,(1+\widetilde{\lambda}_{\text{qf}}^2)^2\,\langle  J_{0}( k_1) J_{0}( k_2) J_0( k_3)J_0( k_4)\rangle_\text{FF}
 \end{align}
 We will now present the results for the four-point correlator $\langle T_{\mu\nu}(k_1) J_0(k_2) J_0(k_3) J_0(k_4)\rangle_\text{QF}$. Let us rescale back our scalar operator as $J_0\longrightarrow\frac{J_0}{1+\widetilde\lambda^2}$ aligning with \cite{Maldacena:2012sf} . We then have :
 \begin{align}
\langle T_{\mu\nu}(k_1) J_0(k_2) J_0(k_3) J_0(k_4)\rangle_\text{QF}&=\frac{\widetilde{N}}{2(1+\widetilde{\lambda}_{\text{qf}}^2)^2}\langle T_{\mu\nu}(k_1) J_0(k_2) J_0(k_3) J_0(k_4)\rangle_\text{FF}\nonumber\\[5pt]
&\hspace{.5cm}+\frac{\widetilde{N} \widetilde{\lambda}_{\text{qf}}}{2(1+ \widetilde{\lambda}_{\text{qf}}^2)^2} \langle  T_{\mu\nu}(k_1) J_0(k_2) J_0(k_3) J_0(k_4)\rangle_\text{CB}
\end{align}
Written explicitly in terms of $\widetilde\lambda_{\text{qf}}$ and $\widetilde N_{\text{qf}}$ makes the duality manifest. 
In the $\widetilde\lambda_{\text{qf}}\rightarrow 0$ limit one gets the free fermion results :
\begin{align}
\label{dualitytranex}
\langle T_{\mu\nu}(k_1)J_0(k_2)J_0(k_3)J_0(k_4)\rangle_\text{QF}&=N_f\left[\langle T_{\mu\nu}(k_1)J_0(k_2)J_0(k_3)J_0(k_4)\rangle_{\text{FF}}\right]
\end{align}

\subsection*{Quasi-bosonic theory}\
The two-point function of spin-one currents in the quasi-bosonic theory is given by :
\begin{align}
\langle J_\mu(k_1) J_\nu(-k_1)\rangle_\text{QB}&=\frac{\widetilde N}{2}\bigg[ \langle J_\mu(k_1) J_\nu(-k_1)\rangle_\text{FB}+\frac{\widetilde\lambda_{\text{qb}}}{32} ~\epsilon_{\mu\nu\rho}\,k_{1}^\rho\bigg]
\end{align}
%
%
The four-point function of the scalar operator in the quasi-bosonic theory is given by :
\begin{align}
&\langle J_{0}( k_1) J_{0}( k_2) J_0(k_3)J_0(k_4)\rangle_\text{QB}\notag\\&~~=\widetilde{N}^2(1+\widetilde{\lambda}_{\text{qb}}^2)^2\bigg[\frac{1}{\widetilde{N}}\langle\langle J_{0}( k_1) J_{0}( k_2) J_0( k_3)J_0( k_4)\rangle\rangle_\text{FB}\notag\\&\hspace{3cm}-\frac{8}{\pi^{3}}\frac{\widetilde{\lambda}_{q b}^{2}}{\widetilde{N}(1+\widetilde{\lambda}_{q b}^{2})} 
\bigg(\widetilde{\Psi}^{1}_{1,0}(k_1,k_2,k_3,k_4)+(2\leftrightarrow3)+(2\leftrightarrow4)\bigg)\bigg]
\end{align}
where
\begin{align}
\widetilde{\Psi}^{1}_{1,0}(p_i)
= \frac{\pi^2}{|p_3+p_4||p_1-p_2| |p_3-p_4| |2p_1+p_2||2p_2+p_1||2p_3+p_4||2p_4+p_3|},.
\end{align}
We give the results for the $\langle TJ_0J_0\rangle$ correlator with the scalar operator rescaled as $J_0\longrightarrow\frac{J_0}{1+\widetilde\lambda^2}$ aligning with \cite{Maldacena:2012sf} :
\begin{align}
\langle T_{\mu\nu}(k_1)J_0(k_2)J_0(k_3)\rangle_{\text{QB}}=\frac{\widetilde N}{2(1+\widetilde\lambda_{\text{qb}}^2)}\,\langle T_{\mu\nu}(k_1)J_0(k_2)J_0(k_3) \rangle_{\text{FB}}
\end{align}
In the limit $\widetilde\lambda_{\text{qb}}\rightarrow 0$ we get the free boson result :
\begin{align}
\langle T_{\mu\nu}(k_1)J_0(k_2)J_0(k_3)\rangle_{\text{QB}}=N_b\,\langle T_{\mu\nu}(k_1)J_0(k_2)J_0(k_3) \rangle_{\text{FB}}
\end{align}

\subsection*{Free massive bosonic theory}
The scalar two-point function is given by :
\begin{align}
\label{j02ptmassive}
\langle J_{0}(k) J_0(-k)\rangle_{\text{FMB}}=\frac{1}{4\pi k}\arctan\left(\frac{k}{2m}\right)
\end{align}
The two-point function of the spin-one current is given by :
\begin{align}
\label{jmujnumassiveboson}
\langle J_{\mu}(k) J_\nu(-k)\rangle_{\text{FMB}}&=\langle J_{\mu}(k) J_\nu(-k)\rangle^\text{transverse}_{\text{FMB}}+\langle J_{\mu}(k) J_\nu(-k)\rangle^\text{local}_{\text{FMB}}
\end{align}
where
\begin{align}
\langle J_{\mu}(k) J_\nu(-k)\rangle^\text{transverse}_{\text{FMB}}&=A\left(g_{\mu\nu}-\frac{k_\mu k_\nu}{k^2}\right)\\[5pt]
\langle J_{\mu}(k) J_\nu(-k)\rangle^\text{local}_{\text{FMB}}&=2 \frac{k_{\mu}k_{\nu}}{k^2}\langle J_0(k)\rangle_{\text{FMB}}\notag\\[5pt]
&= -\frac{m}{2k^2\pi}k_{\mu}k_{\nu}
\end{align}
and
\begin{align}
A=-\frac{1}{8\pi k}\bigg[(k^2+4m^2)~\arctan\left(\frac{k}{2m}\right)+2km\bigg]
\end{align}
The two-point function of the stress tensor with the scalar operator is given by :
\begin{align}
\label{j0tmmmassive}
\langle T_{--}(k) J_0(-k)  \rangle
=-\frac{k_{-}^2m\left(k-2m\,\text{arctan}\left(\frac{k}{2m}\right)\right)}{16k^3\pi}
\end{align}
The correlators given in \eqref{j02ptmassive}, \eqref{jmujnumassiveboson} and \eqref{j0tmmmassive} are consistent with the higher spin equation.

\section{Conclusion and discussion}
\label{Summary}
In this paper we presented momentum space higher spin Ward identities as an efficient technique to determining spinning correlators in interacting theories such as the quasi-bosonic and quasi-fermionic theories. Starting from two-point spinning correlators, we demonstrated that one can solve higher spin equations to obtain the parity odd contribution to spinning correlators. We then solved higher spin equations to obtain three- and four-point spinning correlators and matched earlier results obtained by other means. It would be interesting to extend our analysis to correlators with more spinning operators whose computation via Feynman diagrams is rather intricate (see for example \cite{Kalloor:2019xjb} in which the computation of $\langle J_1J_1J_0J_0\rangle$ was performed in a specific kinematic region for certain components). We emphasize that higher spin equations in momentum space are sensitive to contact terms in correlators. These include new contact terms in addition to those that come from conservation Ward identities in momentum space. Interestingly, if one explicitly computes three-point functions such as $\langle J_1J_1J_0\rangle$ in Chern-Simons matter theories, one obtains such additional contact terms \cite{Aharony:2012nh,GurAri:2012is}.
  
It will be interesting  to determine momentum space correlators in terms of  analogues of higher spin invariants in position space examined in \cite{Vasiliev:1990en,Vasiliev:1999ba,Klebanov:2002ja,Sezgin:2002rt,Giombi:2010vg,Didenko:2012tv,Didenko:2013bj}. Furthermore  it will be interesting to study the higher spin constraints beyond the large-$N$ approximation
\cite{Karch:2016sxi,Seiberg:2016gmd,Metlitski:2016dht,Gomis:2017ixy,Cordova:2017kue,Aharony:2018npf}.


As the higher spin equations are not restricted to conformal fixed points, one could imagine them to be useful in theories obtained by marginal and relevant deformations of conformal fixed points. We examined this interesting new direction briefly through the example of the massive free boson theory. It would be interesting to extend this analysis to solve for higher-point correlators in such theories. This would pave way to a new understanding of conformal perturbation theory.

Furthermore, it would be exciting to study the constraints imposed by higher spin symmetry on correlators in the presence of chemical potential or at finite temperature when the conformal symmetry is broken. In such cases the correlators are once again extremely hard to evaluate directly. Such theories exhibit interesting phases such as the Higgs phase, Fermi sea etc. at both zero and finite temperature \cite{Shirazrecent}. At finite temperature one deals with thermal expectation values as opposed to vacuum expectation values. It will be interesting to understand how one applies the higher spin equations in these contexts.

We saw in our computations that higher spin equations can be solved to obtain correlators that are conformally invariant.  Hence it would be an interesting direction of study to understand how higher spin equations subsume conformal invariance.  It will be interesting to study spin-raising and weight-shifting operators in conjunction with higher spin equations. We hope to come back to these exciting issues in the near future.

\section*{Acknowledgments}
We thank Naveen S. Prabhakar for several useful comments on a previous version of the draft. We thank an anonymous referee for several useful comments that helped improve the presentation of the paper.
The work of S.J and V.M is supported by Ramanujan Fellowship. The work of R.R.J is supported by the MIUR PRIN Contract 2015 MP2CX4 ``Non-perturbative Aspects Of Gauge Theories And Strings''. The work of R.R.J is also partially supported by ``Fondi Ricerca Locale dell’Universit\`a del Piemonte Orientale''.  R.R.J thanks IISER-Pune for hospitality. S.J and V.M would like to acknowledge their debt to the people of India for their steady support of research in basic sciences.
\appendix

\section{Notations and Normalizations}
\label{notation}
In this section of the appendix we explain the normalizations and notations we will be utilizing in this article.
Through out this article we have used the following notation for the three and four-point  momentum space correlators of operators $O_i$	
\begin{align}
\langle O_1 (k_1)O_2(k_2)O_3(k_3)O_4(k_4)\rangle	&=(2 \pi)^{3} \delta\left(\boldsymbol{k_1}+\boldsymbol{k_2}+\boldsymbol{k_3}+\boldsymbol{k_4}\right)\langle\langle O_1 (k_1)O_2(k_2)O_3(k_3)O_4(k_4)\rangle\rangle\notag\\[5pt]
\langle O_1 (k_1)O_2(k_2)O_3(k_3)\rangle&=(2 \pi)^{3} \delta\left(\boldsymbol{k_1}+\boldsymbol{k_2}+\boldsymbol{k_3}\right)\langle\langle O_1 (k_1)O_2(k_2)O_3(k_3)\rangle\rangle
\end{align}
where the bold symbol $\boldsymbol{k}$ represent four vectors.
It is well known from \cite{Maldacena:2011jn,Maldacena:2012sf} that the slightly broken higher spin symmetry constraints the structure of three-point functions in the quasi-fermionic and quasi-bosonic  theory to be of the following form :
\begin{align}
\langle J_{s_{1}} J_{s_{2}} J_{s_{3}}\rangle=& \alpha_{s_{1} s_{2} s_{3}}\langle J_{s_{1}} J_{s_{2}} J_{s_{3}}\rangle_\text{FB}+\beta_{s_{1} s_{2} s_{3}}\langle J_{s_{1}} J_{s_{2}} J_{s_{3}}\rangle_\text{FF}+\gamma_{s_{1} s_{2} s_{3}}\langle J_{s_{1}} J_{s_{2}} J_{s_{3}}\rangle_\text{odd}\label{QF3ptstr}
\end{align}
where for the coefficients $\alpha$ and $\beta$ are determined completely in terms of two parameter $\widetilde{N}$ and $\widetilde{\lambda}_{qb/qf}$. 
%
We will utilize the normalization used in \cite{Maldacena:2012sf} for all the higher spin operators $J_s$ for $s>0$ i.e the two-point functions in the interacting theory are related to the free theory as follows
\begin{align}
\langle J_{s}(k)J_{s}(-k)\rangle_\text{QF}&=\frac{\widetilde{N}}{2}\langle J_{s}(k)J_{s}(-k)\rangle_\text{FF}\label{2ptJSQF}\\[5pt]
\langle J_{s}(k)J_{s}(-k)\rangle_\text{QB}&=\frac{\widetilde{N}}{2}\langle J_{s}(k)J_{s}(-k)\rangle_\text{FB}\label{2ptJSQB}
\end{align}
The difference in the overall factor of $\frac 12$ from \cite{Maldacena:2012sf} is because we will be working with a complex scalar. Also observe  that in the $\lambda\to 0$ limit the quasi-fermionic and quasi-bosonic theory results go to the free fermionic and free bosonic two-point functions respectively.
The normalization of the free bosonic and free fermion theory  two-point functions of the spin one current will be relevant to our computations and these are given as
\begin{align}
\langle J_\mu(k) J_\nu(-k)\rangle_\text{FB}=\langle J_\mu(k) J_\nu(-k)\rangle_\text{FF}=\frac{1}{16k}(k_{\mu}k_{\nu}-g_{\mu\nu}k^2)\label{2ptJmuF}
\end{align}
On  the other hand for the scalar operators $J_0$, we utilize the normalization
\begin{align}
\langle J_0(k)J_0(-k)\rangle_\text{QF}&=\frac{\widetilde{N}}{2}(1+\tilde{\lambda}_{\text{qf}}^2)\langle J_0(k)J_0(-k)\rangle_\text{FF}\label{2ptJ0QF}\\
\langle J_0(k)J_0(-k)\rangle_\text{QB}&=\frac{\widetilde{N}}{2}(1+\tilde{\lambda}_{\text{qb}}^2)\langle J_0(k)J_0(-k)\rangle_\text{FB}\label{2ptJ0QB}
\end{align}
Note that the normalization of the scalar operator $J_0$ which we will utilize here differs from \cite{Maldacena:2012sf} by $J_0=J_0^{MZ}(1+\tilde{\lambda}^2)$.
We will be using the following  normalization for the two functions for the scalar operators in free bosonic and free fermion theory
\begin{align}
\langle J_0(k)J_0(-k)\rangle_\text{FB}&=\frac{1}{8 k}\label{2ptJ0FF}\\[5pt]
\langle J_0(k)J_0(-k)\rangle_\text{FF}&=-\frac{k}{8 }\label{2ptJ0FB}
\end{align}
The three-point functions of the scalar operators in the free theories are given as
\begin{align}
\langle \langle J_0(k_1)J_0(k_2)J_0(k_3)\rangle\rangle_\text{FB}&=\frac{1}{4 k_1 k_2 k_3}\label{3ptJ0FB}\\[5pt]
\langle \langle J_0(k_1)J_0(k_2)J_0(k_3)\rangle\rangle_\text{FF}&=0\label{3ptJ0FF}
\end{align}
The reconstruction formulas for the three-point functions of the higher-spin operators fix the local and transverse parts of the correlators upto a form factor as shown below \cite{Bzowski_2014}
\begin{align}
\hspace{-.5cm}\left\langle\left\langle T_{\mu\nu}( k_1)\,J_0( k_2)\,J_0( k_3)\right\rangle \right\rangle_{\text{FB/FF}}&=\langle T_{\mu\nu}( k_1)\,J_0( k_2)\,J_0( k_3)\rangle_{\text{transverse}}+
\langle T_{\mu\nu}( k_1)\,J_0( k_2)\,J_0( k_3)\rangle_{\text{local}}\label{Tj0j0reconstructionF}\\[5pt]
\langle \langle J_{\mu}( k_1)\,J_{\nu}( k_2)\,J_0( k_3)\rangle\rangle_{\text{FB/FF}}&=\langle J_{\mu}( k_1)\,J_\nu( k_2)\,J_0( k_3)\rangle_{\text{transverse}}+\langle J_{\mu}( k_1)\,J_\nu( k_2)\,J_0( k_3)\rangle_{\text{local}}\label{jmujnu3pful}
\end{align}
The transverse parts are given as
\begin{align}
\langle J_{\mu}( k_1) J_\nu( k_2) J_0(k_3)\rangle_{\text{FB}}^{\text{transverse}}&=\pi_{\alpha}^{\mu}\left(k_1\right) \pi_{\beta}^{\nu}\left(k_{2}\right)\left[A_{1}(k_1,k_2,k_3) k_{2}^{\alpha} k_{3}^{\beta}+A_{2}(k_1,k_2,k_3) g^{\alpha \beta}\right]\notag\\[5pt]
\langle J_{\mu}( k_1) J_\nu( k_2) J_0(k_3)\rangle_{\text{FF}}^{\text{transverse}}&=2 \epsilon_{k_{1} \alpha}^{\mu} \epsilon_{k_{2} \beta}^{\nu}\left(k_{1}, k_{2}\right)\left[A\left(k_{1}, k_{2}, k_{3}\right) k_{1 \rho}+B\left(k_{1}, k_{2}, k_{3}\right) k_{2 \rho}\right] \epsilon^{\alpha \beta \rho}\notag\\[5pt]
\langle T_{\mu\nu}( k_1)\,J_0( k_2)\,J_0( k_3)\rangle_{\text{FB/FF}}^{\text{transverse}}&=A_{1(FB/FF)}\left(k_{1}, k_{2}, k_{3}\right) \Pi_{\mu \nu}^{\alpha \beta}\left(k_{1}\right) k_{2}^{\alpha} k_{2}^{\beta}\label{jmujnu3p}
\end{align}
The explicit form  of the form factors and the local parts for the free theories may be found in \cite{Jain:2020rmw} where they were determined by direct computation as well as by solving momentum space higher spin equations in the free boson and the free fermion theories.

\section{Schouten identity}
\label{SchoutenIdentity}
In our computation the following Schouten identity turned out to be helpful :
\begin{align}
\epsilon^{\mu\nu k_1}(k_1\cdot k_2)+\epsilon^{\mu k_1 k_2}k_1^\nu&=\epsilon^{\mu\nu k_2}k_1^2+\epsilon^{\nu k_1 k_2}k_1^\mu\cr
\epsilon^{\mu\nu k_1}k_2^2+\epsilon^{\mu k_1 k_2}k_2^\nu&=\epsilon^{\mu\nu k_2}(k_1\cdot k_2)+\epsilon^{\nu k_1 k_2}k_2^\mu
\end{align}
The Schouten identity can be used to eliminate two of the $\epsilon$ structures in favour of the other two :
\begin{align}
\epsilon_{\mu\nu k_1}&=\frac{k_1^2\,\left(-\epsilon_{\mu k_1k_2}k_{2\nu}+\epsilon_{\nu k_1k_2}k_{2\mu}\right)+k_1\cdot k_2\,\left(-\epsilon_{\nu k_1k_2}k_{1\mu}+\epsilon_{\mu k_1k_2}k_{1\nu}\right)}{k_1^2\,k_2^2-(k_1\cdot k_2)^2}\cr
\epsilon_{\mu\nu k_2}&=\frac{k_2^2\,\left(-\epsilon_{\nu k_1k_2}k_{1\mu}+\epsilon_{\mu k_1k_2}k_{1\nu}\right)+k_1\cdot k_2\,\left(\epsilon_{\nu k_1k_2}k_{2\mu}-\epsilon_{\mu k_1k_2}k_{2\nu}\right)}{k_1^2\,k_2^2-(k_1\cdot k_2)^2}
\end{align}

\section{Five-point function:  $\langle \boldsymbol{J_0J_0J_0J_0 J_0}\rangle_\text{QF} $ }

In this section we will derive an expression for the momentum space five-point function of the scalar operator in the quasi-fermionic theory. We will demonstrate that this five-point correlator may expressed in terms of three, four and five-point scalar correlators in the free fermionic and the critical bosonic theory.Note that the three-point function of the scalar correlator in the quasi-fermionic theory is a contact term. To begin with let us first demonstrate that the free fermionic scalar five-point point correlator $\langle J_{0}J_{0}J_{0}J_0 J_0\rangle_\text{FF}$ is non-zero whereas the three-point correlator $\langle J_{0}J_{0}J_{0}\rangle_\text{FF}$ vanishes.

\subsection*{Three-point : $\boldsymbol{\langle J_{0}J_{0}J_{0}\rangle_\text{FF}}$ }
The three-point function of $J_0$ operators in free fermion theory is given by :
\begin{align}
\langle J_{0}(k_1)J_{0}(k_2)J_{0}(k_3)\rangle=\int_{l_1l_2l_3}\langle  \bar{\psi}(l_1) \psi(k_1-l_1)\bar{\psi}(l_2) \psi(k_2-l_2)\bar{\psi}(l_3) \psi(k_3-l_3)\rangle
\end{align}
\subsubsection*{Wick Contraction: $1\bar{2}2\bar{3}3\bar{1}$+$1\bar{3}3\bar{2}2\bar{1}$}

\begin{align}
G_1&=-\int_{l_1l_2l_3}\frac{ \text{Tr}(\slashed{l_1}\slashed{l_2}\slashed{l_3})}{l_1^2l_2^2l_3^2}\delta(k_1-l_1+l_2)\delta(k_2-l_2+l_3)\delta(k_3-l_3+l_1)\notag\\
&=-\int_{l} \frac{Tr(\gamma^\mu\gamma^\nu\gamma^\rho)(l+k_1)_{\mu}l_\nu( l-k_2)_{\rho}}{l^2(l+k_1)^2(l-k_1)^2}
\end{align}
The numerator simplifies upon 
\begin{align}
G_1=\int_{l} \frac{2 i \epsilon^{\mu \nu\rho} k_{1\mu}l_\nu k_{2\rho}}{l^2(l+k_1)^2(l-k_2)^2}
\end{align}
Now  we know that the integral is of the form
\begin{align}
\int_{l} \frac{l_\nu}{l^2(l+k_1)^2(l-k_2)^2}=A k_{1\nu}+ B k_{2\nu}
\end{align}
where $A$ and $B$ are functions of $k_1,k_2$ and $k_3$.
Due to the above form of the required integral, $G_1$ vanishes due to the presence of $\epsilon^{\mu \nu\rho}$ as follows
\begin{align}
G_1=2 i \epsilon^{\mu \nu\rho} k_{1\mu}k_{2\rho}(A k_{1\nu}+ B k_{2\nu})=0
\end{align}
Similarly it can be shown that the  term arising due to other Wick contraction also vanishes. Hence,
\begin{align}
\langle J_{0}(k_1)J_{0}(k_2)J_{0}(k_3)\rangle_\text{FF}=0
\end{align}

\subsection*{Five-point function : $\boldsymbol{\langle J_{0}J_{0}J_{0} J_0 J_0\rangle_\text{FF}}$}
\label{5ptscalar}
The five-point function of $J_0$ operators in free fermion theory is given by
\begin{align}
\langle& J_{0}(k_1)J_{0}(k_2)J_{0}(k_3)J_{0}(k_4)J_{0}(k_5)\rangle_\text{FF}\notag\\&=\int_{l_1l_2l_3}\langle  \bar{\psi}(l_1) \psi(k_1-l_1)\bar{\psi}(l_2) \psi(k_2-l_2)\bar{\psi}(l_3) \psi(k_3-l_3)\bar{\psi}(l_4)\psi(k_4-l_4)\bar{\psi}(l_5) \psi(k_5-l_5)\rangle
\end{align}
Firstly we will check that this correlator does not straightforwardly vanish as three-point function. Subsequently we will derive the form of the five-point correlator without doing an explicit computation.

Consider the Wick contraction $1\bar{2}2\bar{3}3\bar{4}4\bar{5}$
\begin{align}
H_1&=-\int_{l_1l_2l_3}\frac{\text{Tr}(\slashed{l_1}\slashed{l_2}\slashed{l_3}\slashed{l_4}\slashed{l_5})}{l_1^2l_2^2l_3^2l_4^2l_5^2 }\delta(k_1-l_1+l_2)\delta(k_2-l_2+l_3)\delta(k_3-l_3+l_4)\delta(k_4-l_4+l_5)\delta(k_5-l_5+l_1)\notag\\
&=-\int_{l} \frac{Tr(\gamma^\mu\gamma^\nu\gamma^\rho\gamma^\sigma\gamma^\delta)(l+k_1)_{\mu}l_\nu( l-k_2)_{\rho}(l-k_2-k_3)_{\sigma}(l+k_1+k_5)_{\delta}}{l^2(l+k_1)^2(l-k_2)^2(l-k_2-k_3)^2(l+k_1+k_5)^2}\label{5FWC}
\end{align}
The trace of the five gamma matrices in three dimensions is given as follows
\begin{align}
\operatorname{Tr}\left(\gamma_{\mu} \gamma_{\nu} \gamma_{\rho} \gamma_{\sigma} \gamma_{\delta}\right)=2 i \epsilon_{\mu \nu \rho} g_{\sigma \delta}+2 i \epsilon_{\sigma \delta m}\left(g_{\mu \nu} g_{\rho m}-g_{\mu \rho}g_{\nu m}+g_{\mu m} g_{\nu \rho}\right)
\end{align}
Upon utilizing the above identity in \eqref{5FWC}) it is easy to check that the trace of the gamma matrices contracted with the momentas is non-zero unlike the  three-point function.
\subsection*{Five-point function $ \langle \boldsymbol{J_0J_0J_0J_0 J_0}\rangle_\text{QF} $ }\label{5ptJ0}
In this section we will compute the momentum space  five-point correlator of the scalar primary $J_0$ in the quasi-fermionic theory. It is well known from the study of conformal blocks that a given five-point function could be expanded in terms of the four-point function as depicted in the diagram below for the scalar  primary which we are interested in.
\begin{figure}[h]
	\begin{center}
		\includegraphics[scale=0.4]{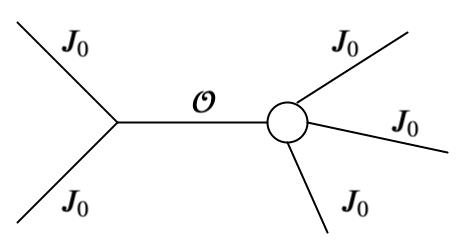}
	\end{center}
	\caption{Schematic for the conformal block expansion of the 5-point correlator }
\end{figure}
This can also be understood as an insertion of a complete set of states as expressed below
\begin{align}
\langle& J_{0}( k_1) J_{0}( k_2) J_0(k_3)J_0(k_4)J_0(k_5)\rangle\notag\\[5pt]&=\sum_{\mathcal{O}}\langle J_{0}( k_1) J_{0}( k_2)\mathcal{O}( -k_1- k_2)\rangle \langle \mathcal{O}( k_1+ k_2)J_0(k_3)J_0(k_4)J_0(k_5)\rangle\label{cblock}
\end{align}
Note that the sum  $\sum_{\mathcal{O}}$ runs over all the primary and descendant operators in the theory. However, it is known that the descendants do not contribute in the momentum space. Hence the sum is over all the primary operators which in Chern Simons matter theories are the scalar and the almost conserved spinning operators which we denote here as $J_s$. Therefore the above sum for the momentum space correlators reduced to
\begin{align}
\langle& J_{0}( k_1) J_{0}( k_2) J_0( k_3)J_0(k_4)J_0( k_5)\rangle_\text{QF}\notag\\&=\sum_{s=0}^{\infty}\langle J_{0}( k_1) J_{0}( k_2)J_s( -k_1- k_2)\rangle_\text{QF} \langle J_s( k_1+ k_2)J_0( k_3)J_0( k_4)J_0( k_5)\rangle_\text{QF}
\end{align}
Let us now separate the contributions to the five-point function from the scalar and the higher spin intermediate operators
\begin{align}
\langle& J_{0}( k_1) J_{0}( k_2) J_0(k_3)J_0( k_4)J_0(k_5)\rangle_\text{QF}\notag\\[7pt]&=\langle J_{0}( k_1) J_{0}( k_2)J_0( -k_1- k_2)\rangle_\text{QF} \langle J_0(k_1+ k_2)J_0( k_3)J_0( k_4)J_0( k_5)\rangle_\text{QF}\notag\\[5pt]&~~~+\sum_{s=1}^{\infty}\langle J_{0}( k_1) J_{0}( k_2)J_{2s}( -k_1- k_2)\rangle_\text{QF} \langle J_{2s}( k_1+ k_2)J_0( k_3)J_0(k_4)J_0(k_5)\rangle_\text{QF}
\end{align}
Note that the contributions form the odd spin correlators are zero. 
We can now substitute the result we obtained for the scalar four-point function $\langle J_0J_0J_0J_0 \rangle_\text{QF}$ for quasi-fermionic theory  in \eqref{J04ptqf} and  $\langle J_{2s}J_0J_0J_0 \rangle_\text{QF}$ given in footnote \ref{footnote10} in terms of the corresponding correlators in free fermion theory and critical bosonic theories as follows
\begin{align}
\langle& J_{0}( k_1) J_{0}( k_2) J_0(k_3)J_0(k_4)J_0(k_5)\rangle_\text{QF}\notag\\[7pt]&=\beta_{0000}\langle J_{0}( k_1) J_{0}( k_2)J_0( -k_1- k_2)\rangle_\text{QF} \langle J_0( k_1+ k_2)J_0( k_3)J_0( k_4)J_0( k_5)\rangle_\text{FF}\notag\\[5pt]&~~~+\sum_{s=1}^{\infty}\langle J_{0}( k_1) J_{0}( k_2)J_{2s}(- k_1- k_2)\rangle_\text{QF} \bigg[\beta_{2000}\langle J_{2s}( k_1+ k_2) J_0(k_3) J_0(k_4) J_0(k_5)\rangle_\text{FF}\notag\\&~~~~~~~~~~~~~~~~~~~~~~~~~~~~~~~~~~~~~~~~~~~~~~~~~~~~~+\gamma_{2000}\langle J_{2s}( k_1+ k_2)  J_0(k_3) J_0(k_4)J_0(k_5)\rangle_\text{CB}\bigg]\label{Tst1}
\end{align}
From \cite{Maldacena:2011jn,Maldacena:2012sf} we know that the three-point functions in the quasi fermionic theory are given by
\begin{align}
\langle J_{0}( k_1) J_{0}( k_2)J_{2s}(- k_1- k_2)\rangle_\text{QF}&=\beta_{s00} \langle J_{0}( k_1) J_{0}( k_2)J_{2s}(- k_1- k_2)\rangle_\text{FF}\notag\\[5pt]
&=\beta_{s00}  \langle \tilde{J}_{0}( k_1) \tilde{J}_{0}( k_2)J_{2s}(- k_1- k_2)\rangle_\text{CB}\label{J2sJ0J0QF}
\end{align}
where $\beta_{s00} $
Substituting the above three-point functions in \eqref{Tst1} we get
\begin{align}
\langle& J_{0}( k_1) J_{0}( k_2) J_0(k_3)J_0(k_4)J_0(k_5)\rangle_\text{QF}\notag\\[7pt]&=\beta_{0000}\langle J_{0}( k_1) J_{0}( k_2)J_0( k_1+ k_2)\rangle_\text{QF} \langle J_0(- k_1- k_2)J_0(k_3)J_0(k_4)J_0(k_5)\rangle_\text{FF}\notag\\[5pt]
&+\sum_{s=1}^{\infty}\beta_{s00} \langle J_{0}( k_1) J_{0}( k_2)J_{2s}( k_1+ k_2)\rangle_\text{FF} \bigg[\beta_{2000}\langle J_{2s}(- k_1- k_2) J_0(k_3) J_0(k_4) J_0(k_5)\rangle_\text{FF}\notag\\&~~~~~~~~~~~~~~~~~~~~~~~~~~~~~~~~~~~~~~~~~~~~~~~~~~~~~+\gamma_{2000}\langle J_{2s}(- k_1- k_2)  \widetilde{J}_0(k_3) \widetilde{J}_0(k_4)\widetilde{J}_0(k_5)\rangle_\text{CB}\bigg]
\end{align}
where we have used our expectation for the four-point correlator $\langle J_{2s}  J_0J_0J_0\rangle_\text{QF}$ given in footnote \ref{footnote10}. 
We re-express the above equation using \eqref{J2sJ0J0QF} as follows :
\begin{align}
\langle& J_{0}( k_1) J_{0}( k_2) J_0(k_3)J_0(k_4)J_0(k_5)\rangle_\text{QF}\notag\\[7pt]&=\beta_{0000}\langle J_{0}( k_1) J_{0}( k_2)J_0( k_1+ k_2)\rangle_\text{QF} \langle J_0(- k_1- k_2)J_0( k_3)J_0( k_4)J_0( k_5)\rangle_\text{FF}\notag\\[5pt]&+\beta_{s00}\bigg[\beta_{2000} \sum_{s=1}^{\infty}\langle J_{0}( k_1) J_{0}( k_2)J_{2s}(- k_1- k_2)\rangle_\text{FF}  \langle J_{2s}( k_1+ k_2) J_0(k_3) J_0(k_4) J_0(k_5)\rangle_\text{FF}\notag\notag\\[5pt]&~~~+\gamma_{2000} \sum_{s=1}^{\infty}\langle \widetilde J_{0}( k_1) \widetilde J_{0}( k_2)J_{2s}( -k_1- k_2)\rangle_\text{CB}\langle J_{2s}( k_1+ k_2)  \widetilde J_0(k_3)  \widetilde J_0(k_4) \widetilde J_0(k_5)\rangle_\text{CB}\bigg]\label{Tst2}
\end{align}
We will now utilize the  conformal block expansion in \eqref{cblock}  to re express the summation in the last two lines of \eqref{Tst2} in terms of the free fermion theory and the critical bosonic theory  five-point correlators. To do so consider the conformal block expansion for scalar five-point correlators in critical bosonic theory and free fermion theory
\begin{align}
\langle& \widetilde J_{0}( k_1) \widetilde J_{0}( k_2) \widetilde J_0(k_3)\widetilde J_0(k_4) \widetilde J_0( k_5)\rangle_\text{CB}\notag\\[5pt]&=\langle \widetilde J_{0}( k_1)  \widetilde J_{0}( k_2)\widetilde J_0( -k_1- k_2)\rangle_\text{CB} \langle \widetilde  J_{0}( k_1+ k_2)\widetilde J_0(k_3)\widetilde J_0(k_4)\widetilde J_0(k_5)\rangle_\text{CB}\notag\\[5pt]&
\hspace{0.5cm}+\sum_{s=1}^{\infty}\langle \widetilde J_{0}( k_1) \widetilde J_{0}( k_2)J_{2s}( -k_1- k_2)\rangle_\text{CB} \langle J_{2s}( k_1+ k_2)\widetilde J_0( k_3)\widetilde J_0(k_4)\widetilde J_0( k_5)\rangle_\text{CB}\notag\\[5pt]
\langle& J_{0}( k_1) J_{0}( k_2) J_0(k_3)J_0(k_4)J_0(k_5)\rangle_\text{FF}\notag\\[5pt]&=\sum_{s=1}^{\infty}\langle J_{0}( k_1) J_{0}( k_2)J_{2s}( -k_1- k_2)\rangle_\text{FF} \langle J_{2s}( k_1+ k_2)J_0( k_3)J_0(k_4)J_0(k_5)\rangle_\text{FF}
\end{align}
where in the last line we have used the fact that $\langle J_{0} J_0J_0\rangle_\text{FF}=0$.
Upon using the above conformal block expansions for the scalr five-point function of the free fermionic and the critical bosonic in \eqref{Tst2} we arrive at our final  expression for the five-point scalar operator in quasi-fermionic theory in terms of the correlators of free fermionic and critical bosonic theories which is as follows :
\begin{align}
\langle& J_{0}( k_1) J_{0}( k_2) J_0(k_3)J_0(k_4)J_0(k_5)\rangle_\text{QF}\notag\\[5pt]&=\beta_{0000}\langle J_{0}( k_1) J_{0}( k_2)J_0( k_1+ k_2)\rangle_\text{QF} \langle J_0(- k_1- k_2)J_0( k_3)J_0(k_4)J_0( k_5)\rangle_\text{FF}+\notag\\[5pt]&\hspace{0.5cm}
-\gamma_{2000}\beta_{s00}\langle \widetilde J_{0}( k_1)  \widetilde J_{0}( k_2)\widetilde J_0( -k_1- k_2)\rangle_\text{CB} \langle \widetilde J_{0}( k_1+ k_2)\widetilde J_0(k_3)\widetilde J_0(k_4)\widetilde J_0(k_5)\rangle_\text{CB}
\notag\\[5pt]&\hspace{0.5cm}+\beta_{s00} \bigg[\beta_{2000} \langle J_{0}( k_1) J_{0}( k_2) J_0( k_3)J_0(k_4)J_0( k_5)\rangle_\text{FF}+\gamma_{2000}\langle \widetilde{J}_{0}( k_1) \widetilde{J}_{0}( k_2) \widetilde{J}_0( k_3)\widetilde{J}_0( k_4)\widetilde{J}_0( k_5)\rangle_\text{CB}\bigg]
\end{align}
Substituting the above in \eqref{Tst2} we get
\begin{align}
\label{5ptcorr}
\langle& J_{0}( k_1) J_{0}( k_2) J_0(k_3)J_0(k_4)J_0(k_5)\rangle_\text{QF}\notag\\[5pt]&=s_0 \langle J_{0}( k_1) J_{0}( k_2)J_0( k_1+ k_2)\rangle_\text{QF} \langle J_0(- k_1- k_2)J_0( k_3)J_0(k_4)J_0( k_5)\rangle_\text{FF}\notag\\[5pt]&\hspace{0.5cm}
-s_2\langle \widetilde J_{0}( k_1)  \widetilde J_{0}( k_2)\widetilde J_0( -k_1- k_2)\rangle_\text{CB} \langle \widetilde J_{0}( k_1+ k_2)\widetilde J_0(k_3)\widetilde J_0(k_4)\widetilde J_0(k_5)\rangle_\text{CB}\notag\\[5pt]&\hspace{0.5cm}+s_1 \langle J_{0}( k_1) J_{0}( k_2) J_0(k_3)J_0(k_4)J_0(k_5)\rangle_\text{FF}+s_2 \langle \widetilde{J}_{0}( k_1) \widetilde{J}_{0}( k_2) \widetilde{J}_0( k_3)\widetilde{J}_0( k_4)\widetilde{J}_0( k_5)\rangle_\text{CB}
\end{align}
The coefficients $s_0,s_1$ and $s_2$ in our normalization are given by :
\begin{align}
s_0&=\beta_{0000}=\widetilde{N}(1+\widetilde\lambda_{\text{qf}}^2)^2\notag\\[5pt]
s_1&=\beta_{2000}\beta_{s00}=\frac{\widetilde N^2(1+ \widetilde{\lambda}_{\text{qf}}^2)^2}{2}\\[5pt]
s_2&=\gamma_{2000}\beta_{s00}=\widetilde{\lambda}_{\text{qf}}\frac{\widetilde N^2(1+ \widetilde{\lambda}_{\text{qf}}^2)^2}{2}
\end{align}
The three-point function of the scalar operator in the quasi-fermionic theory vanishes in the limit $\widetilde{\lambda}_{\text{qf}}\rightarrow 0$ \cite{Aharony:2018pjn} \footnote{This is because the scalar operator in the theory is odd under parity and therefore so is any correlator with an odd number of scalar operators. The three-point function in particular is proportional to $\lambda_{f}$ which is also odd under parity and therefore the correlator vanishes in the limit $\lambda_{f}\rightarrow 0$. Explicitly one has \cite{Aharony:2018pjn} $$\langle J_0(q_1)J_0(q_2)J_0(q_3)\rangle_{\text{QF}}=(2\pi)^3\delta^3(q_1+q_2+q_3)8\pi^2N_f\frac{\pi^2}{\widetilde\kappa_f^3}\lambda_f$$ where $\widetilde\kappa_f=\frac{N_f}{\lambda_f}$.} and in this limit one obtains from \eqref{5ptcorr} :
\begin{align}
\langle& J_{0}( k_1) J_{0}( k_2) J_0(k_3)J_0(k_4)J_0(k_5)\rangle_\text{QF}=2N_f^2\langle J_{0}( k_1) J_{0}( k_2) J_0(k_3)J_0(k_4)J_0(k_5)\rangle_\text{FF}
\end{align}
To summarize, in this section of the appendix we have provided a possible derivation for the momentum space scalar five-point function in the quasi-fermionic theory in terms of the momentum space scalar three, four and five-point functions in free fermionic and critical bosonic theory. In this analysis we are not fully sure of the contribution from the double trace operators. We will come back to this in future.

\section{Conformal partial wave in momentum space}\label{CPWS}
In this Appendix we briefly describe the derivation of the conformal partial wave in momentum space. In order to define confromal partial wave we require the concept of the shadow operator  which is defined in the embedding space as follows \cite{SimmonsDuffin:2012uy}
\begin{align}
\widetilde{\mathcal{O}}(X)=\int D^{d} Y \frac{1}{(-2 X \cdot Y)^{d-\Delta}} \mathcal{O}(Y)
\end{align}
where $X$ and $Y$ are coordinates in the embedding space.
In configuration space this may be expressed as
\begin{align}
\widetilde{\mathcal{O}}(x)=\int D^{d} y \frac{1}{(x-y)^{2(d-\Delta)}} \mathcal{O}(y)\label{ShaP}
\end{align}
The conformal partial wave in position space is defined  in terms of the shadow operator as
\begin{align}
\Psi^{\Delta^i}_{\Delta,0}=\int d^dx\langle\mathcal{O}_1(x_1)\mathcal{O}_2(x_2)  \mathcal{O}(x)\rangle\langle \widetilde{\mathcal{O}}(x)\mathcal{O}_3(x_3)\mathcal{O}_4(x_4)\rangle
\end{align}
We will utilize the definition of shadow operator in \eqref{ShaP} in the above equation for conformal partial with $J_0$ in the quasi-bosonic theory (i.e $\Delta=1$) as the external and intermediate operator  to obtain
\begin{align}
\Psi^{\Delta}_{\Delta,0}=\int d^3x d^3y\frac{1}{(x-y)^{2}}\notag  \langle J_0(x_1)J_0(x_2)  J_0(x)\rangle\langle J_0(y)J_0(x_3)J_0(x_4)\rangle
\end{align}
We express the field in the momentum space to obtain
\begin{align}
\Psi^{\Delta}_{\Delta,0}=\int d^3x d^3y d^3k d^3k_1  \Pi_{i=1}^{4}d^3p_i \frac{e^{i(\textbf{k}.\textbf{x}+\textbf{k}_1.\textbf{ y}+\sum_{i=1}^{4} \textbf{p}_i.\textbf{x}_i ) }}{(x-y)^{2}}  \langle J_0(p_1)J_0(p_2)  J_0(k)\rangle\langle J_0(k_1)J_0(p_3)J_0(p_4)\rangle 
\end{align}
Note that this is not  the Fourier transform of the conformal partial wave which we do at the end. 
Following this we perform  a change of variables to $z=\frac{x-y}{2}$ and $\bar{z}=\frac{x+y}{2}$ and perform $\bar{z},k_1$ integration  to obtain
\begin{align}\label{Pfou}
\Psi^{\Delta^i}_{\Delta,0}=\int d^3k  d^3z  \Pi_{i=1}^{4}d^3p_i \frac{e^{-2i \textbf{k}.\textbf{z} +i\sum_{i=1}^{4}  \textbf{p}_i.\textbf{x}_i }}{(z)^{2}}  \langle J_0(p_1)J_0(p_2)  J_0(k)\rangle\langle J_0(-k)J_0(p_3)J_0(p_4)\rangle 
\end{align}
We first perform the $z$ integral which  we know to be
\begin{align}\label{fou}
\int d^3z \frac{e^{-2i \textbf{k}.\textbf{z}}}{z^2} =\frac{\pi^2}{k}
\end{align}
where we have used
\begin{align}
\int \mathrm{d}^{d} \boldsymbol{x} e^{-\mathrm{i} p \cdot \boldsymbol{x}} \frac{1}{x^{2 \Delta}}=\frac{\pi^{d / 2} 2^{d-2 \Delta} \Gamma\left(\frac{d-2 \Delta}{2}\right)}{\Gamma(\Delta)} p^{2 \Delta-d}
\end{align}
with $ \Delta=1$ and $d=3$. Substituting \eqref{fou} in \eqref{Pfou} we obtain
\begin{align}
\Psi^{1}_{1,0}(x_i)=\pi^2\int d^3k\,\Pi_{i=1}^{4}d^3p_i e^{ i\sum_{i=1}^{4}  \textbf{p}_i.\textbf{x}_i }  \langle J_0(p_1)J_0(p_2)  J_0(k)\rangle\langle J_0(-k)J_0(p_3)J_0(p_4)\rangle \frac{1}{k}\label{psixip}
\end{align}
This is the conformal partial wave in position space and hence the fourier transform of it  which we denote as $\widehat{\Psi}^{1}_{1,0}(q_i)$ is given by
\begin{align}
\widehat{\Psi}^{1}_{1,0}(q_i)&=\int  \Pi_{j=1}^{4}d^3x_j \Psi^{1}_{1,0}(x_i)e^{- i\sum_{i=1}^{4}  \textbf{q}_i\cdot \textbf{x}_i }\label{C9}
\end{align}
Substituting  \eqref{psixip} in \eqref{C9}  we obtain
\begin{align}
\widehat{\Psi}^{1}_{1,0}(q_i)=\pi^2\int  d^3k \langle J_0(q_1)J_0(q_2)  J_0(k)\rangle\langle J_0(-k)J_0(q_3)J_0(q_4)\rangle \frac{1}{k}
\end{align}
Utilizing the form of the three-point function in free boson theory to be \eqref{3ptJ0FB} we obtain the required conformal partial wave to be :
\begin{align}
\widehat{\Psi}^{1}_{1,0}(p_i)&=\frac{\pi^2}{16}\int d^3k  \frac{ \delta^3(p_1+p_2+k)\delta^3(p_3+p_4-k)}{k|p_1-p_2| |p_1-k||p_2-k||p_3-p_4| |p_3+k||p_4+k|}\nonumber\\
&= \frac{\pi^2 \delta^3(p_1+p_2+p_3+p_4)}{|p_3+p_4||p_1-p_2| |p_3-p_4| |2p_1+p_2||2p_2+p_1||2p_3+p_4||2p_4+p_3|}. \label{CPWFin}
\end{align}
In the main text we define $\widetilde\Psi^{1}_{1,0}$ using $\widehat\Psi^{1}_{1,0}(p_i)=\widetilde\Psi^{1}_{1,0}(p_i)\,\delta^3(p_1+p_2+p_3+p_4)$.

\section{Parity odd correlators}
\label{parityodd}
\subsection{Quasi-fermionic theory}
It was shown in \cite{Maldacena:2012sf} that if all  the three operators appearing in a three-point correlator have spin $s>0$,  they contain three structures: the free bosonic, the free fermionic, and the parity odd structure. In this sub-section we attempt to constrain such parity odd structures of the correlators  $\langle TJ_1J_1\rangle$ and $\langle J_1J_1J_0J_0\rangle$ through higher spin equations.

\subsubsection{Constraining $\boldsymbol{\langle TJ_1J_1\rangle_\text{QF}^\text{odd}}$ }

Consider the action of $Q_3$ on $\langle J_- J_- J_-\rangle$ :
\begin{align}
\langle&[Q_{3}, J_{-}(x_1)]J_-(x_2)J_-(x_3)]\rangle+1\leftrightarrow 2+1\leftrightarrow 3=\int_x \langle \partial_\mu\,J^\mu_{--}(x) J_-(x_1)J_-(x_2)J_-(x_3)\rangle
\end{align}
Upon utilizing the algebra in \eqref{QFQ3algebra}, the current equation \eqref{DivJ3AF} and large $N$ factorization, we get after a Fourier transform of the above equation :
\begin{align}
\label{HSEJJTint}
&-2\,k_{1-}\langle \langle   T_{--}(k_1)J_-(k_2) J_-(k_3)\rangle \rangle_\text{QF}+1\leftrightarrow 2+1\leftrightarrow 3 \notag
\\[5pt]&\hspace{-1cm}=\frac{16\widetilde{\lambda}_{\text{qf}}}{\widetilde{N}(1+\widetilde{\lambda}_{\text{qf}}^2)} k_{1-} \langle \langle J_0(k_1) J_{-}(k_2) J_-(k_3) \rangle \rangle_\text{QF}\langle \langle J_-(k_1) J_-(-k_1) \rangle \rangle_\text{QF}+1\leftrightarrow 2+1\leftrightarrow 3
\end{align}
Note that  $ \langle J_1J_1J_0 \rangle_{\text{QF}}$ is as given in \eqref{J1J1J0QF} and $\beta_{110}$ and $ \gamma_{110} $ are as given in \eqref{JJJ0QFcoef}.
From the results of \cite{Maldacena:2012sf} the correlator $ \langle TJ_1J_1 \rangle$  appearing in the above equation has the following structure :
\begin{align}
 \langle   T_{--}(k_1)J_-(k_2) J_-(k_3) \rangle_\text{QF}&=\alpha_{211} \langle   T_{--}(k_1)J_-(k_2) J_-(k_3) \rangle_\text{FB}+\beta_{211}  \langle T_{--}(k_1)J_-(k_2) J_-(k_3)  \rangle_\text{FF}\notag\\[5pt]&\hspace{1cm}+\gamma_{211} \langle T_{--}(k_1)J_-(k_2) J_-(k_3)  \rangle_\text{odd}
\end{align}
where
\begin{align}
\alpha_{211}=\frac{\widetilde{N}}{2(1+\widetilde{\lambda}_{\text{qf}}^2)},\quad\beta_{211}=\frac{\widetilde{N}\widetilde{\lambda}_{\text{qf}}^2}{2(1+\widetilde{\lambda}_{\text{qf}}^2)},\quad \gamma_{211}=\frac{\tilde{N}\tilde{\lambda}_{\text{qf}}}{2(1+\tilde{\lambda}_{\text{qf}}^2)}
\end{align}
Substituting in \eqref{HSEJJTint} we get :
\begin{align}
&-2\,\bigg[\alpha_{211}k_{1-}\langle \langle   T_{--}(k_1)J_-(k_2) J_-(k_3)\rangle \rangle_\text{FB}+\beta_{211} k_{1-}\langle \langle T_{--}(k_1)J_-(k_2) J_-(k_3) \rangle \rangle_\text{FF}\notag\\[5pt]&\hspace{1cm}+\gamma_{211}k_{1-}\langle \langle T_{--}(k_1)J_-(k_2) J_-(k_3) \rangle \rangle_\text{odd}\bigg ]+1\leftrightarrow 2+1\leftrightarrow 3 \notag\\[5pt]&
=\frac{16\widetilde{\lambda}_{\text{qf}}}{\widetilde{N}(1+\widetilde{\lambda}_{\text{qf}}^2)}\bigg[\beta_{110}k_{1-} \langle \langle J_0(k_1) J_{-}(k_2) J_-(k_3) \rangle \rangle_\text{FF}\langle \langle J_-(k_1) J_-(-k_1) \rangle \rangle_\text{QF}\notag\\[5pt]&+\gamma_{110}k_{1-} \langle \langle J_0(k_1) J_{-}(k_2) J_-(k_3) \rangle \rangle_\text{odd}\langle \langle J_-(k_1) J_-(-k_1) \rangle \rangle_\text{QF}\bigg]+1\leftrightarrow 2+1\leftrightarrow 3
\end{align}

The free theory equation \eqref{TJJFF} implies that terms proportional to $\beta_{211}$ add up to zero. Using \eqref{TJJFB} terms proportional to $\alpha_{211}$ simplify and we get :
\begin{align}
\label{HSEintQ3JJT}
-2\bigg[-\frac{\alpha_{211}}{8 } k_{1-}^{3}  \langle \langle J_{0}(k_1) J_-(k_2)J_-(k_3)\rangle \rangle_\text{FB}+1\leftrightarrow 2+1\leftrightarrow 3\notag\\[5pt]
+\gamma_{211}k_{1-}\langle \langle T_{--}(k_1)J_-(k_2) J_-(k_3) \rangle \rangle_\text{odd}+1\leftrightarrow 2+1\leftrightarrow 3 \bigg ]\notag\\[5pt]
=\frac{16\widetilde{\lambda}_{\text{qf}}}{\widetilde{N}(1+\widetilde{\lambda}_{\text{qf}}^2)} \bigg[\beta_{110}k_{1-} \langle \langle J_0(k_1) J_{-}(k_2) J_-(k_3) \rangle \rangle_\text{FF}\langle \langle J_-(k_1) J_-(-k_1) \rangle \rangle_\text{QF}\notag\\[5pt]
+\gamma_{110}k_{1-} \langle \langle J_0(k_1) J_{-}(k_2) J_-(k_3) \rangle \rangle_\text{odd}\langle J_-(k_1) J_-(-k_1)  \rangle_\text{QF}\bigg]+1\leftrightarrow 2+1\leftrightarrow 3
\end{align}
We now separate the above equation into terms that contain parity even and parity odd structures. The even part gives the following equation :
\begin{align}
&\bigg[\frac{\alpha_{211}}{4} k_{1-}^{3}  \langle \langle J_{0}(k_1) J_-(k_2)J_-(k_3)\rangle \rangle_\text{FB}\bigg]+1\leftrightarrow 2+1\leftrightarrow 3\notag\\
&=\frac{16\widetilde{\lambda}_{\text{qf}}}{\widetilde{N}(1+\widetilde{\lambda}_{\text{qf}}^2)} \bigg[\gamma_{110}k_{1-} \langle \langle J_0(k_1) J_{-}(k_2) J_-(k_3) \rangle \rangle_\text{odd}\langle J_-(k_1) J_-(-k_1)  \rangle_\text{FF}\bigg]+1\leftrightarrow 2+1\leftrightarrow 3
\end{align}
We verified that the free theory correlators derived in \cite{Jain:2020rmw} and our ansatz for the odd part of $\langle J_1 J_1J_0\rangle_\text{QF}$ \eqref{JJJ0QFodd} solve the above equation.
%
The odd part of \eqref{HSEintQ3JJT} gives :
\begin{align}\label{QFOddE}
&-2\bigg[\gamma_{211}k_{1-}\langle \langle T_{--}(k_1)J_-(k_2) J_-(k_3) \rangle \rangle_\text{odd} \bigg ]+1\leftrightarrow 2+1\leftrightarrow 3\notag\\&
=\frac{16\widetilde{\lambda}_{\text{qf}}}{\widetilde{N}(1+\widetilde{\lambda}_{\text{qf}}^2)}\,\bigg[\beta_{110}k_{1-} \langle \langle J_0(k_1) J_{-}(k_2) J_-(k_3) \rangle \rangle_\text{FF}\langle J_-(k_1) J_-(-k_1)  \rangle_\text{FF} \bigg ]+1\leftrightarrow 2+1\leftrightarrow 3
\end{align}
The above higher spin equation which involves the odd part of the $\langle T J J\rangle $ correlator leads to five equations for the form factors, one from each of the momentum coefficients. We will address the parity odd structures to this correlator in momentum space in future.

\subsubsection{Constraining $\boldsymbol{\langle J_1J_1J_0J_0\rangle_\text{QF}^\text{odd}}$}

In this subsection we attempt to constrain the parity odd structure  of the correlator $\langle J_1J_1J_0J_0\rangle_\text{QF}$. We consider the action of $Q_3$ on $\langle J_-J_0J_0J_0\rangle_\text{QF}$ :
\begin{align}
\langle& \left[Q_{3}, J_{-}(x_1)\right]J_0(x_2)J_0(x_3)J_0(x_4)\rangle_\text{QF}+\langle  J_{-}(x_1)\left[Q_{3}, J_{0}(x_2)\right]J_0(x_3)J_0(x_4)\rangle_\text{QF}+2\leftrightarrow 3+2\leftrightarrow 4\notag\\&\hspace{5cm}=\int_x \langle \partial_{\mu}J^{\mu}_{--}(x)J_-(x_1)J_0(x_2)J_0(x_3)J_0(x_4)\rangle_\text{QF}
\end{align}
Upon utilizing the algebra in \eqref{QFQ3algebra}, the current equation \eqref{DivJ3AF} and large $N$ factorization, we get after Fourier transforming  :
\begin{align}
-2&\,k_{1-} \langle\langle T_{--}(k_1)J_0(k_2)J_0(k_3)J_0(k_4)\rangle\rangle_\text{QF}\notag\\[5pt]&-\epsilon_{-k_2 \nu}k_{2-}\langle \langle  J_{-}(k_1)  J^{\nu}(k_2)J_0(k_3)J_0(k_4)\rangle\rangle_\text{QF} +2\leftrightarrow 3+2\leftrightarrow 4\notag\\[5pt]&=\frac{16\,\widetilde{\lambda}_{\text{qf}}}{\widetilde N(1+\widetilde{\lambda}_{\text{qf}}^2)} \bigg[k_{1-}\langle J_-(k_1)J_{-}(-k_1)\rangle_\text{QF}\langle\langle J_0(k_1)J_0(k_2)J_0(k_3)J_0(k_4)\rangle\rangle_\text{QF}\notag\\[5pt]&-(k_{3-}+k_{4-})\langle\langle J_-(-k_1-k_2)J_{-}(k_1)J_0(k_2)\rangle\rangle_\text{QF}\langle J_0(-k_3-k_4)J_0(k_3)J_0(k_4)\rangle_\text{QF}+2\leftrightarrow 3+2\leftrightarrow 4\notag\\[5pt]&
-k_{2-}\langle\langle J_-(k_1)J_{-}(k_2)J_0(k_3)J_0(k_4)\rangle \rangle_\text{QF}\langle \langle J_0(k_2)J_0(-k_2)\rangle_\text{QF}+2\leftrightarrow 3+2\leftrightarrow 4\bigg]
\end{align}
From recent results \cite{Li:2019twz} and our analysis in Section \ref{Tj0j0j0QF} we know that the correlator $\langle TJ_0J_0J_0\rangle_\text{QF}$ appearing in above equation has two structures given by \eqref{J24ptqf} and \eqref{ansatztj0j0j0QF}. The four-point function of the scalar operator in terms of the corresponding free theory correlator is as in \eqref{J04ptqf}.

Let us consider the following ansatz for the correlator $\langle J_1\,J_1\,J_0\,J_0\rangle$ : 
\begin{align}
\langle\langle& J_{\mu}(k_1)J_\nu(k_2)J_0(k_3)J_0(k_4)\rangle\rangle_\text{QF}\notag\\[5pt]&= \alpha_{1100}   \langle\langle J_{\mu}(k_1)J_\nu(k_2)J_0(k_3)J_0(k_4)\rangle\rangle_\text{FF}+\beta_{1100}   \langle\langle J_{\mu}(k_1)J_\nu(k_2)J_0(k_3)J_0(k_4)\rangle\rangle_\text{FB}\notag\\[5pt]&+\gamma_{1100}   \langle\langle J_{\mu}(k_1)J_\nu(k_2)J_0(k_3)J_0(k_4)\rangle\rangle_\text{odd}
\end{align}
We substitute this ansatz for $\langle J_1 J_1J_0J_0\rangle$ and separate the equations arising from parity odd and even structures. The parity even structures lead to the following equation :
\begin{align}
-2\,&\beta_{2000}k_{1-} \langle\langle T_{\mu\nu}(k_1)J_0(k_2)J_0(k_3)J_0(k_4)\rangle\rangle_\text{CB}\notag\\[5pt]
&\hspace{-.5cm}-\epsilon_{-k_2 \nu}(k_{2-} \gamma_{1100}\langle\langle  J_{-}(k_1)  J^{\nu}(k_2)J_0(k_3)J_0(k_4)\rangle\rangle_\text{odd}) +2\leftrightarrow 3+2\leftrightarrow 4\notag\\[5pt]
&\hspace{-.5cm}=\frac{16\,\widetilde{\lambda}_{\text{qf}}}{\widetilde N(1+\widetilde{\lambda}_{\text{qf}}^2)}\bigg[ \beta_{0000}k_{1-}\langle J_-(k_1)J_{-}(-k_1)\rangle_\text{QF}\langle\langle J_0(k_1)J_0(k_2)J_0(k_3)J_0(k_4)\rangle\rangle_\text{FF}\notag\\[5pt]
&\hspace{2cm}-\alpha_{1100}k_{2-}\langle\langle J_-(k_1)J_{-}(k_2)J_0(k_3)J_0(k_4)\rangle\rangle_\text{FF} \langle J_0(k_2)J_0(-k_2)\rangle_\text{QF}\notag\\[5pt]
&\hspace{-.5cm}-\beta_{1100}k_{2-}\langle\langle J_-(k_1)J_{-}(k_2)J_0(k_3)J_0(k_4)\rangle\rangle_\text{FB} \langle J_0(k_2)J_0(-k_2)\rangle_\text{QF}+2\leftrightarrow 3+2\leftrightarrow 4\bigg]
\end{align}
whereas the odd structures give the following equation :
\begin{align}
&-2\,\alpha_{2000} k_{1-} \langle\langle T_{--}(k_1)J_0(k_2)J_0(k_3)J_0(k_4)\rangle\rangle_\text{FF}\notag\\[5pt]&-\epsilon_{-k_2 \nu}(k_{2-} \alpha_{1100}\langle\langle  J_{-}(k_1)  J^{\nu}(k_2)J_0(k_3)J_0(k_4)\rangle\rangle_\text{FF}+k_{2-} \beta_{1100}\langle\langle  J_{-}(k_1)  J^{\nu}(k_2)J_0(k_3)J_0(k_4)\rangle\rangle_\text{FB})\notag\\[5pt]&\hspace{-1cm}=\frac{16\,\widetilde{\lambda}_{\text{qf}}}{\widetilde N(1+\widetilde{\lambda}_{\text{qf}}^2)} \bigg[ -(k_{3-}+k_{4-})\langle J_-(-k_1-k_2)J_{-}(k_1)J_0(k_2)\rangle_\text{QF}\langle\langle J_0(-k_3-k_4)J_0(k_3)J_0(k_4)\rangle\rangle_\text{QF}+2\leftrightarrow 3+2\leftrightarrow 4\notag\\[5pt]&
-\gamma_{1100}k_{2-}\langle\langle J_-(k_1)J_{-}(k_2)J_0(k_3)J_0(k_4)\rangle\rangle_\text{odd} \langle J_0(k_2)J_0(-k_2)\rangle_\text{QF}+2\leftrightarrow 3+2\leftrightarrow 4\bigg]
\end{align}
Thus we obtain an equation for $\langle J_1 J_1 J_0 J_0\rangle_\text{odd}$ in terms of free theory correlators and $\langle J_1J_1J_0\rangle_{\text{QF}}$ which we studied in detail in Section \ref{3ptQFsection}.
We leave the study of finding the form factors of $\langle J_\mu J_\nu J_0 J_0\rangle_\text{odd}$ to a future work.

\subsection{Quasi-bosonic theory}
As discussed in the Appendix \ref{notation}, it was shown in \cite{Maldacena:2012sf} that a correlator $\langle J_{s_1}J_{s_2}J_{s_3}\rangle$ with $s_1\neq0,s_2\neq0,s_3\neq0$ in a theory with slightly broken higher spin symmetry  has three non-trivial structure the free bosonic, the free ferimionic and the parity odd structure.
In this subsection we derive the higher spin equations involving such a correlator  $\langle TJ_1J_1\rangle_\text{QB}$  and in particular we attempt to solve  for the parity odd structures in this correlator . 

\subsubsection{Constraining $\boldsymbol{\langle TJ_1J_1\rangle_\text{QB}^{\text{odd}}}$ }

Consider the action of $Q_3$ on $\langle J_- J_- J_-\rangle$. This gives :
\begin{align}
\langle&[Q_{3}, J_{-}(x_1)]J_-(x_2)J_-(x_3)]\rangle+1\leftrightarrow 2+1\leftrightarrow 3=\int_x \langle \partial_{\mu}J_{--}^{\mu}(x) J_-(x_1)J_-(x_2)J_-(x_3)\rangle
\end{align}
Utilizing the algebra \eqref{QBQ3algebra}, the current equation \eqref{DivJ3QB} and large $N$ factorization, we obtain after Fourier transforming :
\begin{align}
\label{JJToddQBapp}
& 4 k_{1-}\langle\langle   T_{--}(k_1)J_-(k_2) J_-(k_3)\rangle\rangle_\text{QB}+\frac{k_{1-}^3}{2}\langle\langle J_{0}(k_1)J_-(k_2) J_-(k_3)\rangle\rangle_\text{QB}+1\leftrightarrow 2+1\leftrightarrow 3 \notag\\[5pt]&=\frac{32\,\widetilde\lambda_{\text{qb}}}{\widetilde{N}(1+\widetilde\lambda_{\text{qb}}^2)}\bigg[\epsilon_{\mu k_1-} k_{1-}\langle\langle J_0(k_1)J_-(k_2)J_-(k_3) \rangle\rangle_\text{QB}\langle J_-(k_1)J^{\mu}(-k_1)\rangle_\text{QB}+1\leftrightarrow 2+1\leftrightarrow 3 \bigg ]
\end{align}
We know from the results of \cite{Maldacena:2012sf}, that these correlators have the following structure :
\begin{align}
\langle\langle J_0(k_1)J_-(k_2)J_-(k_3)\rangle\rangle_\text{QB}&=\alpha_{110}\langle\langle J_0(k_1)J_-(k_2)J_-(k_3)\rangle\rangle_\text{FB}+\gamma_{110}\langle\langle J_0(k_1)J_-(k_2)J_-(k_3)\rangle\rangle_{\text{QB}}^{\text{odd}}\notag\\[7pt]
\langle\langle   T_{--}(k_1)J_-(k_2) J_-(k_3)\rangle\rangle_\text{QB}&=\alpha_{211}\langle\langle   T_{--}(k_1)J_-(k_2) J_-(k_3)\rangle\rangle_\text{FB}+\beta_{211} \langle\langle T_{--}(k_1)J_-(k_2) J_-(k_3) \rangle\rangle_\text{FF}\notag\\[7pt]&\hspace{.5cm}+\gamma_{211}\langle\langle T_{--}(k_1)J_-(k_2) J_-(k_3) \rangle\rangle_\text{QB}^\text{odd}\notag\\[7pt]
\langle J_-(k_1)J_\mu(-k_1)\rangle_\text{QB}&=\alpha_{11}\langle J_-(k_1)J_\mu(-k_1)\rangle_\text{FB}+\langle J_-(k_1)J_\mu(-k_1)\rangle_\text{QB}^{\text{contact}}\notag\\[7pt]
\langle J_-(k_1)J_\mu(-k_1)\rangle_\text{QB}^{\text{contact}}&=\alpha_{11} d_{123}\epsilon_{-\mu k_1}
\end{align}
\begin{align}
\alpha_{211}=\frac{\widetilde{N}}{2(1+\widetilde{\lambda}_{\text{qb}}^2)},\quad\beta_{211}=\frac{\widetilde{N}\widetilde{\lambda}_{\text{qb}}^2}{2(1+\widetilde{\lambda}_{\text{qb}}^2)},\quad \gamma_{211}=\frac{\widetilde{N}\widetilde{\lambda}_{\text{qb}}}{2(1+\widetilde{\lambda}_{\text{qb}}^2)}\label{TJJQBabg}\quad
\alpha_{11}=\frac{\widetilde{N}}{2}
\end{align}
Substituting the above expressions in \eqref{JJToddQBapp} and then utilizing the higher spin equations for free fermion and free boson theories derived in \eqref{TJJFB} and \eqref{TJJFF} gives :
\begin{align}
&4 \bigg[- \frac{\alpha_{211}}{8 d_{01}} k_{1-}^{3}  \langle\langle J_{0}(k_1) J_-(k_2)J_-(k_3)\rangle\rangle_\text{FB}+\gamma_{211} k_{1-}\langle\langle T_{--}(k_1)J_-(k_2) J_-(k_3) \rangle\rangle_\text{QB}^\text{odd}+1\leftrightarrow 2+1\leftrightarrow 3\bigg] \notag\\[7pt]&+\frac{1}{2}\bigg[\alpha_{110}k_{1-}^3\langle\langle J_0(k_1)J_-(k_2)J_-(k_3)\rangle\rangle_\text{FB}+\gamma_{110}k_{1-}^3\langle\langle J_{\mu}(k_1)J_\nu(k_2)J_0(k_3)\rangle\rangle_{\text{QB}}^\text{odd}+1\leftrightarrow 2+1\leftrightarrow 3\bigg] \notag\\[7pt]&=\frac{32\,\widetilde\lambda_{\text{qb}}}{\widetilde{N}(1+\widetilde\lambda_{\text{qb}}^2)}\bigg[\alpha_{110}~\epsilon_{\mu k_1-}k_{1-}\langle\langle J_0(k_1)J_-(k_2)J_-(k_3)\rangle\rangle_\text{FB}\langle J_-(k_1)J^{\mu}(-k_1)\rangle_\text{QB}\notag\\[7pt]&\hspace{2.5cm}+\gamma_{110}~\epsilon_{\mu k_1-}k_{1-}\langle\langle J_{0}(k_1)J_-(k_2)J_-(k_3)\rangle\rangle_{\text{QB}}^\text{odd}\langle J_-(k_1)J^{\mu}(-k_1)\rangle_\text{QB}+1\leftrightarrow 2+1\leftrightarrow 3 \bigg ]
\end{align}
The  equation splits into two equations : one for the parity odd and the other for the parity even structures. The parity even part gives :
\begin{align}
&- \frac{\alpha_{211}}{2 }  \bigg[k_{1-}^{3}  \langle\langle J_{0}(k_1) J_-(k_2)J_-(k_3)\rangle\rangle_\text{FB}+1\leftrightarrow 2+1\leftrightarrow 3\bigg] \notag\\[5pt]&+\frac{\alpha_{110}}{2}\bigg[k_{1-}^3\langle\langle J_0(k_1)J_-(k_2)J_-(k_3)\rangle\rangle_\text{FB}+1\leftrightarrow 2+1\leftrightarrow 3\bigg]\notag\\[5pt]&
=\frac{32\,\widetilde\lambda_{\text{qb}}}{\widetilde{N}(1+\widetilde\lambda_{\text{qb}}^2)}\bigg[~\alpha_{110}\epsilon_{\mu k_1-}k_{1-}\langle\langle J_0(k_1)J_-(k_2)J_-(k_3)\rangle\rangle_\text{FB}\langle J_-(k_1)J^{\mu}(-k_1)\rangle_\text{QB}^{\text{contact}}+1\leftrightarrow 2+1\leftrightarrow 3 \bigg ]
\end{align}
where $\alpha_{211}$ and $\alpha_{110}$ are as given in \eqref{TJJQBabg}. Note that the term proportional to $\epsilon_{\mu k_1-}\langle J_-(k_1)J^{\mu}(-k_1)\rangle_\text{FB}$ and their corresponding exchanges on the RHS simply vanish due to contraction.
Upon utilizing the results for the free boson correlator, we have verified that the above higher spin equation is exactly satisfied. 

The parity odd part of the higher spin equation gives :
\begin{align}\label{QBOddE}
4&\gamma_{211}  \bigg[k_{1-}\langle\langle T_{--}(k_1)J_-(k_2) J_-(k_3) \rangle\rangle_\text{QB}^\text{odd}+1\leftrightarrow 2+1\leftrightarrow 3\bigg]\notag\\[5pt]&+\frac{\gamma_{110}}{2}\bigg[k_{1-}^3\langle\langle J_{\mu}(k_1)J_\nu(k_2)J_0(k_3)\rangle\rangle_{\text{QB}}^\text{odd}+1\leftrightarrow 2+1\leftrightarrow 3\bigg]\notag\\[5pt]&=\frac{32\widetilde\lambda_{\text{qb}}}{\widetilde{N}(1+\widetilde\lambda_{\text{qb}}^2)}\bigg[\gamma_{110}~\epsilon_{\mu k_1-}k_{1-}\langle\langle J_{\mu}(k_1)J_\nu(k_2)J_0(k_3)\rangle\rangle_{\text{QB}}^\text{odd}\langle J_-(k_1)J^{\mu}(-k_1)\rangle_\text{QB}^{\text{contact}}+1\leftrightarrow 2+1\leftrightarrow 3 \bigg ]
\end{align}  
We will address the problem of  solving for the parity odd structures in this correlator in a future work.

\section{Correlation functions in  free massive bosonic theory}
\label{massive-app}
In this section of the appendix we provide brief details of the computation of some of the correlators in the free massive bosonic theory and the integrals  required to evaluate them.
\subsubsection*{Required Integrals}
Note that we will use the following identities to derive the required integrals :
\begin{align}
\widetilde{I}_{\mu}&=\int_l \frac{l_\mu}{ ((l-k)^2+m^2)}=D k_{\mu} \\
\widetilde{S}_{\mu}&=\int_l \frac{l_\mu}{ (l^2+m^2)}=0\\
\widetilde{I}_0&=\int_l \frac{1}{((l-k)^2+m^2)}=\int \frac{1}{(l^2+m^2) }\\
I_{\mu\nu}&=\int_l\frac{l_\mu l_\nu}{(l^2+m^2) ((l-k)^2+m^2)}=A k_{\mu} k_{\nu}+B g_{\mu \nu}\\
I_{\mu}&=\int_l\frac{l_\mu}{(l^2+m^2) ((l-k)^2+m^2)}=C k_\nu\\
I_0&=\int_l\frac{1}{(l^2+m^2) ((l-k)^2+m^2)}=\frac{\arctan\left(\frac{k}{2m}\right)}{4\pi k}
\end{align}
where we introduced the following notation
\begin{align}
 \int_l\equiv\frac{\int d^3l}{(2\pi)^3}\,,
 \end{align}
and
\begin{align}
D&=\widetilde{I}_0=-\frac{m}{4\pi}\\
C&=\frac{I_0}{2}\\
A&=\frac{2 \widetilde{I}_0+(3 k^2+m^2)I_0}{8 k^2}\\
B&=\frac{2 \widetilde{I}_0-( k^2+4m^2)I_0}{8 k^2}
\end{align}
\subsubsection*{$\boldsymbol{\langle J_0 \rangle_{\text{FMB}}}$}

\begin{align}
\langle J_{0}(k) \rangle& =\int_{l}\frac{1}{(l^2+m^2)  }=\widetilde{I}_0
\end{align}

\subsubsection*{$\boldsymbol{\langle J_0J_0 \rangle_{\text{FMB}}}$}

\begin{align}
\langle J_{0}(k) J_0(-k)& =\int_{l}\frac{1}{(l^2+m^2) ((l-k)^2+m^2) }=I_0
\end{align}

\subsubsection*{$\boldsymbol{\langle J_{\mu}J_{\nu}\rangle_{\text{FMB}}}$}
\begin{align}
J_{\mu}(k)=\int d^{3} l\,(2 l-k)_{\mu}\,\bar{\phi}(l)\,\phi(k-l)
\end{align}
Hence the Wick contraction gives :
\begin{align}
\langle J_{\mu}(k_1) J_\nu(-k_1)  \rangle&=\int_{l_1 l_2}\frac{(2l_{1\mu}-k_{1\mu})(2l_{2\mu}-k_{2\mu})}{(l_1^2+m^2)(l_2^2+m^2)}\delta(l_1+k_2-l_2)\delta(l_2+k_1-l_1)
\end{align}
This in turn is given by
\begin{align}
\langle J_{\mu}(k) J_\nu(-k)\rangle&=\int_{l}\frac{(2l_\mu-k_\mu)(2l_\nu-k_\nu)}{(l^2+m^2) ((l-k)^2+m^2) }\notag\\[5pt]
&=4\,I_{\mu\nu}-2\,k_\mu\,I_\nu-2\,k_\nu\,l_\mu+k_\mu\,k_\nu\,I_0
\end{align}
\subsubsection*{$\boldsymbol{\langle T_{\mu\nu}J_0}\rangle_{\text{FMB}}$}
\begin{align}
T_{\mu\nu}(k)=&\int d^3l H_{\mu \nu}(l,k)\,\bar{\phi}(l)\,\phi(k-l)\notag\\[5pt]
H_{\mu\nu}=&-\frac{3}{8}(l_{\mu}(k-l)_{\nu}+(k-l)_{\mu}l_{\nu})+\frac{1}{8}((k-l)_{\mu}(k-l)_{\nu}+l_{\mu}l_{\nu})\notag\\[5pt]&+\frac{1}{4}\eta_{\mu \nu}\,l\cdot (k-l)-\frac{\eta_{\mu\nu}}{24}(l^2+(k-l)^2)+\eta_{\mu \nu}\frac{m^2}{3}
\end{align}
Hence the correlator is given as
\begin{align}
\langle T_{\mu\nu}(k_1) J_0(k_2)  \rangle\notag=\int_{l_1l_2}\langle\bar{\phi}(l_1) H_{\mu\nu}\phi(k_1-l_1)\bar{\phi}(l_2)\phi(k_2-l_2)\rangle\notag
\end{align}
The relevant Wick contractions $1\bar{2}~2\bar{3}~3\bar{1}$ and  $1\bar{3}~3\bar{2}~2\bar{1}$ together give :
\begin{align}
\langle T_{\mu\nu}(k) J_0(-k) \rangle&=\frac{m}{48k^3\pi}\left[6g_{\mu\nu}k^3-3k k_\mu k_\nu+(7g_{\mu\nu}k^2+6k_{\mu}k_{\nu})m\,\text{arctan}\left(\frac{k}{2m}\right)\right]
\end{align}
In particular 
\begin{align}
\langle T_{--}(k) J_0(-k)  \rangle
=-\frac{k_{-}^2m\left(k-2m\,\text{arctan}\left(\frac{k}{2m}\right)\right)}{16k^3\pi}
\end{align}
%

\bibliographystyle{JHEP}

\end{document}